\documentclass[pre,showpacs,floatfix,onecolumn,superscriptaddress]{revtex4}

\usepackage{graphicx}
\usepackage{subfigure}
\usepackage{amsmath}
\usepackage{amssymb}
\usepackage{latexsym}
\usepackage{multirow}
\usepackage{color}
\usepackage[hidelinks]{hyperref}
\usepackage{mathtools}
\usepackage{comment}
\usepackage{ulem}
\usepackage{ragged2e}

\graphicspath{{./figs/}}

\let\cosh\relax
\DeclareMathOperator\cosh{cosh}
\let\sinh\relax
\DeclareMathOperator\sinh{sinh}

\begin{document}

\title{Narrow escape problem in two-shell spherical domains}

\author{Matthieu Mangeat}
\email{mangeat@lusi.uni-sb.de}

\author{Heiko Rieger}
\email{h.rieger@mx.uni-saarland.de}

\affiliation{Center for Biophysics \& Department for Theoretical Physics, Saarland University, D-66123 Saarbr{\"u}cken, Germany.}

\begin{abstract}
Intracellular transport in living cells is often spatially inhomogeneous with an accelerated effective diffusion close to the cell membrane and a ballistic motion away from the centrosome due to active transport along actin filaments and microtubules, respectively. Recently it was reported that the mean first passage time (MFPT) for transport to a specific area on the cell membrane is minimal for an optimal actin cortex width. In this paper we ask whether this optimization in a two-compartment domain can also be achieved by passive Brownian particles. We consider a Brownian motion with different diffusion constants in the two shells and a potential barrier between the two and investigate the narrow escape problem by calculating the MFPT for Brownian particles to reach a small window on the external boundary. In two and three dimensions, we derive asymptotic expressions for the MFPT in the thin cortex and small escape region limits confirmed by numerical calculations of the MFPT using the finite element method and stochastic simulations. From this analytical and numeric analysis we finally extract the dependence of the MFPT on the ratio of diffusion constants, the potential barrier height and the width of the outer shell. The first two are monotonous whereas the last one may have a minimum for a sufficiently attractive cortex, for which we propose an analytical expression of the potential barrier height matching very well the numerical predictions.
\end{abstract}

\maketitle

%%%%%%%%%%%%%%%%%%%%
%%%%%%%%%%%%%%%%%%%%
%%% Introduction %%%
%%%%%%%%%%%%%%%%%%%%
%%%%%%%%%%%%%%%%%%%%

\section{Introduction}
\label{section0}

The intracellular transport of various cargo-particles (for example proteins, vesicles, mitochondria) towards a specific area on the cell membrane, like the immunological synapse in T cells and natural killer cells~\cite{janeway1997}, is crucial for the correct functioning of cells and organisms. The cytoskeleton mediated cargo transport in many cells with a centrosome can schematically be described by a two-shell geometry with an inner sphere containing microtubules radiating from a microtubule organizing center (MTOC, involving the centrosome), and an outer shell containing the actin cortex, a thin dense network of actin filaments underneath the cell membrane~\cite{alberts2014}. Molecular motor driven cargo transport along the cytoskeleton filaments is achieved by dynein (from the centrosome to the cortex) and kinesin (from the cortex to the centrosome) on the microtubules, and myosin on the actin filaments. The stochastic attachment and detachment of molecular motors to and from the filaments combined with intermittent diffusion of the detached motors can mathematically be described by an intermittent search process~\cite{loverdo2008, benichou2011a}, alternating randomly between a cytoplasmic diffusive transport and a ballistic transport on filaments~\cite{bressloff2013}. Recently, it was shown the transport time to reach a specific area, the escape area, on the cell membrane by this intermittent search can be minimized with respect to the cortex width~\cite{schwarz2016a, schwarz2016b, hafner2016, hafner2018} due to an accelerated effective diffusion close to the cell membrane and the efficiency of the intermittent search. For small typical size $\varepsilon$ of the escape area, this intracellular transport constitutes the narrow escape problem (NEP) in a spatially inhomogeneous environment.

NEPs are widely studied in various biological and chemical contexts~\cite{schuss2007, bressloff2013, chou2014, holcman2014, iyerbiswas2015}, and consist in calculating the mean-first passage time (MFPT) of a Brownian motion to reach a small escape window from a given starting point~${\bf x}$, denoted hereinafter $t({\bf x})$. This time is also called narrow escape time (NET) in the limit of a small escape region (here $\varepsilon \ll 1$). Mathematically the NEP for non-interacting particles reduces to solve the spatial differential equation written as ${\cal L}^\dagger t({\bf x}) = -1$, where ${\cal L}^\dagger$ is the conjugate of Fokker-Planck operator~\cite{redner2001}, with mixed Neumann-Dirichlet boundary conditions (absorbing on escape region and reflecting elsewhere). Numerical estimates of the MFPT can be obtained by either using the finite element method to solve the partial differential equation (PDE)~\cite{zienkiewicz1977, hecht2013} or by simulating the stochastic process with Kinetic Monte Carlo algorithms~\cite{oppelstrup2009, schwarz2013}, for any studied geometry.

Analytically, the NEP has been widely scrutinized over the last decades for several types of geometry and stochastic processes~\cite{singer2006, condamin2007, benichou2008, chevalier2011}. In two dimensions, the leading order of the MFPT is proportional to $\ln \varepsilon$ whereas in three dimensions, this leading order is proportional to $\varepsilon^{-1}$. These leading orders are derived from a Poisson's equation with a singular source term, presenting the same solution than the electric field for an electron. The sub-leading corrections play an important role to analyze the difference between the starting points having the same leading order~\cite{ward1993, pillay2010, cheviakov2010, cheviakov2012, gomez2015}. Those corrections depend generally on the Green's function of the closed domain~\cite{barton1989}. In particular, we can mention that asymptotic expressions of the NET have been developed for homogeneously diffusive particles for the disk geometry~\cite{singer2006, pillay2010} and for the sphere geometry~\cite{cheviakov2010, cheviakov2012, gomez2015} for many small escape regions ($\varepsilon \ll 1$). An exact solution of the MFPT was recently derived from complex analysis for the disk geometry~\cite{caginalp2012} for one escape region. Moreover, another exact solution of the equation $\nabla^2 t({\bf x}) = -1/D({\bf x})$ has been calculated for an arbitrary simply connected planar domain using a conformal mapping onto the unit disk~\cite{grebenkov2016}. The NEP has also been widely studied for the surface-mediated diffusion problem~\cite{benichou2010, benichou2011b, calandre2012, rupprecht2012a, rupprecht2012b} representing a very thin cortex in the biological context mentioned above. It was shown that the MFPT can be optimized with respect to absorption and desorption rates of Brownian particles on the surface. Additionally the MFPT has been analyzed in presence of radially attractive potentials pushing the particles to the surface~\cite{grebenkov2017} leading to a MFPT optimization due to a competition between bulk and surface events. Finally the first exit time distribution has been investigated for several two-dimensional geometries~\cite{rupprecht2015} or for the sphere geometry~\cite{grebenkov2019} for which it was shown recently that the typical first passage time may be shorter than the MFPT by several decades. This time distribution has also been studied in presence of logarithmic potentials~\cite{ryabov2015}.

In this article we ask whether the MFPT shows a non-monotonous, and hence optimizable, dependence on the width if the outer shell (the actin cortex width in the biological context mentioned above) when the ballistic transport on filaments is neglected, i.e. for passive Brownian particles. Within the two-shell geometry, we consider a Brownian motion in each shell with different diffusion constants presenting a potential barrier between the two shells. We recently studied the impact of a heterogeneous diffusivity in Ref.~\cite{mangeat2019} showing that without potential barrier no optimization of the MFPT can be observed, which motivated the study presented in this article. The mathematical model will be defined in Sec.~\ref{sectionModel}. In Sec.~\ref{sectionThinCortex} we will present the limit of the model for thin cortex. In Sec.~\ref{sectionNumerics} we will present our numerical estimates of the MFPT obtained by the numerical solutions of the PDE and by stochastic simulations. We will also investigate the dependence of the MFPT on all parameters by a qualitative analysis. In Sec.~\ref{sectionNE} we will derive an asymptotic expression of the NET and we will compare it with the numerical solutions. From these expressions we will then derive a condition for the potential barrier height to observe a MFPT optimization. Finally Sec.~\ref{sectionDiscussion} will conclude with a discussion about the results and an outlook.

%%%%%%%%%%%%%%%%%
%%%%%%%%%%%%%%%%%
%%% THE MODEL %%%
%%%%%%%%%%%%%%%%%
%%%%%%%%%%%%%%%%%

\section{The model}
\label{sectionModel}

%%%%%%%%%%%%%%%%%%%%
%%% FIGURE MODEL %%%
%%%%%%%%%%%%%%%%%%%%

\begin{figure}[t]
\begin{center}
\includegraphics[width=12cm]{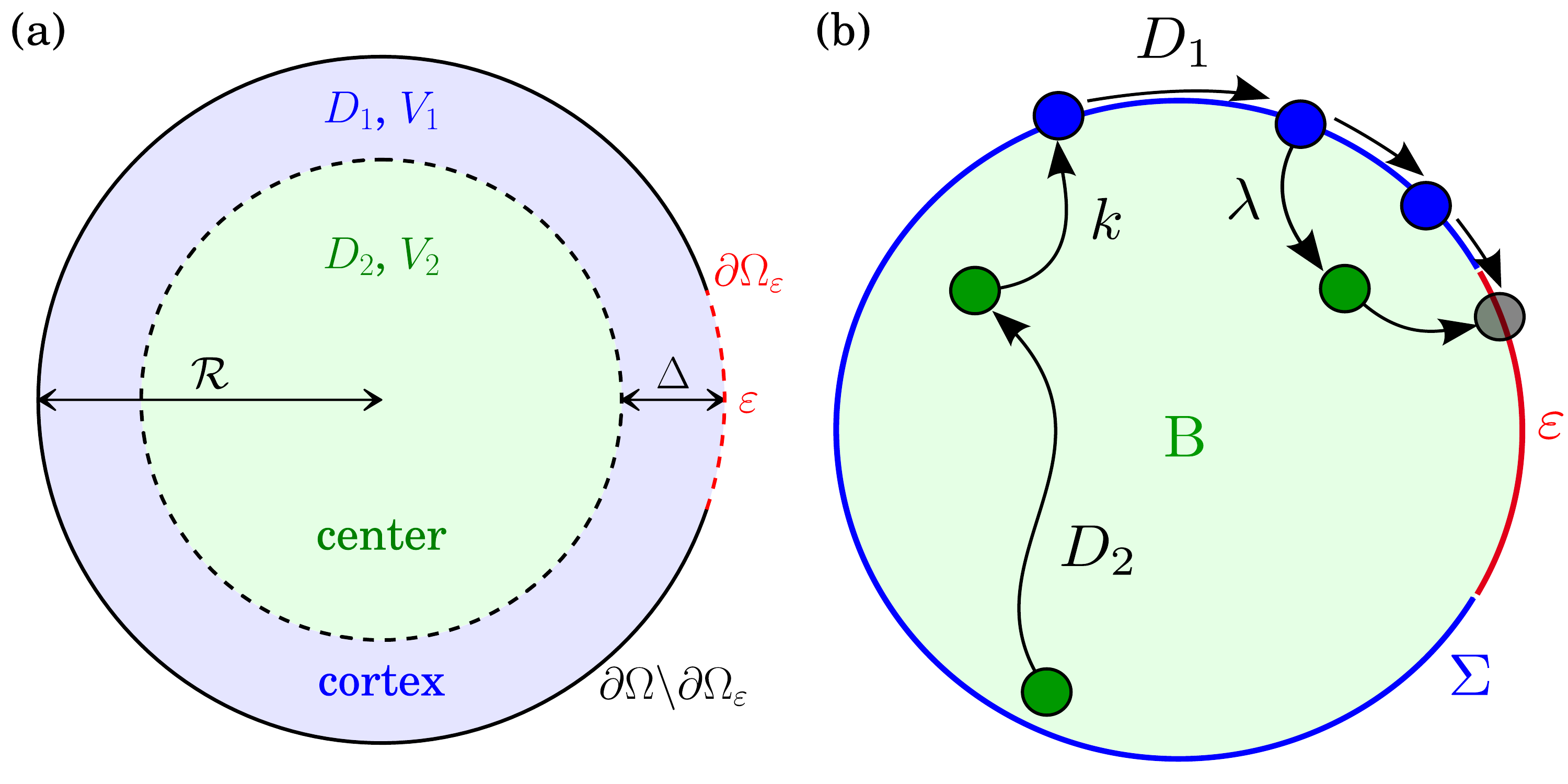}
\caption{{\bf (a)}~Schematic of the two-shell geometry with piecewise diffusivity and potential, respectively equal to $D_1$ and $V_1$ in the outer shell (cortex) and $D_2$ and $V_2$ in the inner shell (center). The width of the outer shell is $\Delta$ for a domain radius ${\cal R}$ while the escape angle is $\varepsilon$. {\bf (b)}~Schematic of the SMDP with surface and bulk diffusion constants $D_1$ and $D_2$. The desorption and absorption rate on the surface are respectively $\lambda$ and $k$ while the escape region angle is $\varepsilon$.} \label{figure_model}
\end{center}
\end{figure}

In this article, we study the NEP in the two-shell geometry sketched in Fig.~\ref{figure_model}(a), which corresponds to a disk/sphere of radius ${\cal R}$ constituted by two concentric regions in which the particles diffuse with different diffusion constants and potentials in each shell. The outer shell (or cortex) corresponds to the actin cortex of the cytoskeleton mentioned in the introduction with a width $\Delta$ while the inner shell (or central region) represents the area of microtubules with a radius ${\cal R}-\Delta$. The diffusion constants and the potentials are denoted as $D_1$ and $V_1$, respectively, in the cortex and $D_2$ and $V_2$ in the central region. The particles experience a potential barrier of height $\Delta V=V_2-V_1$ when transiting from the outer to the inner shell. The external boundary at the radius $|{\bf x}|= {\cal R}$ is denoted as $\partial \Omega$ and represents the cell membrane. The escape window, representing e.g. the synapse (see introduction), is a circular (spherical) arc on this external boundary with an apparent angle $\varepsilon$ and is denoted $\partial \Omega_\varepsilon$. In two and three dimensions, with polar coordinates ($r$,$\theta$) or spherical coordinates ($r$,$\theta$,$\varphi$), respectively, the escape region is defined by the radius $r={\cal R}$ and the polar angle $|\theta|<\varepsilon/2$.

A smooth diffusion constant with a sigmoidal shape will lead to similar solutions, as long as the size of the transition interval (from $D_1$ to $D_2$) is much smaller than the other length scales in the  problem, $\Delta$ and ${\cal R}$. A different model has been studied with diffusion constant switching between two distinct values according to a Markov jump process~\cite{holcman2010}. Here we will focus on $D_1 > D_2$, motivated by molecular motor driven propulsion along actin filaments in the cell cortex (see Introduction) which leads on long time scales to a faster diffusion in the outer shell. The case $D_1 < D_2$ could model a slower membrane diffusion~\cite{cusseddu2019}, which is not the focus of this present work. Moreover, we will focus on an attractive outer shell ($\Delta V > 0$). Due to the actin cortex underneath the cell membrane, the probability to stay in the outer shell is higher than to leave it if no attachment to microtubules is involved. Hence a phenomenological description by a lower potential might be appropriate. A repulsive outer shell ($\Delta V < 0$) probably has no biological relevance, except perhaps for large particles not equipped with molecular motors. Note that a repulsive potential in the outer shell simply increases the MFPT. Different studies of the NEP have also considered more general drift terms and non-localized potential wells~\cite{singer2007, lagache2008}.

The probability density $p({\bf y},\tau|{\bf x})$ to be at the position ${\bf y}$ after a time $\tau$ starting at the position ${\bf x}$ without having reached the escape region satisfies the forward Fokker-Planck equation $\partial_\tau p({\bf y},\tau|{\bf x}) = {\cal L}_{\bf y}  p({\bf y},\tau|{\bf x})$, with ${\cal L}_{\bf y}$ is the forward Fokker-Planck operator, or
\begin{equation}
\label{eqFP}
\partial_\tau p({\bf y},\tau|{\bf x}) = \nabla_{\bf y} \cdot \left\{ D({\bf y}) \exp(-\beta V({\bf y}))  \nabla_{\bf y} \left[ \exp(\beta V({\bf y})) p({\bf y},\tau|{\bf x}) \right]\right\} = - \nabla_{\bf y} \cdot {\bf J}({\bf y},\tau),
\end{equation}
where $D({\bf y})$ and $V({\bf y})$ are radially piecewise constant within the two-shell geometry, and ${\bf J}({\bf y},\tau)$ defines the current of particles composed by the diffusive current $-D({\bf y}) \nabla_{\bf y} p({\bf y},\tau)$ and the convective current $- \beta D({\bf y}) \nabla_{\bf y} V({\bf y}) p({\bf y},\tau)$ in a thermal bath with temperature $\beta^{-1}$. The boundary conditions on the surface are $p({\bf y},\tau|{\bf x}) = 0$ on the absorbing boundary $\partial \Omega_\varepsilon$ and ${\bf n} \cdot {\bf J}({\bf y},\tau) = 0$ on the reflective boundary $\partial \Omega \backslash \partial \Omega_\varepsilon$, where ${\bf n}$ represents the normal vector to the surface. From the Eq.~(\ref{eqFP}), the probability densities $p_i$ defined in each shell $i$ satisfy the diffusion equation $\partial_\tau p_i = D_i \nabla_{\bf y}^2 p_i$ and the boundary conditions between the two shells are $p_1 \exp(\beta V_1) = p_2 \exp(\beta V_2)$ and $D_1 {\bf n} \cdot \nabla_{\bf y} p_1 = D_2 {\bf n} \cdot \nabla_{\bf y} p_2$. The particles then perform a Brownian motion in each shell and need to pass through a potential barrier $\pm \Delta V$ to go from one shell to the other.

The mean first passage time (MFPT) for a particle starting at ${\bf x}$ to reach the escape region~\cite{redner2001} is defined by
\begin{equation}
t({\bf x}) = \int_\Omega d{\bf y} \int_0^\infty d\tau p({\bf y},\tau|{\bf x}),
\end{equation}
and satisfy then the equation ${\cal L}_{\bf x}^\dagger t({\bf x}) = -1 $ with ${\cal L}_{\bf x}^\dagger$ the backward Fokker-Planck operator:
\begin{equation}
\label{eqMFPT}
{\cal L}_{\bf x}^\dagger t({\bf x}) = \exp(\beta V({\bf x})) \nabla_{\bf x} \cdot \left[ D({\bf x}) \exp(-\beta V({\bf x})) \nabla_{\bf x} t({\bf x}) \right] = - 1.
\end{equation}
The corresponding boundary conditions on the surface are:
\begin{equation}
\label{eqMFPTBC}
t({\bf x}) = 0, \ {\bf x} \in \partial \Omega_\varepsilon \qquad {\rm and} \qquad {\bf n} \cdot \nabla_{\bf x} t({\bf x}) = 0, \ {\bf x} \in \partial \Omega \backslash \partial \Omega_\varepsilon,
\end{equation}
where ${\bf n}$ represents the normal vector to the surface. The system of equations satisfied by the MFPTs $t_i({\bf x})$ defined in each shell $i$ is then
\begin{gather}
D_1 \nabla^2_{\bf x} t_1({\bf x}) = -1, \quad {\bf x} \in \Omega_1 \label{MFPT1} \\
D_2 \nabla^2_{\bf x} t_2({\bf x}) = -1, \quad {\bf x} \in \Omega_2 \label{MFPT2} \\
t_1({\bf x}) =  t_2({\bf x}) \quad {\rm and} \quad D_1 \exp(-\beta V_1) {\bf n} \cdot \nabla_{\bf x} t_1({\bf x}) = D_2 \exp(-\beta V_2) {\bf n} \cdot \nabla_{\bf x} t_2({\bf x}) , \quad {\bf x} \in \partial \Omega_{1/2} \label{MFPT3}\\
t_1({\bf x}) = 0, \quad {\bf x} \in \partial \Omega_\varepsilon \quad {\rm and} \quad {\bf n} \cdot \nabla_{\bf x} t_1({\bf x}) = 0, \quad {\bf x} \in \partial \Omega \backslash \partial \Omega_\varepsilon. \label{MFPT4}
\end{gather}

In the following sections we will focus mainly on two quantities: (i) the average MFPT (denoted GMPFT hereafter for global MFPT) which characterizes the MFPT of the particles starting according to the stationary probability density for the closed domain, denoted $P_s({\bf x})$, satisfying the Boltzmann weight for the potential~$V({\bf x})$:
\begin{equation}
\langle t \rangle = \int_\Omega d{\bf x} \ P_s({\bf x}) t({\bf x}) = \frac{\int_\Omega d{\bf x} \ \exp(-\beta V({\bf x})) t({\bf x})}{\int_\Omega d{\bf x} \ \exp(-\beta V({\bf x}))}
\end{equation}
and in particular the dimensionless quantity: $T= D_1 \langle t \rangle / {\cal R}^2$; and (ii) the MFPT for particles starting at the center (denoted CMFPT hereafter for center MFPT): $t({\bf 0})$ for which the dimensionless quantity is $T_0= D_1 t({\bf 0}) / {\cal R}^2$. These two quantities are most important to characterize the average transport efficiency towards the escape region and the transport between the center and the escape region, respectively. We will analyze their dependence on the four main parameters: the escape angle $\varepsilon$, the dimensionless width of the outer shell $\Delta/{\cal R}$, the ratio of diffusion constants $D_1/D_2$ and the dimensionless potential difference $\beta \Delta V = \beta(V_2-V_1)$.

%%%%%%%%%%%%%%%%%%%%%%%%%%%%%%%%%%
%%%%%%%%%%%%%%%%%%%%%%%%%%%%%%%%%%
%%% SURFACE-MEDIATED DIFFUSION %%%
%%%%%%%%%%%%%%%%%%%%%%%%%%%%%%%%%%
%%%%%%%%%%%%%%%%%%%%%%%%%%%%%%%%%%

\section{Thin cortex limit}
\label{sectionThinCortex}

In this section we derive the expression of the MFPT in the thin cortex limit ($\Delta \rightarrow 0$), and we relate it to the surface-mediated diffusion problem (SMDP) defined in Refs.~\cite{rupprecht2012a, rupprecht2012b}. For this SMDP with a desorption rate $\lambda$ and an absorption rate $k$ (see Fig.~\ref{figure_model}(b)), the MFPT equations for a particle starting on the surface and in the bulk, denoted $t_\Sigma$ and $t_{\rm B}$ respectively, are
\begin{gather}
\frac{D_1}{{\cal R}^2} \Delta_\theta t_\Sigma(\theta) - \frac{\lambda}{k} \frac{\partial t_{\rm B}}{\partial r}({\cal R},\theta)  =-1, \label{eqSMSurf}\\
D_2 \nabla^2 t_{\rm B}(r,\theta) = -1 \label{eqSMBulk}
\end{gather}
where $D_1$ and $D_2$ are respectively the diffusion constants on the surface and in the bulk, and $\Delta_\theta = (\sin \theta)^{2-d} \partial_\theta (\sin\theta)^{d-2} \partial_\theta$ is the Laplace operator on the unit hypersphere in $d$ dimension, assuming the invariance over the azimuthal angle $\varphi$ in dimension $d=3$. The boundary conditions are
\begin{equation}
\frac{\partial t_{\rm B}}{\partial r}({\cal R},\theta) = k \left[t_\Sigma(\theta) - t_{\rm B}({\cal R},\theta) \right], \label{eqSMBC}
\end{equation}
on the surface and $t_\Sigma = 0$ on the escape region located at $|\theta|<\varepsilon/2$. This last condition will be omitted in the following analysis since it is trivially satisfied by all NEPs. We may note that the solution of this system of equations depends on $k$ and $\lambda$ independently whereas our two-shell geometry model depends only on the potential difference $\Delta V = V_2-V_1$. Hence the SMDP has one more parameter also in the $\Delta \to 0$ limit. In appendix~\ref{appendixSMD} we show that Eqs.~(\ref{MFPT1})-(\ref{MFPT4}) can be related to the Eqs.~(\ref{eqSMSurf})-(\ref{eqSMBC}) only when the desorption and absorption rates are infinite ($k \to \infty$ and $\lambda \to \infty$) for a constant ratio given by:
\begin{equation}
\label{eqRELSM}
\lim_{\substack{k \to \infty\\\lambda \to \infty}} \frac{kD_2}{\lambda} = \lim_{\Delta \rightarrow 0}\frac{\Delta \exp(-\beta V_1)}{\exp(-\beta V_2)} \equiv \kappa.
\end{equation}
The potential difference must then depend logarithmically on the width of the outer shell as $\beta \Delta V \simeq  \ln(\kappa/\Delta)$ to recover the SMDP from the two-shell geometry NEP.

This relation can be derived by considering the stationary state for the closed domain ($\varepsilon=0$) and by using the continuity of the MFPT (which gives the limit $k\to\infty$). In the two-shell geometry the stationary state satisfies the relation $p_1^{\rm st}/p_2^{\rm st} = \exp[-\beta(V_1-V_2)]$, while the stationary state of the SMDP satisfies $p_\Sigma^{\rm st}/p_{\rm B}^{\rm st} = kD_2/\lambda$. In the limit $\Delta \to 0$ the relation between the density probabilities of the two problems are $p_{\rm B}^{\rm st} = p_2^{\rm st}$ and $p_\Sigma^{\rm st} = \Delta p_1$ after integration over the radius $r\in[{\cal R}-\Delta,{\cal R}]$. By identification of both equilibrium states the Eq.~(\ref{eqRELSM}) is recovered.

In appendix~\ref{appendixDelta0} we derive the limit of Eqs.~(\ref{MFPT1})-(\ref{MFPT4}) when $\Delta \to 0$. As long as $\Delta V<+\infty$ (excluding the annulus geometry) the MFPT equations become $D_2 \nabla^2 t_{\rm B}(r,\theta) = 0$ and $t_{\Sigma}(\theta) = t_{\rm B}({\cal R},\theta)$ with the boundary condition
\begin{equation}
\label{condDelta0}
\frac{\partial t_{\rm B}}{\partial r}({\cal R},|\theta|>\varepsilon/2) = 0.
\end{equation}
These equations correspond to the SMDP with $k\to\infty$ and $kD_2 \ll \lambda$, consistent with Eq.~(\ref{eqRELSM}) for $\kappa=0$, which is strictly the same as the disk geometry NEP with a diffusion constant $D_2$, implying the general relation:
\begin{equation}
\label{MFPTDelta0rel}
\lim_{\Delta \to 0} t({\bf x}) = \frac{D_1}{D_2} \lim_{\Delta \to {\cal R} } t({\bf x})
\end{equation}
for all starting position ${\bf x}$. In passing we note that the limit $\Delta \to {\cal R}$ is equivalent to the special case $D_1=D_2$ and $\beta \Delta V=0$ for which the central region disappears, due to the continuity of the MFPT. In this special case the GMFPT is equal to $\langle t \rangle = \langle t_{\rm B} \rangle$ since the average over $t_\Sigma$ does not contribute when $\Delta=0$ and the CMFPT is $t({\bf 0}) = t_{\rm B}(0)$, leading to the following results for $T=D_1\langle t \rangle/{\cal R}^2$ and $T_0=D_1 t({\bf 0})/{\cal R}^2$ in the narrow escape limit~\cite{singer2006, pillay2010, cheviakov2010}:
\begin{equation}
\label{MFPTDelta0}
T \simeq
\begin{dcases}
\frac{D_1}{D_2} \left( - \ln \frac{\varepsilon}{4} + \frac{1}{8} \right) &\qquad (d=2)\\
\frac{D_1}{D_2} \left( \frac{2\pi}{3\varepsilon} - \frac{1}{3}\ln \varepsilon - \frac{1}{10} \right) &\qquad (d=3)
\end{dcases}
\qquad {\rm and} \qquad T_0 \simeq
\begin{dcases}
\frac{D_1}{D_2} \left( - \ln \frac{\varepsilon}{4} + \frac{1}{4} \right) &\qquad (d=2)\\
\frac{D_1}{D_2} \left( \frac{2\pi}{3\varepsilon} - \frac{1}{3}\ln \varepsilon \right) &\qquad (d=3)
\end{dcases}
\end{equation}

In appendix~\ref{appendixDelta0} we hypothesize that the thin cortex limit, leading to Eq.~(\ref{MFPTDelta0rel}), is valid for
\begin{equation}
\frac{\Delta}{\cal R} \ll \min\left( 1 , \frac{D_2}{D_1} \exp(-\beta \Delta V) \right).
\end{equation}

In the annulus geometry, i.e. $\Delta V = +\infty$ or $\exp(-\beta \Delta V)=0$, the MFPT satisfies the same equations with Eq.~(\ref{condDelta0}) replaced by
\begin{equation}
\frac{D_1}{{\cal R}^2} \Delta_\theta t_{\Sigma}(\theta) =-1.
\end{equation}
These equations correspond to the SMDP with $k\to\infty$ and $\lambda \ll kD_2$, consistent with Eq.~(\ref{eqRELSM}) for $\kappa = \infty$. The general solutions of the surface MFPT $t_\Sigma(\theta)$ are~\cite{rupprecht2012a}:
\begin{equation}
\frac{D_1}{{\cal R}^2} t_{\Sigma}(\theta) =
\begin{dcases}
\frac{(\theta-\varepsilon/2)(2\pi-\varepsilon/2 -\theta)}{2} & \qquad (d=2)\\
\ln\left( \frac{1-\cos\theta}{1-\cos(\varepsilon/2)} \right) & \qquad (d=3)
\end{dcases}.
\end{equation}
The GMFPT is equal to $\langle t \rangle = \langle t_{\Sigma} \rangle$ since the average over $t_{\rm B}$ does not contribute when $\exp(-\beta \Delta V)=0$ (i.e. annulus geometry) and the CMFPT is $t({\bf 0}) = t_{\rm B}(0)$. The results for $T$ and $T_0$ are then~\cite{rupprecht2012a}
\begin{equation}
\label{MFPTDelta0annulus}
T=
\begin{dcases}
\frac{(2\pi-\varepsilon)^3}{24} & \qquad (d=2)\\
- \frac{1+\cos(\varepsilon/2)}{2} - 2 \ln \sin(\varepsilon/4) & \qquad (d=3)
\end{dcases},
\qquad {\rm and} \qquad T_0 = T + \frac{D_1}{2dD_2}.
\end{equation}

Herewith the $\Delta \to 0$ limit of the two-shell geometry NEP in 2d and 3d is well understood, and is generally related to the SMDP for particular values of absorption and desorption rates.

%%%%%%%%%%%%%%%%%%%%%%%%%%%
%%%%%%%%%%%%%%%%%%%%%%%%%%%
%%% NUMERICAL SOLUTIONS %%%
%%%%%%%%%%%%%%%%%%%%%%%%%%%
%%%%%%%%%%%%%%%%%%%%%%%%%%%

\section{Numerical solutions}
\label{sectionNumerics}

This section elaborates how to obtain numerical estimates of the MFPT with the finite element method to solve the Eqs.~(\ref{MFPT1})-(\ref{MFPT4}) and, alternatively, with stochastic simulations. Then we present numerical results and analyze qualitatively the parameter dependencies of the MFPT. We solve numerically the system of equations~(\ref{MFPT1})-(\ref{MFPT4}) with the finite element method~\cite{zienkiewicz1977} using the software package FreeFem++~\cite{hecht2013}. Since this software allows to define piecewise constant functions (as the diffusion constant and the potential), the Eq.~(\ref{eqMFPT}) can be directly solved in the two-shell geometry. Using the stationary probability density for the closed domain $P_s({\bf x}) \propto \exp(-\beta V({\bf x}))$, Eq.~(\ref{eqMFPT}) becomes
\begin{equation}
\nabla_{\bf x} \cdot \left[ D({\bf x}) P_s({\bf x}) \nabla_{\bf x} t({\bf x}) \right] = - P_s({\bf x}).
\end{equation}
The weak formulation of this equation is the functional equation over the full space $\Omega$:
\begin{equation}
\int_\Omega d{\bf x} \ v({\bf x}) \nabla_{\bf x} \cdot \left[ D({\bf x}) P_s({\bf x}) \nabla_{\bf x} t({\bf x}) \right] = - \int_\Omega d{\bf x} \ v({\bf x}) P_s({\bf x}),
\end{equation}
where $v({\bf x})$ is an arbitrary smooth function. Integration by parts of the left hand side and the boundary conditions~(\ref{eqMFPTBC}) yields
\begin{equation}
\label{eqFEM2d}
\int_\Omega d{\bf x} \ \nabla_{\bf x} v({\bf x})  \cdot \left[ D({\bf x}) P_s({\bf x}) \nabla_{\bf x} t({\bf x}) \right] =  \int_\Omega d{\bf x} \ v({\bf x}) P_s({\bf x}),
\end{equation}
by imposing the condition $v({\bf x}) = 0$ on the absorbing boundary $\partial \Omega_\varepsilon$. In 3d, cylindrical coordinates $(\rho,\varphi,Z)$ with $x=Z$, $y=\rho \cos \varphi$ and $z=\rho \sin \varphi$ are used. Since the system is invariant under rotations around the $Z$-axis by the azimuthal angle~$\varphi$, Eq.~(\ref{eqFEM2d}) becomes
\begin{equation}
\label{eqFEM3d}
\int_\Omega d{\bf X} \ \rho \nabla_{\bf X} v({\bf X})  \cdot \left[ D({\bf X}) P_s({\bf X}) \nabla_{\bf X} t({\bf X}) \right] =  \int_\Omega d{\bf X} \ \rho v({\bf X}) P_s({\bf X}),
\end{equation}
where ${\bf X} = (Z,\rho)$. The projection on the plane $\varphi=0$ (i.e. $z=0$) yields ${\bf X} = (x,y)$. From Eqs.~(\ref{eqFEM2d}) and~(\ref{eqFEM3d}) the equation to solve in $d$ dimensions is hence
\begin{equation}
\int_\Omega d{\bf x} \ y^{d-2} \nabla_{\bf x} v({\bf x})  \cdot \left[ D({\bf x}) P_s({\bf x}) \nabla_{\bf x} t({\bf x}) \right] =  \int_\Omega d{\bf x} \ y^{d-2} v({\bf x}) P_s({\bf x}),
\end{equation}
where ${\bf x} = (x,y)$ are the two dimensional Cartesian coordinates. This equation can be written as $b(v,t) = l(v)$ where $b(v,t)$ and $l(v)$ are bilinear and linear operators, respectively, on the space of integrable functions~$L^1$.

%%%%%%%%%%%%%%%%%%%%%%
%%% FIGURE FREEFEM %%%
%%%%%%%%%%%%%%%%%%%%%%
\begin{figure}[t]
\begin{center}
\includegraphics[width=12cm]{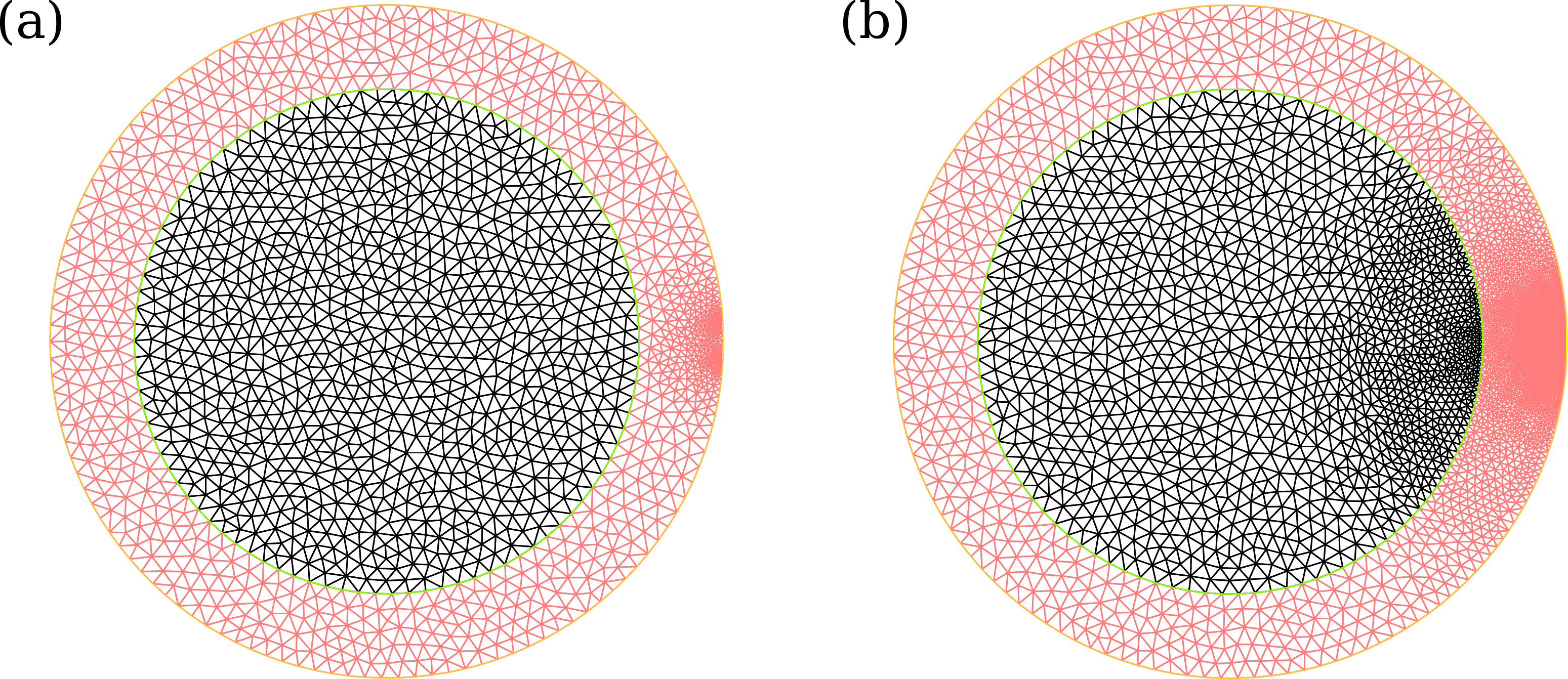}
\caption{{\bf (a)}~Mesh-grid created initially by the software package FreeFem++ with $300$ vertices on the boundary. {\bf (b)}~Mesh-grid after four refinements made by FreeFem++ according to the MFPT solution.} \label{figure_FreeFem}
\end{center}
\end{figure}

To solve this functional equation, the space is discretized into a triangular mesh-grid which is in general not regular. We may note at this point that the width of the outer shell studied in this paper is limited to $\Delta> 0.0003 {\cal R}$ to avoid computationally inconvenient mesh sizes. The MFPT is then calculated at the nodes of a mesh-grid as sketched in Fig.~\ref{figure_FreeFem} (labelled $\alpha \in [1,{\cal N}]$) and interpolated over the complete space with second order Lagrange polynomials $e_\alpha({\bf x})$, forming a basis on the discretized space and defined by its value on the nodes: $1$ at node $\alpha$ and $0$ at other nodes. The MFPT is then $t({\bf x}) = \sum_\alpha \tau_\alpha e_\alpha({\bf x})$ and the functional equation becomes $\sum_\beta \tau_\beta b(e_\alpha,e_\beta) = l(e_\alpha)$ due to linearity and expanding the function $v({\bf x})$ in $e_\alpha({\bf x})$. This expression can be rewritten in matrix form ${\cal B}\tau = {\cal L}$, where ${\cal B}$ is a matrix with elements ${\cal B}_{\alpha\beta} = b(e_\alpha,e_\beta)$, and $\tau$ and ${\cal L}$ are vectors with components $\tau_\alpha$ and $l(e_\alpha)$ respectively. The solution is hence $\tau = {\cal B}^{-1} {\cal L}$. The computational time is mainly determined by the inversion of the ${\cal N}\times{\cal N}$ matrix, i.e. of order ${\cal O}({\cal N}^3)$ with classical inversion methods. The results shown hereinafter will be calculated by starting from a not too narrow mesh-grid with ${\cal N} \simeq 10^3$ nodes by fixing $300$ vertices on the boundary (see Fig.~\ref{figure_FreeFem}(a)), and refining the mesh-grid where the MFPT varies strongly until reaching ${\cal N} \simeq 10^4$ nodes (see Fig.~\ref{figure_FreeFem}(b) after four refinements). We can mention that the mesh-grid is mostly refined close to the escape region where the MFPT is approximatively proportional to $\ln D$ and $D^{-1}$, where $D$ is the distance to the center of the escape region in two and three dimensions respectively, while the mesh-grid is not refined far from the escape region. Following this method, the computation time to obtain a numerical estimate for $t({\bf x})$ for all starting position ${\bf x}$ is less than 10 seconds on a $4$ GHz CPU. In appendix~\ref{appendixF}, the numerical error of this method is evaluated by comparison to an analytically solvable special case: the fully-absorbing limit when $\varepsilon=2\pi$. We estimate therein that the numerical error is of order $10^{-4}$.

This fast numerical method to have an estimate of the MFPT can be compared with the conventional stochastic simulations of a Brownian particle to reach the escape region. The Langevin equation written with the It\^o convention for the position ${\bf x}_t$ at time $t$ is
\begin{equation}
\label{eqIto}
{\bf x}_{t+dt} = {\bf x}_t +  \left[ - D({\bf x}_t) \beta \nabla_{\bf x} V({\bf x}_t) + \nabla_{\bf x} D({\bf x}_t) \right] dt + \sqrt{2 D({\bf x}_t)} d{\bf B}_t
\end{equation}
where $d{\bf B}_t$ is distributed according to a zero mean Gaussian distribution of variance $dt$. The convective term is not well defined when the Brownian particle crosses the inner boundary at $|{\bf x}|= {\cal R}-\Delta$ for the two-shell geometry. To obtain an accurate algorithm we consider that the particles perform a Brownian motion in each shell with different diffusion constants, with an impermeability condition at the external boundary and with a special rule at the interface between the two shells. To obtain a faster solution than simulating the Brownian motion step by step, we consider the Kinetic Monte Carlo (KMC) method~\cite{oppelstrup2009, schwarz2013} as long as the particle stays in the same shell and return to a small hop simulation close to the boundaries (with a distance smaller than $\eta$). The KMC method consists to generate a random time at which the particle hits the boundary for the first time, according to the FPT distribution known for homogeneous diffusion constant in simple geometries (e.g. disk, sphere) with fully-absorbing boundary.

To satisfy the boundary condition between the two shells, the algorithm shown in Ref.~\cite{lejay2013} (without potential barrier) is used to have the correct repartition of particles in the inner and outer shells. The probability ${\cal P}_k$ to be in the shell $k$ is proportional to $\sqrt D_k$ (from Ref.~\cite{lejay2013}) and $\exp(-\beta V_k)$ (from stationary probability density), implying that
\begin{equation}
{\cal P}_1 = \frac{\sqrt{D_1}\exp(-\beta V_1)}{\sqrt{D_1}\exp(-\beta V_1)+\sqrt{D_2}\exp(-\beta V_2)} \qquad {\rm and} \qquad {\cal P}_2 = \frac{\sqrt{D_2}\exp(-\beta V_2)}{\sqrt{D_1}\exp(-\beta V_1)+\sqrt{D_2}\exp(-\beta V_2)}.
\end{equation}
When the particle crosses this interface, the particle is stopped on the interface, the crossing time is calculated and the next time position is strictly taken in the shell $k$ with probability ${\cal P}_k$. The time-step is taken as $dt = \eta^2/(2D_k)$ for having only a few steps without using the KMC method, and $\eta = 10^{-4} {\cal R}$.

The computation time is proportional to the number of stochastic realizations and increases for narrow escape region exactly like the MFPT value. For $10^6$ different stochastic trajectories and an escape angle $\varepsilon=0.2$, the computation time is about six and fourteen hours for the two and three dimensional problems, respectively. We can remark that we have access to only one estimate of the MFPT for a given starting point (GMFPT or CMFPT) due to the initialization of stochastic trajectories. The codes of the numerical methods discussed in this section are available in Ref.~\cite{zenodo}.

%%%%%%%%%%%%%%%%%%%%
%%% FIGURE SIMUS %%%
%%%%%%%%%%%%%%%%%%%%

\begin{figure}[t]
\begin{center}
\includegraphics[width=17cm]{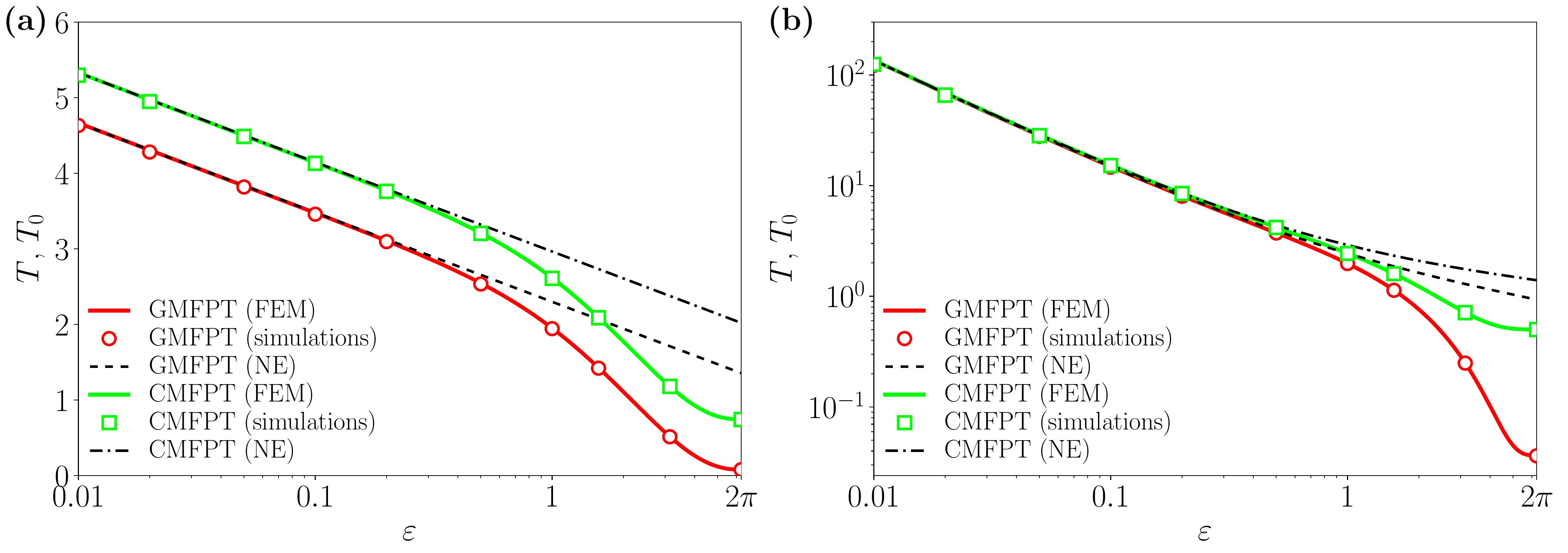}
\caption{(color online) Dependence of the dimensionless GMFPT and CMFPT, denoted respectively as $T=D_1 \langle t \rangle / {\cal R}^2$ (GMFPT) and $T_0 = D_1 t({\bf 0})  / {\cal R}^2$ (CMFPT), on the escape angle $\varepsilon$ for fixed parameters $\Delta= 0.25 {\cal R}$, $D_1/D_2=5$ and $\beta \Delta V=2$ for {\bf (a)}~the two dimensional and {\bf (b)}~the three dimensional NEPs. The lines (FEM) correspond to numerical solutions obtained with the finite element method using FreeFem++ assumed to be the most accurate with a relative error smaller than $10^{-4}$. The circle and square symbols (simulations) represent the stochastic average of these MFPTs obtained with numerical simulations for $10^6$ different samples of escape trajectories. The dashed lines (NE) display the narrow escape limit $\varepsilon \ll 1$ for these two MFPTs.} \label{figure_simus}
\end{center}
\end{figure}

In Fig.~\ref{figure_simus} the dependence of the dimensionless GMFPT and CMFPT on the escape angle $\varepsilon$ is shown for 2d and 3d NEPs. There all other parameters are fixed: $\Delta=0.25{\cal R}$, $D_1/D_2 = 5$ and $\beta \Delta V = 2$. Fig.~\ref{figure_simus} presents also a comparison between the numerical solutions for $T$ and $T_0$ obtained with the finite element method using the software package FreeFem++ (lines) and the stochastic simulations for $10^6$ different stochastic trajectories (symbols). Analyzing the values of the GMFPT and the CMFPT after several refining of the mesh-grid, the relative error on these quantities is of order $10^{-4}$ (see appendix~\ref{appendixF}) while the stochastic error is of order $10^{-3}$ for that number of realizations. Regarding the computation time elaborated above the finite element method is much more efficient than stochastic simulations. Moreover, the numerical values are consistent with the analytical approximations (dotted lines) which will be derived in the following section~\ref{sectionNE}. The difference between CMFPT and GMFPT is of order ${\cal O}(\varepsilon^0)$ while the leading order is of order ${\cal O}(\ln \varepsilon)$ in 2d (Fig.~\ref{figure_simus}(a)) and of order ${\cal O}(\varepsilon^{-1})$ in 3d (Fig.~\ref{figure_simus}(b)). In the following we will exclusively use the finite element method to compute numerical estimates of the MFPTs.

%%%%%%%%%%%%%%%%%%%
%%% FIGURE MFPT %%%
%%%%%%%%%%%%%%%%%%%

\begin{figure}[t]
\begin{minipage}[t]{.49\linewidth}
\begin{center}
\includegraphics[width=8cm]{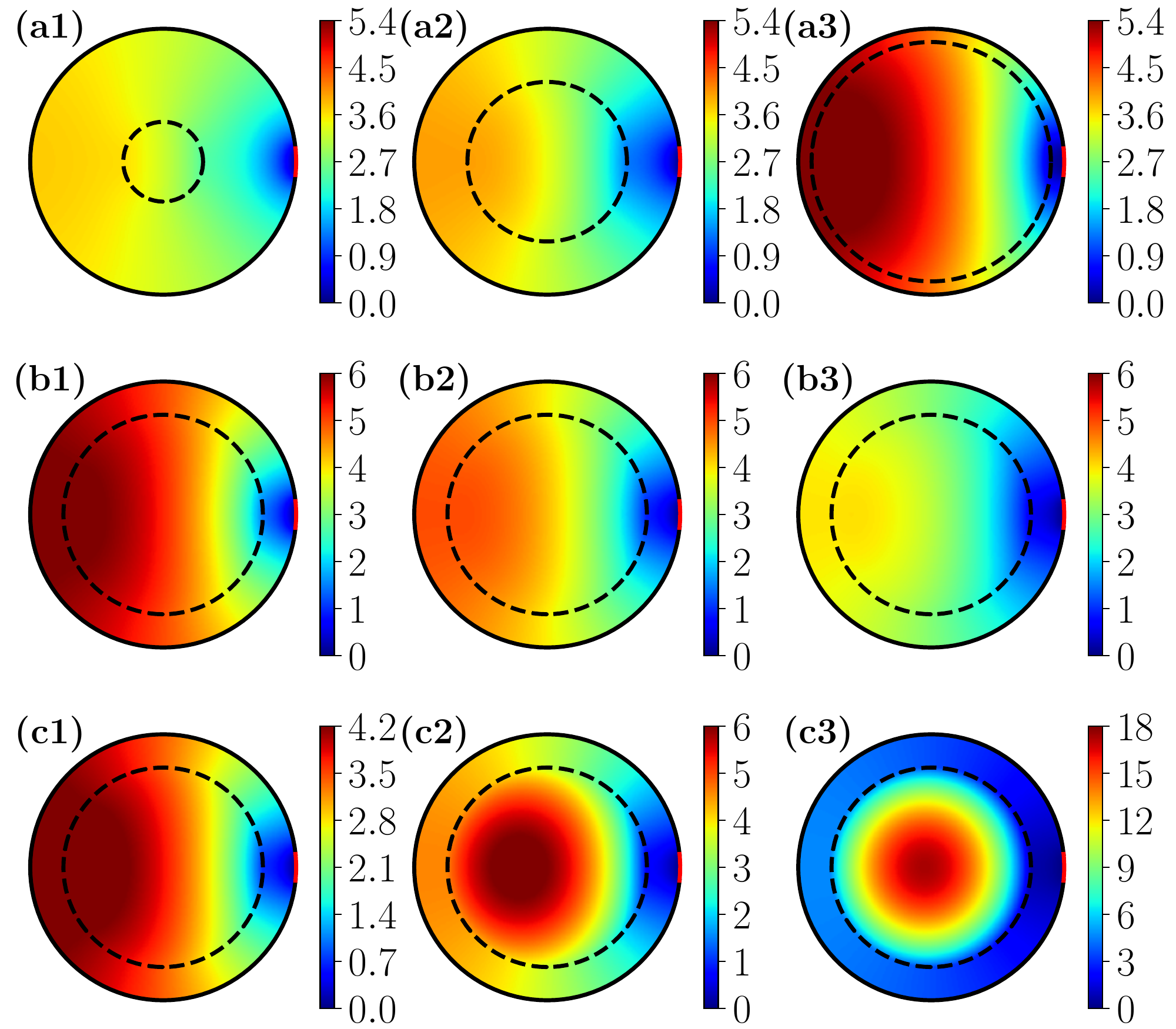}
\end{center}
\end{minipage}
\begin{minipage}[t]{.49\linewidth}
\begin{center}
\includegraphics[width=8cm]{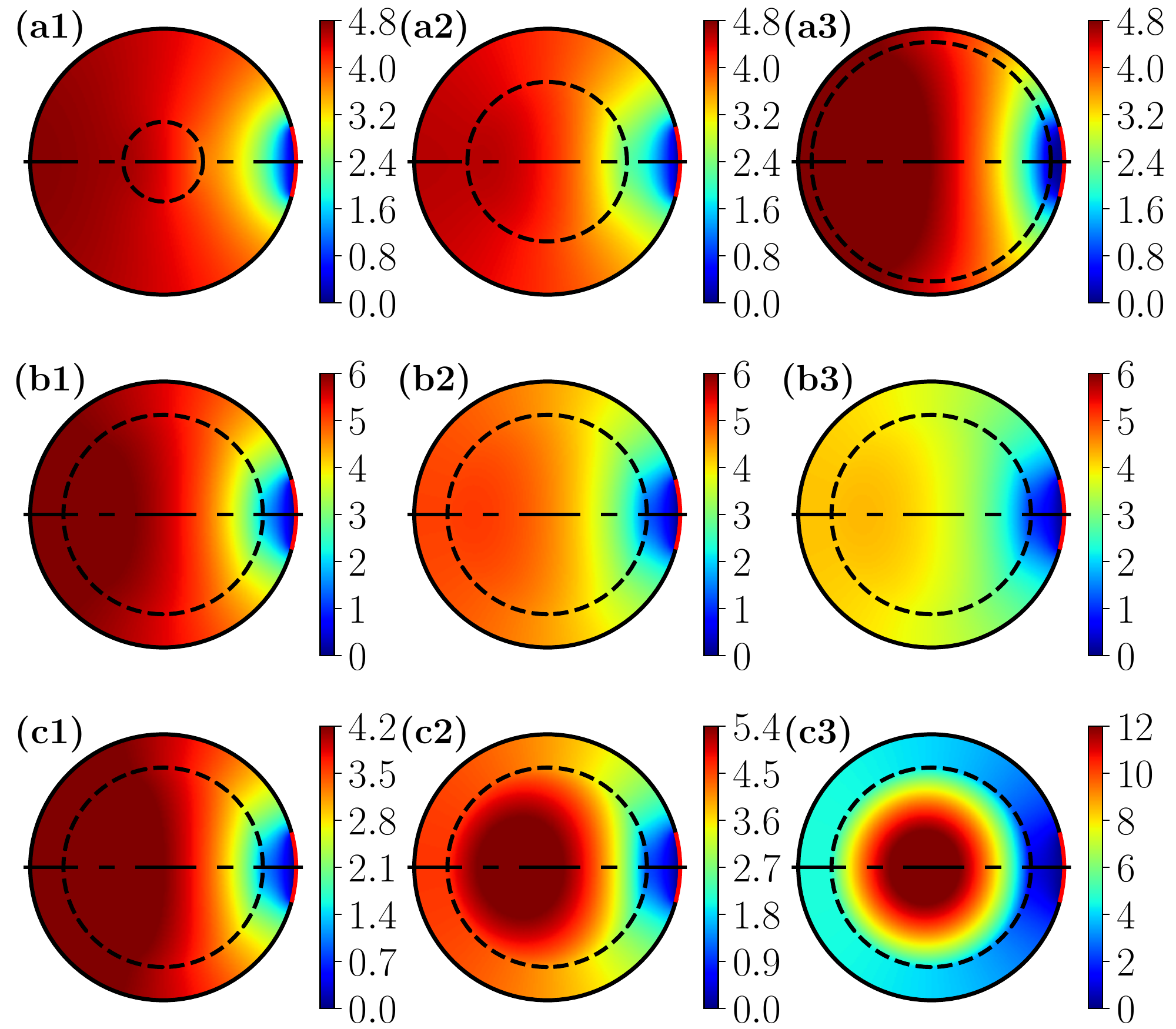}
\end{center}
\end{minipage}
\caption{(color online) Numerical solution of the MFPT obtained with the finite element method using FreeFem++ for the 2d NEP (left panel) and the 3d NEP (right panel) which is axis-symmetric around the abscissa~$x$ (dashed-dotted line) for fixed escape angles $\varepsilon=0.2$ and $\varepsilon=0.5$ respectively. The first line displays the solution for fixed parameters $D_1/D_2=5$ and $\beta \Delta V = 2$, and decreasing values of $\Delta/{\cal R}$: {\bf (a1)}~$\Delta/{\cal R}=0.7$, {\bf (a2)}~$\Delta/{\cal R}=0.4$ and {\bf (a3)}~$\Delta/{\cal R}=0.1$. The second line shows the solution for fixed parameters $\Delta/{\cal R}=0.25$ and $D_1/D_2 = 5$, and increasing potential difference $\beta \Delta V$: {\bf (b1)}~$\beta \Delta V=0.1$, {\bf (b2)}~$\beta \Delta V=1$ and {\bf (b3)}~$\beta \Delta V=10$. The third line displays the solution for fixed parameters $\Delta/{\cal R}=0.25$ and $\beta \Delta V = 2$, and increasing ratio of diffusion constants $D_1/D_2$: {\bf (c1)}~$D_1/D_2=5$, {\bf (c2)}~$D_1/D_2=20$ and {\bf (c3)}~$D_1/D_2=100$.} \label{figure_MFPT}
\end{figure}

Fig.~\ref{figure_MFPT} displays the MFPT for the escape angle $\varepsilon = 0.2$ in 2d (left panel) and $\varepsilon=0.5$ in 3d (right panel). The 3d case is restricted to the $(x,y)$ plane by considering the rotational symmetry around the $x$-axis. In each line only one parameter is varied: $\Delta/{\cal R}$, $D_1/D_2$ or $\beta \Delta V$. The MFPT behaves similarly in 2d and 3d, but quantitative differences can be observed. In the first line of Fig.~\ref{figure_MFPT} the dependence of the MFPT on the width of the outer shell, $\Delta$, is displayed for fixed parameters $D_1/D_2 =5$ and $\beta \Delta V=2$. When $\Delta$ decreases [from~(a1) to~(a3)], the MFPT decreases for particles starting in a region ($x>x_*$) close to the escape window while it increases for other starting positions. This behavior may lead to the presence of a minimum for the GMFPT by averaging over all starting positions, but also for the CMFPT, since $x_* > 0$. From Fig.~\ref{figure_MFPT}(a) we estimate $x_* \simeq 0.3 {\cal R}$ for $d=2$ and $x_* \simeq 0.5 {\cal R}$ for $d=3$. Without showing data we note that the $\Delta$-dependence of the MFPT is different for repulsive cortex ($\beta \Delta <0$) and/or slow cortex ($D_1<D_2$). For the majority of starting points a decreasing MFPT is observed for $D_1>D_2$ and $\Delta V<0$ and an increasing MFPT for $D_1<D_2$ and $\Delta V>0$, which follows the behavior between the two extreme limits given by Eq.~(\ref{MFPTDelta0rel}). For $D_1<D_2$ and $\Delta V<0$ the $\Delta$-dependency of the MFPT is non-monotonous: it is increasing for small width and decreasing for large width, leading to a maximum for the GMFPT and the CMFPT as a function of $\Delta$.

In the second line of Fig.~\ref{figure_MFPT} the dependence of the MFPT on the potential difference, $\beta \Delta V$, is presented for fixed parameters $\Delta = 0.25 {\cal R}$ and $D_1/D_2 = 5$. When $\Delta V$ increases [from~(b1) to~(b3)], the MFPT decreases due to a stronger attraction of the outer shell where the escape region is located, with slightly larger decrease for starting positions close to the absorbing window. Therefore, the GMFPT and the CMFPT are strictly decreasing functions of $\beta \Delta V$.

In the third and last line of Fig.~\ref{figure_MFPT} the dependence of the MFPT on the ratio of the diffusion constants, $D_1/D_2$, is shown for fixed parameters $\Delta = 0.25 {\cal R}$ and $\beta \Delta V=2$. When this ratio increases [from~(c1) to~(c3)], the particles starting in the central region are slowed down implying that the MFPT increases. Hence the GMFPT and the CMFPT are strictly increasing functions of $D_1/D_2$. Moreover, the starting position of the maximal MFPT (MMFPT) is located at the center when $D_1 \gg D_2$ due to this slowing down. Hence the distance $r_{\rm max}$ to the center of the MMFPT starting position decreases discontinuously with the ratio $D_1/D_2$ from $r_{\rm max} = {\cal R}$ to $r_{\rm max} = 0$. For a ratio of diffusion constants $D_1/D_2 < \kappa_c$, the MMFPT starting position is located in the cortex, at the maximum distance to the escape region, i.e. $r_{\rm max} = {\cal R}$. When $D_1/D_2$ becomes larger than $\kappa_c$, the MMPFT starting position jumps to the inner shell with $r_{\rm max} \lesssim {\cal R} - \Delta$, and is located at the maximum distance to the escape region at the transition ($D_1/D_2 = \kappa_c$). Finally for $D_1/D_2 > \kappa_c$, $r_{\rm max}$ decreases continuously from ${\cal R} - \Delta$ to $0$. Note that $\kappa_c$ depends on the outer shell width $\Delta$, and the potential difference $\beta \Delta V$. $\kappa_c$ increases with $\Delta$ and decreases with $\beta \Delta V$. This behavior has been studied in the Ref.~\cite{mangeat2019} without a potential difference ($\beta \Delta V =0$), and the main conclusions stay qualitatively the same for both attractive and repulsive cortex (see appendix~\ref{appendixG}).

With these qualitative observations in mind we will in the following focus on the dependence of the GMFPT and CMFPT on the width of the outer shell $\Delta / {\cal R}$ for which the behavior is clearly non-monotonous for small escape angles.

%%%%%%%%%%%%%%%%%%%%%%%%%%%
%%%%%%%%%%%%%%%%%%%%%%%%%%%
%%% NARROW ESCAPE LIMIT %%%
%%%%%%%%%%%%%%%%%%%%%%%%%%%
%%%%%%%%%%%%%%%%%%%%%%%%%%%

\section{Narrow escape limit and MFPT optimization}
\label{sectionNE}

In this section we derive the GMFPT and the CMFPT in the narrow escape limit ($\varepsilon \ll 1$). We also derive a condition for $\beta \Delta V$ and $D_1/D_2$ for which the MFPT displays a minimum as a function of the outer shell width $\Delta$. In the appendix~\ref{appendixGeneralNE}, we show how the MFPT can be calculated in the narrow escape limit for general diffusivity $D({\bf x})$ and potential $V({\bf x})$~\cite{ward1993, pillay2010, cheviakov2010, cheviakov2012, chevalier2011}. We denote the coordinate of the center of the escape region by ${\bf x_0}$ and assume that the limit of the diffusivity and the potential are sufficiently smooth close to this point, which is clearly the case for the two-shell geometry. The MFPT depends only on the solution of the pseudo Green's function $G({\bf x}|{\bf x_0})$~\cite{barton1989} for the closed domain ($\varepsilon=0$):
\begin{equation}
t({\bf x}) = \langle t \rangle - \frac{G({\bf x}|{\bf x_0})}{P_s({\bf x_0})},
\end{equation}
where $\langle t \rangle$ is the average MFPT and $P_s({\bf x_0})$ is the stationary probability density for the closed domain. The Green's function can be written in terms of the probability density $p({\bf y},\tau | {\bf x})$ as
\begin{equation}
G({\bf x}|{\bf x_0}) = P_s({\bf x_0}) \int_\Omega d{\bf y} \int_0^\infty d\tau \left[ P_s({\bf y}) - p({\bf y},\tau | {\bf x}) \right],
\end{equation}
and is therefore the solution of the system
\begin{gather}
\exp(\beta V({\bf x})) \nabla_{\bf x} \cdot \left[ D({\bf x}) \exp(-\beta V({\bf x})) \nabla_{\bf x}  G({\bf x}|{\bf x_0})  \right] = P_s({\bf x_0}) - \delta({\bf x} - {\bf x_0}), \quad {\bf x} \in \Omega \label{eqGreen1MT}\\
{\bf n} \cdot \nabla G({\bf x}|{\bf x_0}) = 0 , \quad {\bf x} \in \partial \Omega \label{eqGreen2MT}\\
\int_{\Omega} d{\bf x}\ G({\bf x}|{\bf x_0}) P_s({\bf x}) = 0 \label{eqGreen3MT},
\end{gather}
but diverges close to the escape region differently in 2d and 3d. Therefore we treat both cases, $d=2$ and $d=3$, separately in the following.

%%%%%%%%%%%%%%%%%%%%%%%%
%%%%%%%%%%%%%%%%%%%%%%%%
%%% NARROW ESCAPE 2D %%%
%%%%%%%%%%%%%%%%%%%%%%%%
%%%%%%%%%%%%%%%%%%%%%%%%

\subsection{Narrow escape limit in two dimensions}

In appendix~\ref{appendixGeneralNE} we show that close to the escape region the 2d Green's function behaves as
\begin{equation}
G({\bf x} \rightarrow {\bf x_0} |{\bf x_0}) = -\frac{1}{\pi D({\bf x_0})} \ln \frac{|{\bf x} - {\bf x_0}|}{{\cal R}} + R_0,
\end{equation}
where $R_0$ is the regular part of the pseudo Green's function at the center of the escape region and the average MFPT satisfies the expression
\begin{equation}
\label{expNEGMFPT}
\langle t \rangle =  \frac{1}{P_s({\bf x_0})} \left[ - \frac{1}{\pi D({\bf x_0})}\ln \frac{\varepsilon}{4} + R_0 \right].
\end{equation}

In the appendix~\ref{appendix2d} we derive the pseudo Green's function and identify $R_0$. The dimensionless GMFPT is
\begin{gather}
T(\chi) = \frac{D_1 \langle t \rangle}{{\cal R}^2} = \left[1-\chi^2(1-\xi)\right] \left\{- \ln \frac{\varepsilon}{4} - 2 \sum_{k=1}^\infty \left[ \frac{D_1-D_2\xi}{D_1 + D_2 \xi} \right]^k \ln(1-\chi^{2k}) \right\} \nonumber \\
+ \frac{1}{8} + \frac{(D_1-D_2)\chi^2 \xi + 3 D_2 (\xi-1)(1-\chi^2)}{8D_2 \left[1-\chi^2(1-\xi)\right]} \chi^2 - \frac{(1-\xi)^2 \chi^4\ln \chi}{2\left[1-\chi^2(1-\xi)\right]},\label{eqGMFPT2d}
\end{gather}
with $\chi = 1 - \Delta/{\cal R}$ and $\xi = \exp(-\beta \Delta V)$, and the dimensionless CMFPT is
\begin{gather}
T_0(\chi) = \frac{D_1 t({\bf 0})}{{\cal R}^2} = \left[1-\chi^2(1-\xi)\right] \left\{- \ln \frac{\varepsilon}{4} - 2 \sum_{k=1}^\infty \left[ \frac{D_1-D_2\xi}{D_1 + D_2 \xi} \right]^k \ln(1-\chi^{2k}) \right\} \nonumber \\
+ \frac{1}{4} + \frac{D_1-D_2}{4D_2} \chi^2 - \frac{\xi -1}{2}\chi^2\ln \chi.\label{eqCMFPT2d}
\end{gather}

Therefore, the GMFPT and the CMFPT can be written as the sum of two terms: (i) the first line of each expression is common and contains the leading term $-\left[1-\chi^2(1-\xi)\right] \ln(\varepsilon/4)$ and (ii) the second line is the corresponding MFPT for a fully absorbing external boundary ($\varepsilon=2\pi$), see the derivation in appendix~\ref{appendixF}. Note that the series in the common term is zero when $\beta \Delta V= \ln(D_2/D_1)$ corresponding to a zero convection in the It\^o formulation given by Eq.~(\ref{eqIto}) and discussed in more detail in Sec.~\ref{sectionDiscussion}. Moreover, from Eqs.~(\ref{eqGMFPT2d}) and~(\ref{eqCMFPT2d}), we recover the expressions for some special cases which were reported earlier: (i) the narrow escape limit for the disk geometry~\cite{singer2006} when $D_1=D_2$ and $\Delta V=0$:
\begin{gather}
T = - \ln \frac{\varepsilon}{4} + \frac{1}{8}, \qquad T_0 = - \ln \frac{\varepsilon}{4} + \frac{1}{4};
\end{gather}
(ii) the narrow escape limit for the annulus geometry~\cite{rupprecht2015, mangeat2019} when $\Delta V \to +\infty$:
\begin{gather}
T(\chi) = (1-\chi^2) \left[- \ln \frac{\varepsilon}{4} - 2 \sum_{k=1}^\infty \ln(1-\chi^{2k}) \right] + \frac{1}{8} - \frac{ 3 \chi^2}{8} - \frac{\chi^4\ln \chi}{2(1-\chi^2)},\\
T_0(\chi) = (1-\chi^2) \left[- \ln \frac{\varepsilon}{4} - 2 \sum_{k=1}^\infty \ln(1-\chi^{2k}) \right] + \frac{1}{4} + \frac{D_1-D_2}{4D_2} \chi^2 + \frac{1}{2}\chi^2\ln \chi ;
\end{gather}
and (iii) the narrow escape limit for the two-shell geometry with heterogeneous diffusivity~\cite{mangeat2019} when $\Delta V =0$:
\begin{gather}
T(\chi) = - \ln \frac{\varepsilon}{4} - 2 \sum_{k=1}^\infty \left[ \frac{D_1-D_2}{D_1 + D_2} \right]^k \ln(1-\chi^{2k}) + \frac{1}{8} + \frac{D_1-D_2}{8D_2} \chi^4, \\
T_0(\chi) = - \ln \frac{\varepsilon}{4} - 2 \sum_{k=1}^\infty \left[ \frac{D_1-D_2}{D_1 + D_2} \right]^k \ln(1-\chi^{2k}) + \frac{1}{4} + \frac{D_1-D_2}{4D_2} \chi^2.
\end{gather}

\begin{figure}[t]
\begin{center}
\includegraphics[width=17cm]{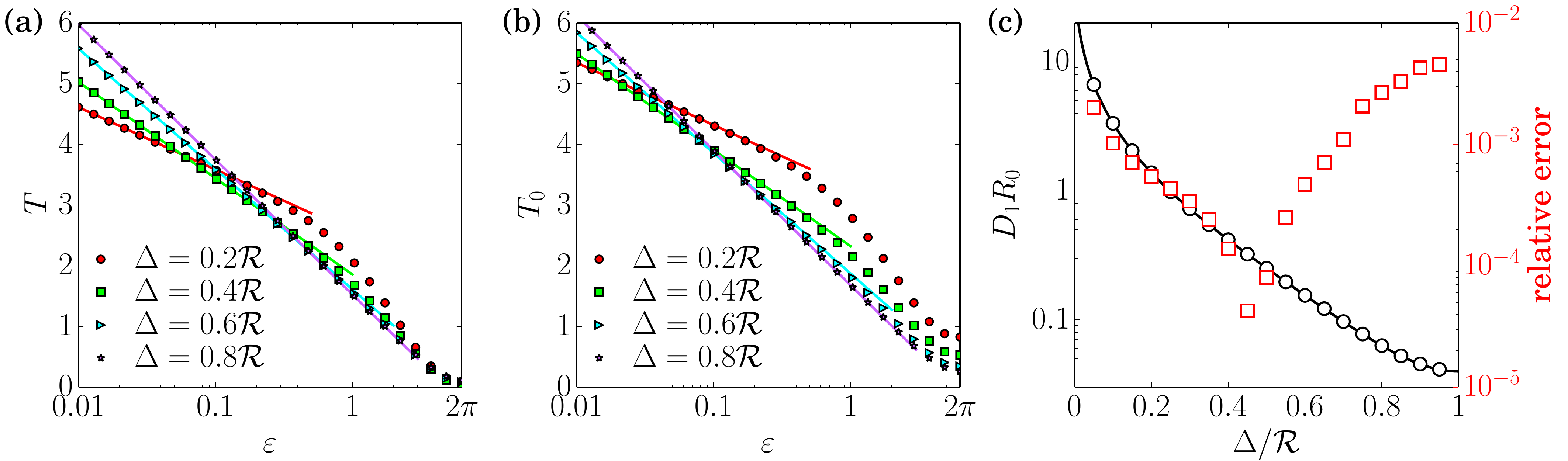}
\caption{Dependence of the dimensionless GMFPT~{\bf (a)} and CMFPT~{\bf (b)} on the escape angle $\varepsilon$ in 2d for several widths $\Delta/{\cal R}$ and fixed parameters $D_1/D_2=5$ and $\beta \Delta V = 2$. The symbols show the numerical solution obtained with the finite element method using FreeFem++ and the lines display the analytical expressions given by Eqs.~(\ref{eqGMFPT2d}) and~(\ref{eqCMFPT2d}) in the narrow escape limit. {\bf (c)}~Correction to the leading term $D_1 R_0$ as a function of $\Delta / {\cal R}$ for fixed parameters $D_1/D_2=5$ and $\beta \Delta V = 2$. The circles represent the fitted value, extracted from numerical solutions with Eq.~(\ref{expNEGMFPT}), while the line shows the analytical solution from Eq.~(\ref{eqGMFPT2d}). The squares display the relative difference between the numerical fit and the analytical expression.} \label{figure2d_NE}
\end{center}
\end{figure}

Figs.~\ref{figure2d_NE}(a) and~\ref{figure2d_NE}(b) display the dependence of the GMFPT and the CMFPT, respectively, on the escape angle for several widths of the outer shell $\Delta$ and the fixed parameters $D_1/D_2 =5$ and $\beta \Delta V = 2$. The analytical expressions (lines) given by Eqs.~(\ref{eqGMFPT2d}) and~(\ref{eqCMFPT2d}) match the numerical solutions (symbols) in the narrow escape limit. The two MFPTs show the same logarithmic decrease with $\varepsilon$ and an absolute slope increasing with $\Delta$, already observed in Fig.~\ref{figure_simus}(a). The accuracy of the narrow escape expressions increases with $\Delta$. Fig.~\ref{figure2d_NE}(c) shows the correction to the leading order $D_1 R_0$ defined as the regular part of the Green's function at the center of the escape region and extracted numerically (circles) from a linear regression analysis of numerical solutions of GMFPT as a function of $\ln\varepsilon$ by using Eq.~(\ref{expNEGMFPT}). Comparing it with the analytical expression derived in appendix~\ref{appendix2d} (line), the maximal relative error is smaller than 1\% (squares) for $D_1/D_2 =5$ and $\beta \Delta V = 2$, except for $\Delta \sim 0.45 {\cal R}$ when $R_0$ crosses $0$. Note that the sub-leading corrections are compared and the accumulated numerical error is less than $10^{-4}$.

\begin{figure}[t]
\begin{center}
\includegraphics[width=17cm]{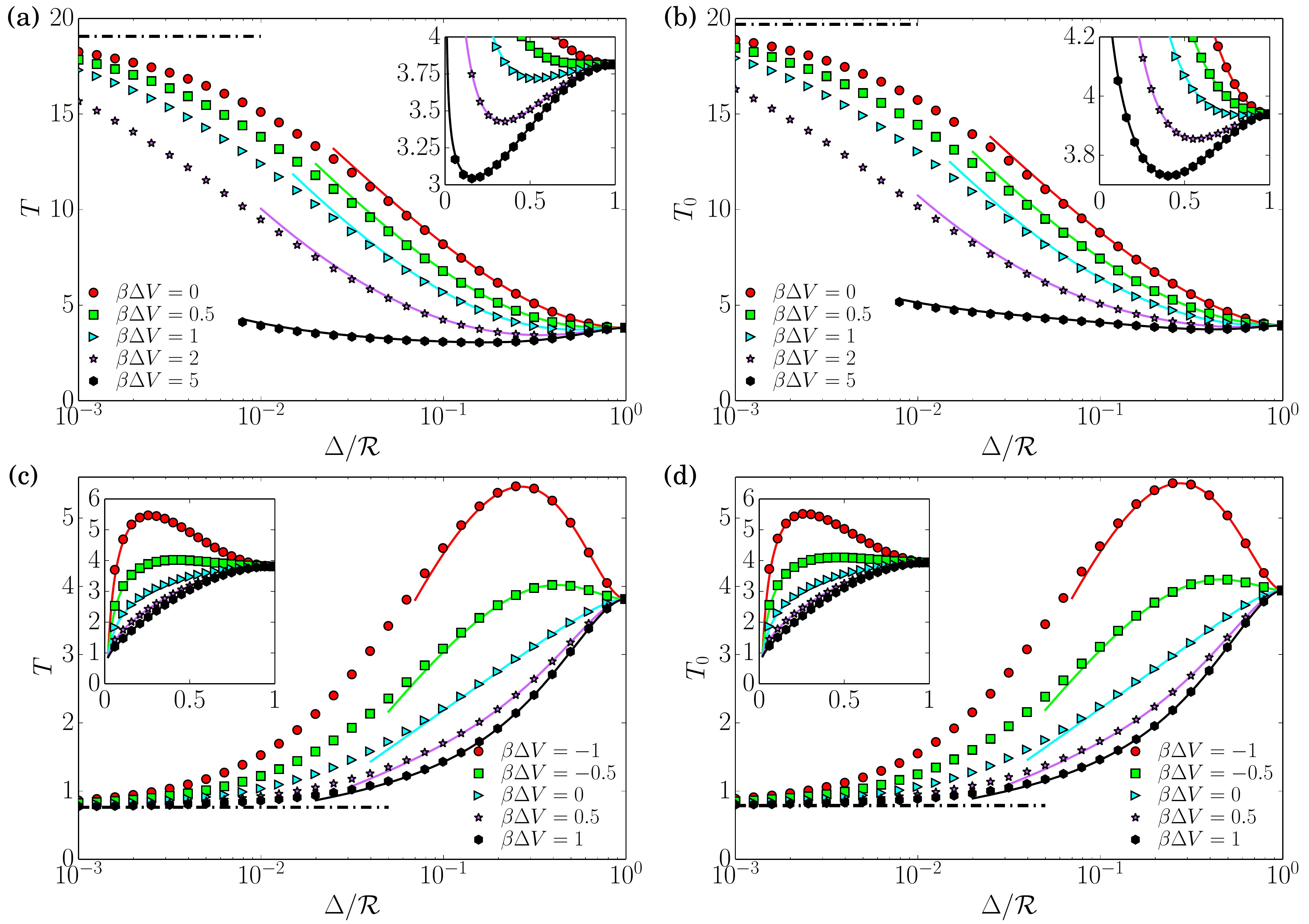}
\caption{Dependence of the dimensionless GMFPT~[{\bf (a)} and {\bf (c)}] and CMFPT~[{\bf (b)} and {\bf (d)}] on the width $\Delta/{\cal R}$ in 2d for several potential differences $\beta \Delta V$, the escape angle $\varepsilon=0.1$ and the ratio of diffusion constants $D_1/D_2=5$~[{\bf (a)} and {\bf (b)}] and $D_1/D_2=0.2$~[{\bf (c)} and {\bf (d)}]. The symbols display the numerical solutions obtained with the finite element method using FreeFem++ and the lines show the analytical expressions given by Eqs.~(\ref{eqGMFPT2d}) and~(\ref{eqCMFPT2d}) for the GMFPT and the CMFPT respectively, in the narrow escape limit (here for $\Delta$ close to ${\cal R}$). The dash-dotted lines display the $\Delta \to 0$ limit given by Eq.~(\ref{MFPTDelta0}). In insets, the same curves are shown in linear scale for a better observation of the minima exhibited by the GMFPT and the CMFPT.} \label{figure2d_DELTA}
\end{center}
\end{figure}

Fig.~\ref{figure2d_DELTA} shows the dependence of the GMFPT and the CMFPT on the width of the outer shell $\Delta$ for fixed escape angle $\varepsilon=0.1$. The numerical solutions (symbols) agree well with the narrow escape expressions (lines) given by Eqs.~(\ref{eqGMFPT2d}) and~(\ref{eqCMFPT2d}) for $\Delta$ close to ${\cal R}$ and approach the thin cortex expressions (dash-dotted lines) given by Eq.~(\ref{MFPTDelta0}) in the $\Delta \to 0$ limit. In figs.~\ref{figure2d_DELTA}(a) and~\ref{figure2d_DELTA}(b) for a larger diffusion constant in the cortex ($D_1/D_2 = 5$), the GMFPT has a minimum for $\beta \Delta V>0.48$ and the CMFPT for $\beta \Delta V>0.85$ compatible with the inset. The minima are perfectly described by the narrow escape expressions. However, in Figs.~\ref{figure2d_DELTA}(c) and~\ref{figure2d_DELTA}(d) for a smaller diffusion constant in the cortex ($D_1/D_2 = 0.2$), the value of the GMFPT and the CMFPT for $\Delta = 0$ is always smaller than for $\Delta = {\cal R}$ due to the fast motion in the central region, consistent with the relation~(\ref{MFPTDelta0rel}) derived in Sec.~\ref{sectionThinCortex}. This implies that the MFPT is always minimal for $\Delta = 0$ when $D_1<D_2$.

We focus now on $D_1>D_2$ for which the MFPT can be optimized for a width $0<\Delta<{\cal R}$. For the GMFPT, the Taylor expansion of Eq.~(\ref{eqGMFPT2d}) at $\chi=0$ gives
\begin{equation}
T(\chi) = -\ln \frac{\varepsilon}{4} + \frac{1}{8} + \chi^2 \left\{ (\xi-1)\left[-\ln \frac{\varepsilon}{4}+\frac{3}{8} \right]+2  \frac{D_1-D_2 \xi}{D_1 + D_2 \xi} \right\} + {\cal O}(\chi^4).
\end{equation}
The optimization is then possible if and only if $T(\chi) - T(0) <0$ or 
\begin{equation}
(\xi-1) T_* +2  \frac{D_1-D_2 \xi}{D_1 + D_2 \xi} <0, \qquad {\rm with} \qquad T_* = -\ln \frac{\varepsilon}{4}+\frac{3}{8}.
\end{equation}
This condition is satisfied for 
\begin{equation}
\label{conditionT2d}
\xi < \frac{(D_2-D_1)T_* + 2D_2 + \sqrt{[(D_2-D_1)T_*+2D_2]^2-4D_1D_2(2-T_*)T_*}}{2D_2T_*}.
\end{equation}
In the limit $\varepsilon \ll 1$, $T_*$ diverges and the last condition can be rewritten as
\begin{equation}
\beta \Delta V > \frac{2(D_1-D_2)}{(D_1+D_2)T_*} + {\cal O}(T_*^{-2}).
\end{equation}

\begin{figure}[t]
\begin{center}
\includegraphics[width=17cm]{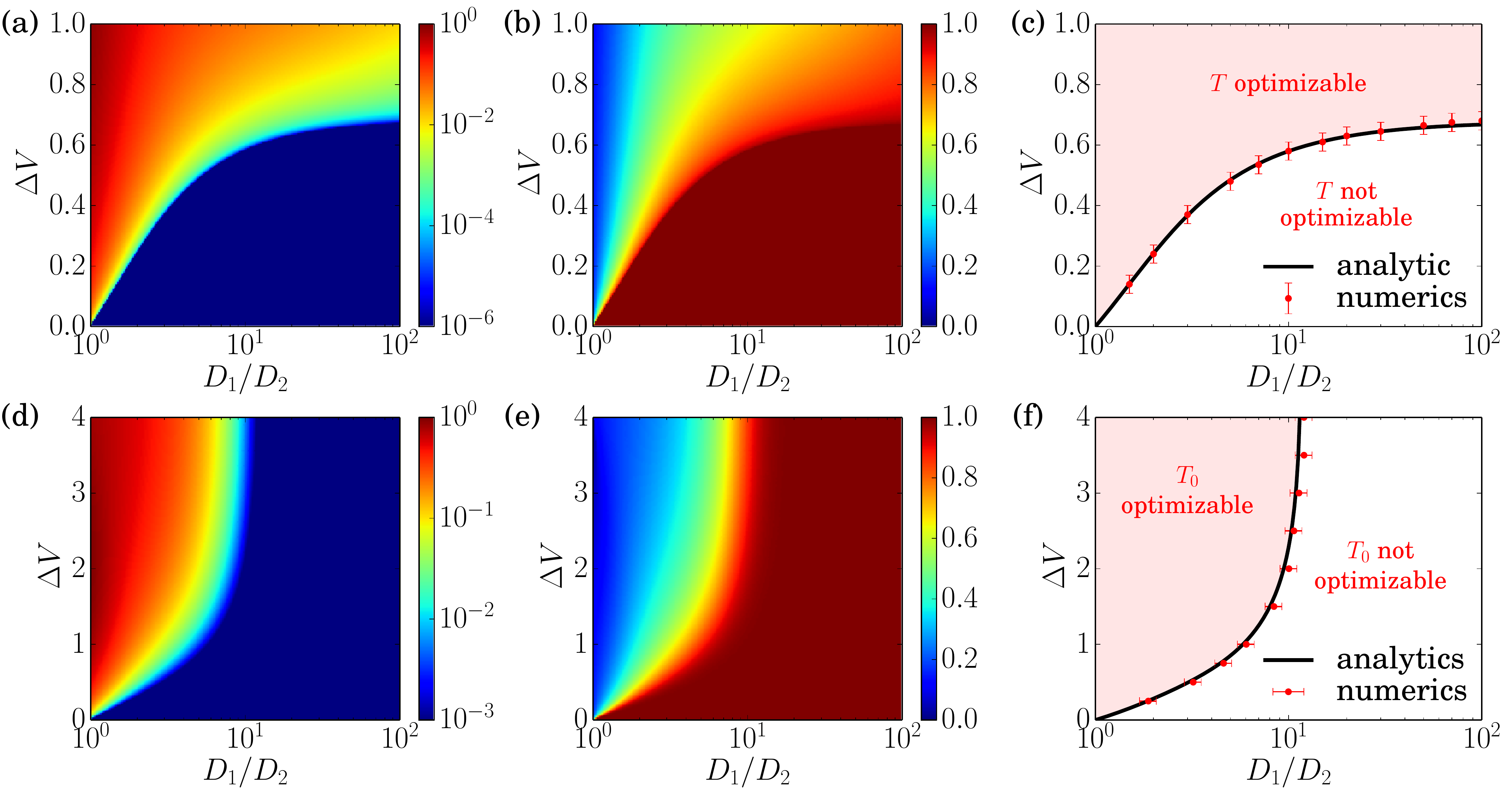}
\caption{MFPT optimization in 2d for $\varepsilon=0.1$. {\bf (a)}~and {\bf (d)}~Difference between the minimal GMFPT/CMFPT and the GMFPT/CMFPT for the disk geometry: $T(\chi_*) - T(0)$ (resp. $T_0(\chi_*) - T_0(0)$) as a function of $D_1/D_2$ and $\beta \Delta V$, calculated with the narrow escape expression given by Eqs.~(\ref{eqGMFPT2d}) and~(\ref{eqCMFPT2d}) respectively. {\bf (b)}~and {\bf (e)}~Corresponding widths $\Delta_*/{\cal R} = 1-\chi_*$ for which the GMFPT/CMFPT is minimal, as a function of $D_1/D_2$ and $\beta \Delta V$. {\bf (c)}~and {\bf (f)}~GMFPT/CMFPT optimization diagrams deduced from the numerical solutions obtained with the finite element method using FreeFem++ (dots) and the analytical condition given by Eqs.~(\ref{conditionT2d}) and~(\ref{conditionT02d}) respectively.} \label{figure2d_optimization}
\end{center}
\end{figure}

Fig.~\ref{figure2d_optimization}(a) shows the difference between the minimal GMFPT $T(\chi_*)$ and the GMFPT for the disk geometry $T(0) = - \ln (\varepsilon/4) + 1/8$ as a function of $D_1/D_2>1$ and $\beta \Delta V > 0$ for an escape angle $\varepsilon=0.1$. In other quadrants this minimal value is zero for $D_1>D_2$ and $\Delta V<0$, or $T(1)-T(0)=(D_1/D_2-1) T(0)$ for $D_1<D_2$, considering the Eq.~(\ref{MFPTDelta0rel}). The GMFPT can be minimized for $D_1=D_2$ and $\Delta V>0$ leading to a discontinuity of this minimal MFPT between the different quadrants. Fig.~\ref{figure2d_optimization}(b) displays the corresponding width $\Delta_* / {\cal R}$ for which the minimum of the GMFPT is attained. Finally Fig.~\ref{figure2d_optimization}(c) shows the GMFPT optimization diagram derived from Eq.~(\ref{conditionT2d}) and the numerical solutions (dots). For $\beta \Delta V > 0.68$ the GMFPT is always optimizable, a value which depends on $\varepsilon$ as $\ln \left| T_*/(T_*-2) \right|$.

For the CMFPT, the Taylor expansion of Eq.~(\ref{eqCMFPT2d}) at $\chi=0$ gives
\begin{gather}
T_0(\chi) = -\ln \frac{\varepsilon}{4} + \frac{1}{4} + \chi^2 \left\{ (\xi-1)\left[-\ln \frac{\varepsilon}{4} - \frac{1}{2} \ln \chi \right] +2  \frac{D_1-D_2 \xi}{D_1 + D_2 \xi} + \frac{D_1-D_2}{4D_2}  \right\} + {\cal O}(\chi^4).
\end{gather}
The optimization is then possible if and only if $T_0(\chi)-T_0(0)<0$ considering all the leading terms in the narrow escape limit and corresponding terms in the Taylor expansion, i.e. assuming that $\ln \varepsilon$ and $\ln \chi$ are of same order for a value $\chi = \chi_0 \ll 1$. We obtain then the condition
\begin{equation}
(\xi - 1) T_* +2  \frac{D_1-D_2 \xi}{D_1 + D_2 \xi} + \frac{D_1-D_2}{4D_2} <0, \qquad T_* = -\ln \frac{\varepsilon}{4} - \frac{1}{2} \ln \chi_0,
\end{equation}
where $\chi_0$ is to be determined. This condition is satisfied for 
\begin{equation}
\label{conditionT02d}
\xi < \frac{(D_2-D_1)T_* + 2D_2 + \frac{D_2-D_1}{4} + \sqrt{\left[(D_2-D_1)T_* + 2D_2 + \frac{D_2-D_1}{4}\right]^2-4D_1D_2\left(2-T_*+ \frac{D_1-D_2}{4D_2}\right)T_*}}{2D_2T_*}.
\end{equation}
In the limit $\varepsilon \ll 1$, $T_*$ diverges and the last condition can be rewritten as
\begin{equation}
\beta \Delta V > \frac{(D_1-D_2)(D_1+9D_2)}{4D_2(D_1+D_2)T_*} + {\cal O}(T_*^{-2}). 
\end{equation}

Fig.~\ref{figure2d_optimization}(d) shows the difference between the minimal CMFPT $T_0(\chi_*)$ and the CMFPT for the disk geometry $T_0(0) = - \ln (\varepsilon/4) + 1/4$ as a function of $D_1/D_2>1$ and $\beta \Delta V > 0$ for an escape angle $\varepsilon=0.1$. In other quadrants this minimal value is $(D_1/D_2-1)T_0(0)$ (for $D_1<D_2$) or zero. Fig.~\ref{figure2d_optimization}(e) displays the corresponding width $\Delta_* / {\cal R}$ for which the minimum of the CMFPT is attained. Finally Fig.~\ref{figure2d_optimization}(f) shows the CMFPT optimization diagram derived from Eq.~(\ref{conditionT02d}) and the numerical solutions (dots). The numerical solution yields $\chi_0 \simeq \exp(-2) \simeq 0.13$ (independent of $\varepsilon$) which implies $T_* = -\ln (\varepsilon/4) +1$. Additionally for $D_1/D_2 > 11.75$ the CMFPT is never optimizable, a value which depends on $\varepsilon$ as $-7+4T_*$. In fact the right hand side of Eq.~(\ref{conditionT02d}) becomes negative implying that $\xi$ needs to be negative to have a CMFPT optimization which cannot be satisfied.

%%%%%%%%%%%%%%%%%%%%%%%%
%%%%%%%%%%%%%%%%%%%%%%%%
%%% NARROW ESCAPE 3D %%%
%%%%%%%%%%%%%%%%%%%%%%%%
%%%%%%%%%%%%%%%%%%%%%%%%

\subsection{Narrow escape limit in three dimensions}

In appendix~\ref{appendixGeneralNE} we show that close to the escape region the 3d Green's function behaves as
\begin{equation}
G({\bf x} \rightarrow {\bf x_0} |{\bf x_0}) = \frac{1}{2\pi D({\bf x_0})} \frac{1}{|{\bf x} - {\bf x_0}|} + \gamma \ln \frac{|{\bf x} - {\bf x_0}|}{2{\cal R}} + R_0,
\end{equation}
where $\gamma$ and $R_0$ are respectively the logarithmic diverging and regular parts of the pseudo Green's function at the center of the escape region and the average MFPT verifies the expression
\begin{equation}
\label{expNEGMFPT3d}
\langle t \rangle =  \frac{1}{P_s({\bf x_0})} \left[ \frac{1}{2 D({\bf x_0}) {\cal R} \varepsilon} + \gamma \ln \varepsilon + R_0 - \frac{3}{2} \gamma \right].
\end{equation}

In the appendix~\ref{appendix3d} we derive the pseudo Green's function and identify  $\gamma$ and $R_0$. The dimensionless GMFPT is
\begin{gather}
T(\chi) = \frac{D_1 \langle t \rangle}{{\cal R}^2} = \left[1-\chi^3(1-\xi)\right] \left\{ \frac{2\pi}{3\varepsilon} - \frac{1}{3} \ln \varepsilon + \frac{1}{3} \sum_{n=1}^\infty \frac{(2n+1)^2}{n(n+1)} \frac{u_n \chi^{2n+1}}{1-u_n \chi^{2n+1}} - \frac{1}{6}  \right\} \nonumber \\
+ \frac{D_2 + [(D_1-D_2)\xi + 4D_2 (1-\xi) + 5 D_2(1-\xi)^2]\chi^5}{15 D_2 \left[1-\chi^3(1-\xi)\right]} - \frac{(1-\xi)\left[1+\chi^3(1-\xi)\right]}{3 \left[1-\chi^3(1-\xi)\right]}\chi^3,\label{eqGMFPT3d}
\end{gather}
with $\chi = 1 - \Delta / {\cal R}$, $\xi = \exp(-\beta \Delta V)$ and the sequence
\begin{equation}
u_n = \frac{(n+1)(D_1-D_2\xi)}{(n+1) D_1 + n D_2 \xi},
\end{equation}
and the dimensionless CMFPT is
\begin{gather}
T_0(\chi) = \frac{D_1 t({\bf 0})}{{\cal R}^2} = \left[1-\chi^3(1-\xi)\right] \left\{\frac{2\pi}{3\varepsilon} - \frac{1}{3} \ln \varepsilon + \frac{1}{3} \sum_{n=1}^\infty \frac{(2n+1)^2}{n(n+1)} \frac{u_n \chi^{2n+1}}{1-u_n \chi^{2n+1}} - \frac{1}{6} \right\} \nonumber \\
+ \frac{1}{6} + \frac{D_1-D_2}{6D_2} \chi^2 - \frac{1-\xi}{3} \chi^2(1-\chi).\label{eqCMFPT3d}
\end{gather}

Consequently, the GMFPT and the CMFPT can be written as the sum of two terms: (i) the first line of each expression is common and contains the leading order $[1-\chi^3(1-\xi)] [ 2\pi/3\varepsilon - \ln \varepsilon/3 -1/6 ]$ and (ii) the second line is the corresponding MFPT for a fully absorbing external boundary ($\varepsilon=2\pi$), see the derivation in the appendix~\ref{appendixF}. Note that the series in common term is zero when $\beta \Delta V= \ln(D_2/D_1)$ as for 2d NEP. From Eqs.~(\ref{eqGMFPT3d}) and~(\ref{eqCMFPT3d}), we recover the expressions for the sphere geometry~\cite{cheviakov2010} when $D_1=D_2$ and $\Delta V=0$:
\begin{equation}
T = \frac{2\pi}{3\varepsilon} - \frac{1}{3} \ln \varepsilon - \frac{1}{10}, \qquad T_0 = \frac{2\pi}{3\varepsilon} - \frac{1}{3} \ln \varepsilon;
\end{equation}
and we mention two other interesting cases studied in 2d in Ref.~\cite{mangeat2019}: (i) the narrow escape limit for the annulus geometry when $\Delta V \to +\infty$:
\begin{gather}
T(\chi) = (1-\chi^3) \left[ \frac{2\pi}{3\varepsilon} - \frac{1}{3} \ln \varepsilon + \frac{1}{3} \sum_{n=1}^\infty \frac{(2n+1)^2}{n(n+1)} \frac{\chi^{2n+1}}{1-\chi^{2n+1}} \right] - \frac{1-6\chi^5+5\chi^6}{10(1-\chi^3)},\\
T_0(\chi) = (1-\chi^3) \left[ \frac{2\pi}{3\varepsilon} - \frac{1}{3} \ln \varepsilon + \frac{1}{3} \sum_{n=1}^\infty \frac{(2n+1)^2}{n(n+1)} \frac{\chi^{2n+1}}{1-\chi^{2n+1}} \right] + \frac{D_1-D_2}{6D_2} \chi^2 - \frac{\chi^2}{6}(2-3\chi);
\end{gather}
and (ii) the narrow escape limit for the two-shell geometry with heterogeneous diffusivity when $\Delta V =0$:
\begin{gather}
T(\chi) =\frac{2\pi}{3\varepsilon} - \frac{1}{3} \ln \varepsilon + \frac{1}{3} \sum_{n=1}^\infty \frac{(2n+1)^2}{n(n+1)} \frac{u_n \chi^{2n+1}}{1-u_n \chi^{2n+1}} - \frac{1}{6} + \frac{D_2+(D_1-D_2)\chi^5}{15D_2}, \\
T_0(\chi) = \frac{2\pi}{3\varepsilon} - \frac{1}{3} \ln \varepsilon + \frac{1}{3} \sum_{n=1}^\infty \frac{(2n+1)^2}{n(n+1)} \frac{u_n \chi^{2n+1}}{1-u_n \chi^{2n+1}} + \frac{D_1-D_2}{6D_2} \chi^2.
\end{gather}

\begin{figure}[t]
\begin{center}
\includegraphics[width=17cm]{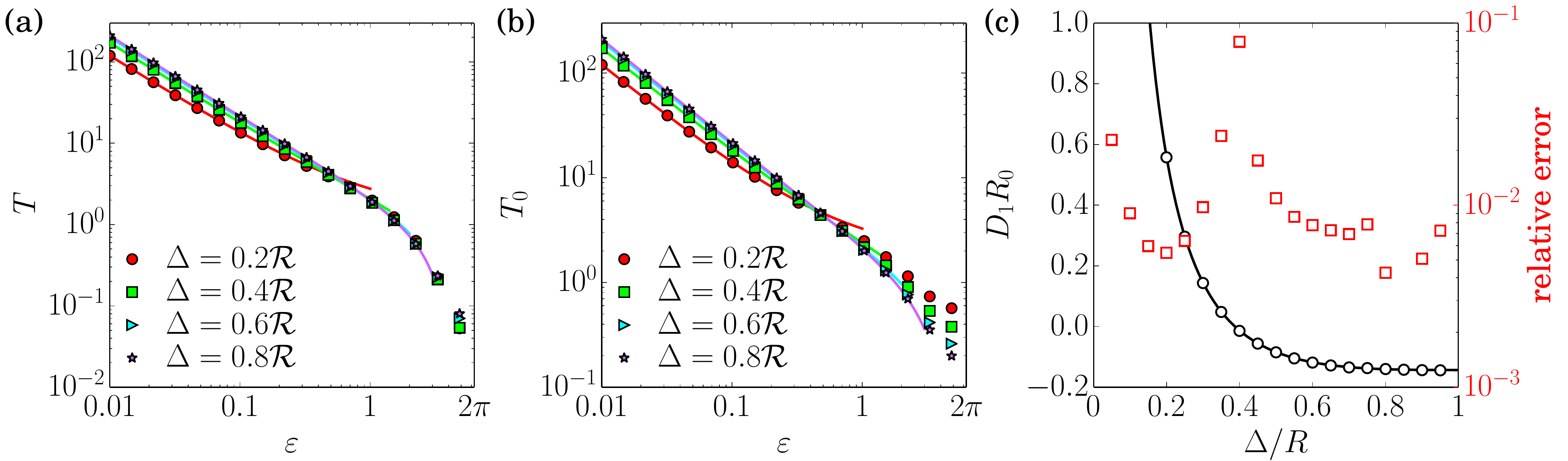}
\caption{Dependence of the dimensionless GMFPT~{\bf (a)} and CMFPT~{\bf (b)} on the escape angle $\varepsilon$ in 3d for several widths $\Delta/{\cal R}$ and fixed parameters $D_1/D_2=5$ and $\beta \Delta V = 2$. The symbols show the numerical solution obtained with the finite element method using FreeFem++ and the lines display the analytical expressions given by Eqs.~(\ref{eqGMFPT3d}) and~(\ref{eqCMFPT3d}) in the narrow escape limit. {\bf (c)}~Correction to the leading term $D_1 R_0$ for the unit sphere ${\cal R}=1$ as a function of $\Delta / {\cal R}$ for fixed parameters $D_1/D_2=5$ and $\beta \Delta V = 2$. The circles represent the fitted value, extracted from numerical solutions with Eq.~(\ref{expNEGMFPT3d}), while the line shows the analytical solution from Eq.~(\ref{eqGMFPT3d}). The squares display the relative difference between the numerical fit and the analytical expression.} \label{figure3d_NE}
\end{center}
\end{figure}

Figs.~\ref{figure3d_NE}(a) and~\ref{figure3d_NE}(b) display the dependence of the GMFPT and the CMFPT, respectively, on the escape angle for several widths of the outer shell $\Delta$ and the fixed parameters $D_1/D_2 =5$ and $\beta \Delta V = 2$. The analytical expressions (lines) given by Eqs.~(\ref{eqGMFPT3d}) and~(\ref{eqCMFPT3d}) match the numerical solutions (symbols) in the narrow escape limit. The two MFPTs show the same hyperbolic decrease in $\varepsilon$, already observed in Fig.~\ref{figure_simus}(b), and the accuracy of the narrow escape expressions increases with $\Delta$. Fig.~\ref{figure3d_NE}(c) shows the correction to the leading order $D_1 R_0$ defined as the regular part of the Green's function (for the unit sphere) at the center of the escape region and extracted numerically (circles) from a regression analysis of numerical solutions of GMFPT as a function of $\varepsilon$ by using Eq.~(\ref{expNEGMFPT3d}). Comparing it with analytical expression derived in appendix~\ref{appendix3d} (line), the maximal relative error is smaller than 3\% (squares), except for $\Delta \sim 0.4 {\cal R}$ when $R_0$ crosses $0$. Note that the second order sub-leading corrections (instead of first order in two dimensions) are compared and the accumulated numerical error is less than $10^{-4}$.

\begin{figure}[t]
\begin{center}
\includegraphics[width=17cm]{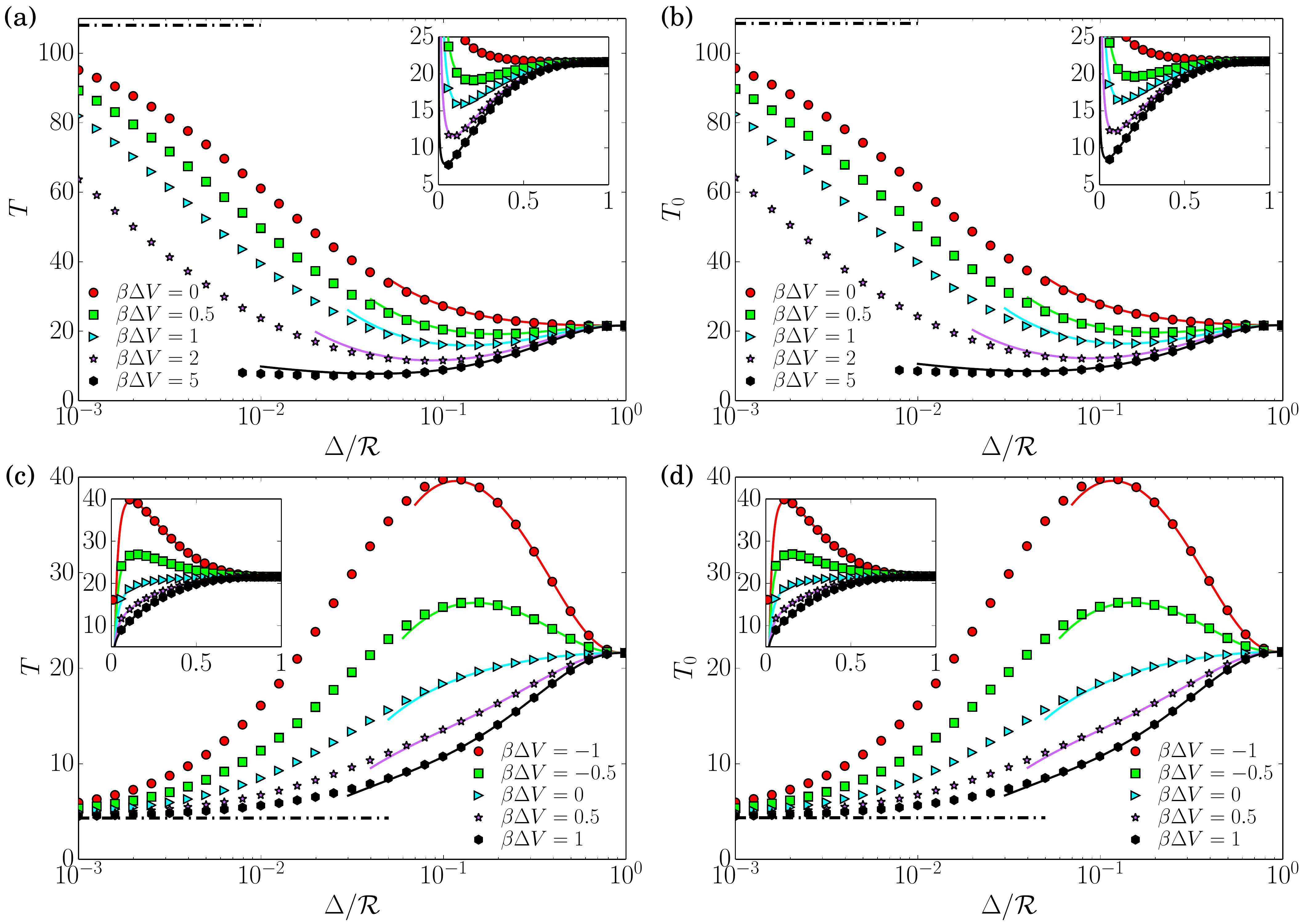}
\caption{Dependence of the dimensionless GMFPT~[{\bf (a)} and {\bf (c)}] and CMFPT~[{\bf (b)} and {\bf (d)}] on the width $\Delta/{\cal R}$ in 3d for several potential differences $\beta \Delta V$, the escape angle $\varepsilon=0.1$ and the ratio of diffusion constants $D_1/D_2=5$~[{\bf (a)} and {\bf (b)}] and $D_1/D_2=0.2$~[{\bf (c)} and {\bf (d)}]. The symbols display the numerical solutions obtained with the finite element method using FreeFem++ and the lines show the analytical expressions given by Eqs.~(\ref{eqGMFPT3d}) and~(\ref{eqCMFPT3d}) for the GMFPT and the CMFPT respectively, in the narrow escape limit (here for $\Delta$ close to ${\cal R}$). The dash-dotted lines display the $\Delta \to 0$ limit given by Eq.~(\ref{MFPTDelta0}). In insets, the same curves are shown in linear scale for a better observation of the minima exhibited by the GMFPT and the CMFPT.} \label{figure3d_DELTA}
\end{center}
\end{figure}

Fig.~\ref{figure3d_DELTA} shows the dependence of the GMFPT and the CMFPT on the width of the outer shell $\Delta$ for fixed escape angle $\varepsilon=0.1$. The numerical solutions (symbols) agree well with the narrow escape expressions (lines) given by Eqs.~(\ref{eqGMFPT3d}) and~(\ref{eqCMFPT3d}) for $\Delta$ close to ${\cal R}$ and approach the thin cortex expressions (dash-dotted lines) given by Eq.~(\ref{MFPTDelta0}) in the $\Delta \to 0$ limit. In Figs.~\ref{figure3d_DELTA}(a) and~\ref{figure3d_DELTA}(b) for a larger diffusion constant in the cortex $D_1/D_2 = 5$, the GMFPT has a minimum for $\beta \Delta V>0.05$ and the CMFPT for $\beta \Delta V>0.14$ compatible with the inset. The minima are perfectly described by the narrow escape expressions as for 2d NEP. However, in Figs.~\ref{figure3d_DELTA}(c) and~\ref{figure3d_DELTA}(d) for a smaller diffusion constant in the cortex $D_1/D_2 = 0.2$ the value of the GMFPT and the CMFPT for $\Delta = 0$ is always smaller than for $\Delta = {\cal R}$ due to the fast motion in the central region, consistent with the relation~(\ref{MFPTDelta0rel}) derived in Sec.~\ref{sectionThinCortex}. This implies that the MFPT is always minimal for $\Delta = 0$ when $D_1<D_2$.

As for the 2d case, we focus on $D_1>D_2$ for which the MFPT can be optimized for a width $0<\Delta<{\cal R}$. For the GMFPT, the Taylor expansion of Eq.~(\ref{eqGMFPT3d}) at $\chi=0$ gives
\begin{equation}
T(\chi) = \frac{2\pi}{3\varepsilon} - \frac{1}{3} \ln \varepsilon - \frac{1}{10} + \chi^3 \left\{ (\xi-1)\left[\frac{2\pi}{3\varepsilon} - \frac{1}{3} \ln \varepsilon + \frac{1}{10} \right]+ 3 \frac{D_1-D_2 \xi}{2D_1 + D_2 \xi} \right\} + {\cal O}(\chi^5).
\end{equation}
The optimization is then possible if and only if $T(\chi) - T(0) <0$ or 
\begin{equation}
(\xi-1) T_* +3  \frac{D_1-D_2 \xi}{2D_1 + D_2 \xi} <0, \qquad T_* = \frac{2\pi}{3\varepsilon} - \frac{1}{3} \ln \varepsilon + \frac{1}{10}.
\end{equation}
This condition is satisfied for 
\begin{equation}
\label{conditionT3d}
\xi < \frac{(D_2-2D_1)T_*+3D_2 + \sqrt{[(D_2-2D_1)T_*+3D_2]^2-4D_1D_2(3-2T_*)T_*}}{2D_2T_*}.
\end{equation}
In the limit $\varepsilon \ll 1$, $T_*$ diverges and the last condition can be rewritten as
\begin{equation}
\beta \Delta V > \frac{3(D_1-D_2)}{(2D_1+D_2)T_*} + {\cal O}(T_*^{-2}).
\end{equation}

\begin{figure}[t]
\begin{center}
\includegraphics[width=17cm]{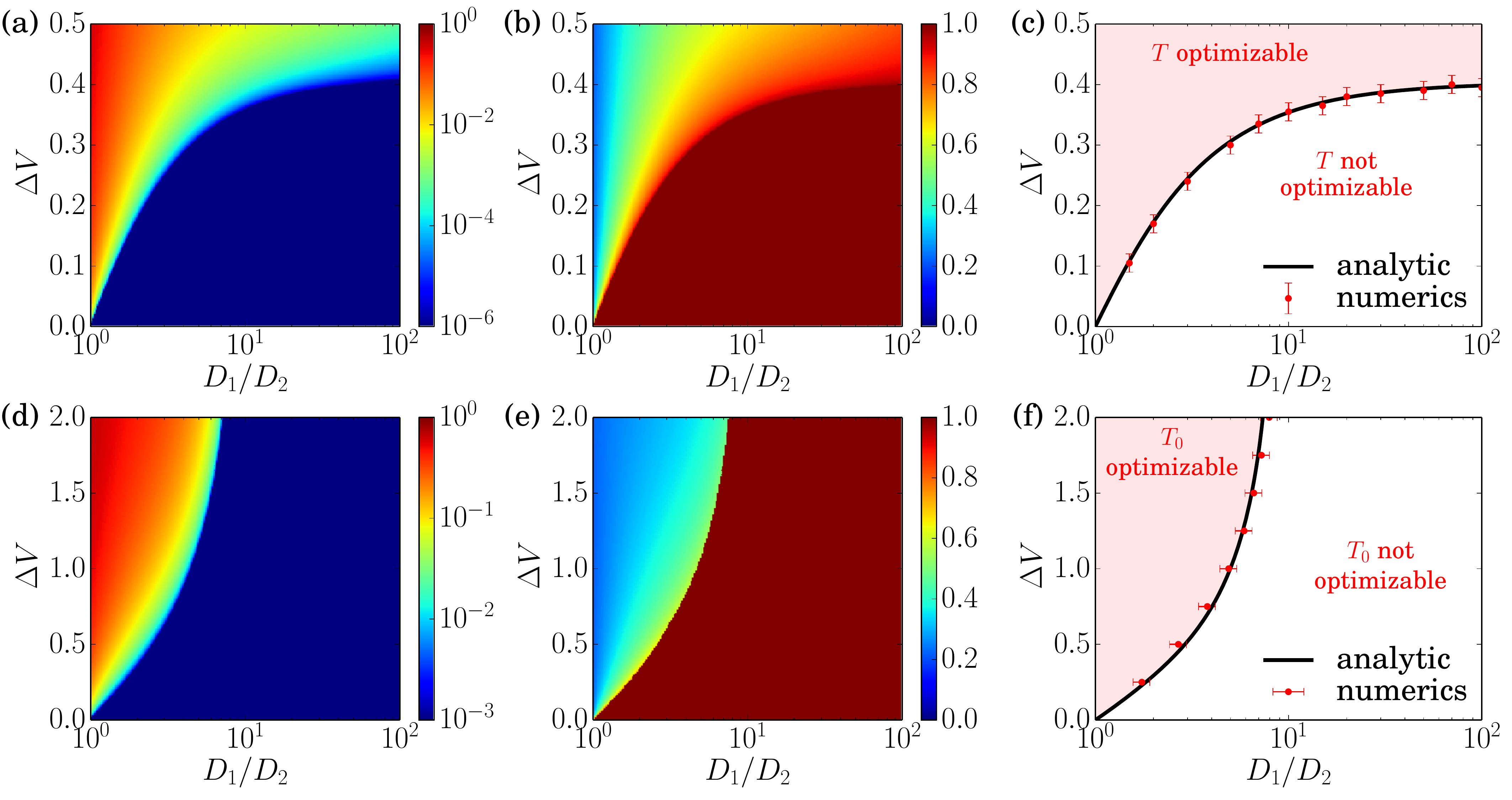}
\caption{MFPT optimization in 3d for $\varepsilon=0.5$. {\bf (a)}~and {\bf (d)}~Difference between the minimal GMFPT/CMFPT and the GMFPT/CMFPT for the disk geometry: $T(\chi_*) - T(0)$ (resp. $T_0(\chi_*) - T_0(0)$) as a function of $D_1/D_2$ and $\beta \Delta V$, calculated from the narrow escape expression given by Eqs.~(\ref{eqGMFPT3d}) and~(\ref{eqCMFPT3d}) respectively. {\bf (b)}~and {\bf (e)}~Corresponding widths $\Delta_*/{\cal R} = 1-\chi_*$ for which the GMFPT/CMFPT is minimal, as a function of $D_1/D_2$ and $\beta \Delta V$. {\bf (c)}~and {\bf (f)}~GMFPT/CMFPT optimization diagrams deduced from the numerical solutions obtained with the finite element method using FreeFem++ (dots) and the analytical condition given by Eqs.~(\ref{conditionT3d}) and~(\ref{conditionT03d}) respectively.} \label{figure3d_optimization}
\end{center}
\end{figure}

Fig.~\ref{figure3d_optimization}(a) displays the difference between the minimal GMFPT $T(\chi_*)$ and the GMFPT for the disk geometry $T(0) = 2\pi/3\varepsilon - \ln \varepsilon/3 - 1/10$ as a function of $D_1/D_2>1$ and $\beta \Delta V > 0$ for an escape angle $\varepsilon=0.5$. In other quadrants this minimal value is $(D_1/D_2-1)T(0)$ (for $D_1<D_2$) or zero. Fig.~\ref{figure3d_optimization}(b) shows the corresponding width $\Delta_* / {\cal R}$ for which the minimum of the GMFPT is attained. Finally Fig.~\ref{figure3d_optimization}(c) displays the GMFPT optimization diagram derived from Eq.~(\ref{conditionT3d}) and the numerical solutions (dots). For $\beta \Delta V > 0.40$ the GMFPT is always optimizable, a value which depends on $\varepsilon$ as $\ln \left| 2T_*/(2T_*-3) \right|$.

For the CMFPT, the Taylor expansion of Eq.~(\ref{eqCMFPT3d}) at $\chi=0$ gives
\begin{gather}
T_0(\chi) = \frac{2\pi}{3\varepsilon} - \frac{1}{3} \ln \varepsilon + \frac{\chi^2}{3} \left[ \frac{D_1-D_2}{2D_2} - 1+\xi \right]
+ \chi^3 \left\{ (\xi-1)\left[\frac{2\pi}{3\varepsilon} - \frac{1}{3} \ln \varepsilon - \frac{1}{2} \right]+ 3 \frac{D_1-D_2 \xi}{2D_1 + D_2 \xi} \right\} + {\cal O}(\chi^5).
\end{gather}
The optimization is then possible if and only if $T_0(\chi)-T_0(0)<0$ considering all the leading terms in the narrow escape limit and corresponding terms in the Taylor expansion, i.e. assuming that $\varepsilon^{-1}$ and $\chi^{-1}$ are of same order for a value $\chi = \chi_0 \ll 1$. We obtain then the condition
\begin{equation}
(\xi-1) T_* +  \frac{D_1-D_2}{6 D_2 \chi_0} + 3  \frac{D_1-D_2 \xi}{2D_1 + D_2 \xi} <0, \qquad T_* = \frac{2\pi}{3\varepsilon} - \frac{1}{3} \ln \varepsilon - \frac{1}{2} + \frac{1}{3\chi_0}.
\end{equation} 
where $\chi_0$ is to be determined. This condition is satisfied for 
\begin{equation}
\label{conditionT03d}
\xi < \frac{(D_2-2D_1)T_* + 3D_2 -\frac{D_1-D_2}{6\chi_0} + \sqrt{\left[(D_2-2D_1)T_*+3D_2-\frac{D_1-D_2}{6\chi_0}\right]^2-4D_1D_2\left(3-2T_*+\frac{D_1-D_2}{3D_2\chi_0}\right)T_*}}{2D_2T_*}.
\end{equation}
In the limit $\varepsilon \ll 1$, $T_*$ diverges and the last condition can be rewritten as
\begin{equation}
\beta \Delta V > \frac{3(D_1-D_2)}{(2D_1+D_2)T_*} + \frac{D_1-D_2}{6D_2 T_* \chi_0} + {\cal O}(T_*^{-2}).
\end{equation}
Numerically we have identified the constant $\chi_0 \simeq 0.4$.

Fig.~\ref{figure3d_optimization}(d) shows the difference between the minimal CMFPT $T_0(\chi_*)$ and the CMFPT for the disk geometry $T_0(0) = 2\pi/3\varepsilon - \ln \varepsilon/3$ as a function of $D_1/D_2>1$ and $\beta \Delta V > 0$ for an escape angle $\varepsilon=0.5$. In other quadrants this minimal value is $(D_1/D_2-1)T_0(0)$ (for $D_1<D_2$) or zero. Fig.~\ref{figure3d_optimization}(e) displays the corresponding width $\Delta_* / {\cal R}$ for which the minimum of the CMFPT is attained. Note that in contrast to the 2d case $\Delta_*$ changes abruptly, apparently discontinuously, when going from the left region (blue) to the right region (red), where $\Delta_* = {\cal R}$. Finally Fig.~\ref{figure3d_optimization}(f) shows the CMFPT optimization diagram from Eq.~(\ref{conditionT03d}) and the numerical solutions (dots). The numerical solution yields $\chi_0 \simeq 0.4$ (independent of $\varepsilon$). Additionally for $D_1/D_2 > 0.81$ the CMFPT is never optimizable, a value which depends on $\varepsilon$ as $1+(-9+6T_*)\chi_0$. In fact the right hand side of Eq.~(\ref{conditionT03d}) becomes negative implying that $\xi$ needs to be negative to have a CMFPT optimization which cannot be satisfied.

%%%%%%%%%%%%%%%%%%
%%%%%%%%%%%%%%%%%%
%%% DISCUSSION %%%
%%%%%%%%%%%%%%%%%%
%%%%%%%%%%%%%%%%%%

\section{Discussion}
\label{sectionDiscussion}

\begin{figure}[t]
\begin{center}
\includegraphics[width=17cm]{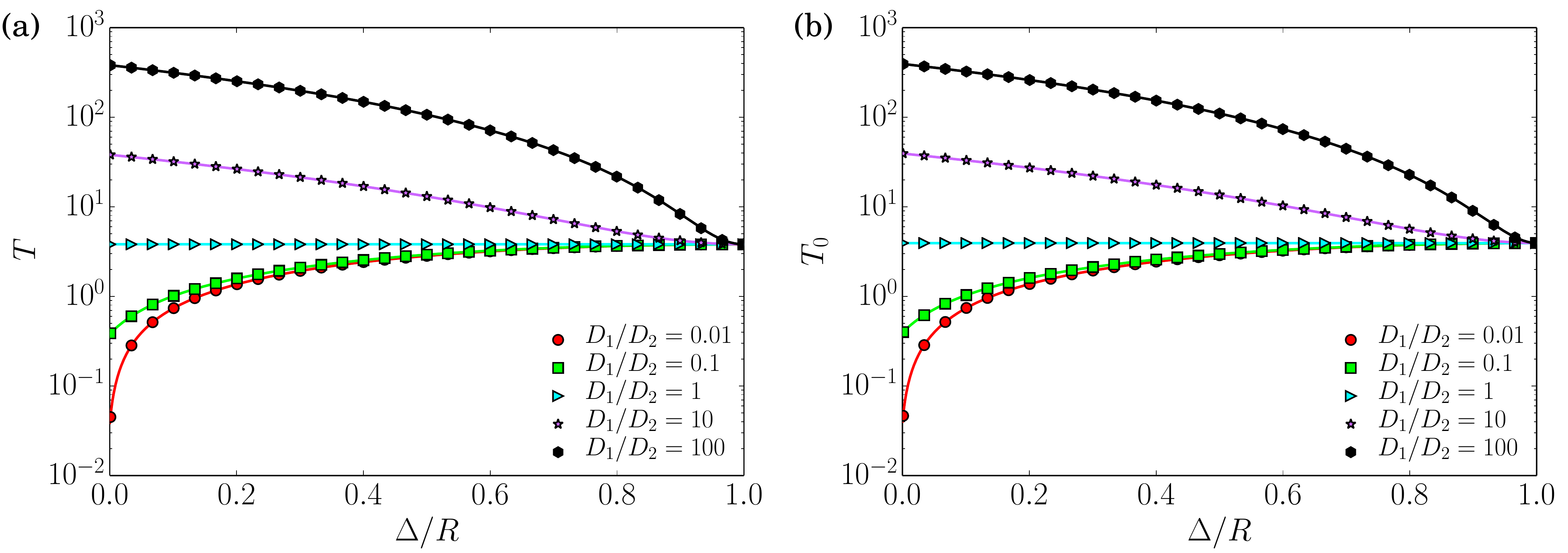}
\caption{Dependence of the dimensionless GMFPT~{\bf (a)} and CMFPT~{\bf (b)} on the width $\Delta/{\cal R}$ in 2d for the escape angle $\varepsilon=0.1$, different ratios of diffusion constants $D_1/D_2$ and a difference potential $\beta \Delta V = \ln(D_2/D_1)$. The symbols show the numerical solutions obtained with the finite element method using FreeFem++ and the lines display the exact analytical expressions given by Eqs.~(\ref{eqGMFPTGreb}) and~(\ref{eqCMFPTGreb}) for the GMFPT and the CMFPT respectively.} \label{figure2d_Grebenkov_DELTA}
\end{center}
\end{figure}

In Sec.~\ref{sectionThinCortex} we have derived the thin cortex limit of the MFPT given by Eqs.~(\ref{MFPTDelta0}) and~(\ref{MFPTDelta0annulus}) for respectively $\Delta V<+\infty$ and $\exp(-\beta\Delta V)=0$ (annulus geometry), which are both particular solutions of the SMDP. In Secs.~\ref{sectionNumerics} and~\ref{sectionNE} we have analyzed the evolution of the GMFPT and the CMFPT according to four parameters: the escape angle $\varepsilon$, the width of the outer shell $\Delta/{\cal R}$, the ratio of diffusion constants $D_1/D_2$ and the potential difference $\Delta V$. The dimensionless MFPT $D_1 t({\bf x})/{\cal R}^2$ is a decreasing function of $\varepsilon$ and $\Delta V$ and an increasing function of $D_1/D_2$, from simple arguments. However the behavior of the MFPT (as well as GMFPT or CMFPT) can be non-monotonous with $\Delta/{\cal R}$. For diffusion constants $D_1 < D_2$, the particles move faster to the escape region when the central region is as large as possible, which implies a minimum of GMFPT/CMFPT when $\Delta=0$ compatible with the relation~(\ref{MFPTDelta0rel}) in the thin cortex limit. For the opposite situation ($D_1 > D_2$) the presence of an optimization depends on the potential difference. If the cortex is repulsive ($\Delta V<0$) the particle moves faster when it is not energetically trapped in the central region, which implies a minimum of GMFPT/CMFPT when $\Delta={\cal R}$. Finally in the last quadrant ($D_1 > D_2$ and $\Delta V>0$) the GMFPT/CMPFT can be optimized due to the competition between the attractiveness of cortex (energetic trap) and the slower diffusion in the central region (diffusive trap). In Sec.~\ref{sectionNE} we have derived an analytical expression for both the GMFPT and the CMFPT in the narrow escape limit ($\varepsilon \ll 1$) in 2d and 3d. These expressions yield a quantitative condition for the potential difference for the MFPT optimization with a width $0 < \Delta_* < {\cal R}$. Nevertheless we did not derive an analytical expression for $\Delta_* = (1-\chi_*) {\cal R}$ as well as $T(\chi_*)$ and $T_0(\chi_*)$ due to the complexity of the narrow escape expressions and the rapidity to obtain them numerically (see Figs.~\ref{figure2d_optimization} and~\ref{figure3d_optimization}).

Finally we look more precisely at the special case $\beta V({\bf x}) = \ln D({\bf x})$. With this potential, the flux of Brownian particles is equal to $-\nabla_{\bf x} \left[ D({\bf x}) p({\bf x},t) \right]$ and the convective term in the It\^o convention, given by Eq.~(\ref{eqIto}), is zero. The MFPT then obeys the Poisson's equation $\nabla_{\bf x}^2 t({\bf x}) = - 1/D({\bf x})$. A recent study reported an exact solution for this equation in arbitrary two dimensional connected domains~\cite{grebenkov2016}. In appendix~\ref{appGreb} we derive from this result an exact solution of the MFPT in the two-shell geometry for which $\beta \Delta V = \ln(D_2/D_1)$. In particular, the GMFPT is
\begin{equation}
\label{eqGMFPTGreb}
T = -\left( 1 + \frac{D_1-D_2}{D_2} \chi^2 \right) \ln \sin(\varepsilon/4) + \frac{1}{8} + \frac{D_1-D_2}{8D_2} \frac{3D_2 +(D_1-3D_2)\chi^2}{D_2+(D_1-D_2)\chi^2} \chi^2 - \frac{(D_1-D_2)^2}{2D_2} \frac{\chi^4 \ln \chi}{D_2+(D_1-D_2)\chi^2},
\end{equation}
and the CMFPT is
\begin{equation}
\label{eqCMFPTGreb}
T_0 = - \left( 1 + \frac{D_1-D_2}{D_2} \chi^2 \right) \ln \sin(\varepsilon/4) + \frac{1}{4} + \frac{D_1-D_2}{4D_2} \chi^2 + \frac{D_2-D_1}{2D_2} \chi^2 \ln \chi.
\end{equation}
These two expressions are exact for all values of $\varepsilon$, $\chi=1-\Delta/{\cal R}$ and $D_1/D_2$, and are compatible with Eqs.~(\ref{eqGMFPT2d}) and~(\ref{eqCMFPT2d}) for $\beta \Delta V = \ln(D_2/D_1)$ in the limit $\varepsilon \ll 1$ (where $\sin(\varepsilon/4) \simeq \varepsilon/4$). Note that the series in Eqs.~(\ref{eqGMFPT2d}) and~(\ref{eqCMFPT2d}) is zero only for this potential difference. The thin cortex limits (i.e. $\chi=1$) are consistent with Eqs.~(\ref{MFPTDelta0rel}) and~(\ref{MFPTDelta0}).

The GMFPT and CMPFT are decreasing functions of the escape angle $\varepsilon$ and of the ratio of diffusion constants $D_1/D_2$. Fig.~\ref{figure2d_Grebenkov_DELTA} displays the dependence of the GMFPT and the CMFPT on the width of the outer shell $\Delta$ for $\varepsilon=0.1$ and several ratios $D_1/D_2$. The analytical expressions given by Eqs.~(\ref{eqGMFPTGreb}) and~(\ref{eqCMFPTGreb}) match perfectly the numerical solutions for all values of $\Delta$. For $D_1>D_2$, the potential difference is negative and the GMFPT/CMFPT is a decreasing function of $\Delta$ leading to a minimum MFPT located at $\Delta={\cal R}$. For $D_1<D_2$, the potential difference is positive and the GMFPT/CMFPT is an increasing function of $\Delta$ resulting in a minimal MFPT for $\Delta=0$. Unfortunately, the interesting range of parameters ($D_1>D_2$ and $\Delta V>0$), which allows an optimization of the MFPT with $\Delta>0$, is not attained for this particular potential difference.

To conclude we have characterized the MFPT for passive Brownian particles to reach a small escape window within the two-shell geometry in which the particles diffuse in each shell with different diffusive constant, with a potential barrier between the two shells. We have derived asymptotic expressions of the MFPT in the thin cortex and the narrow escape limits and we have obtained an exact solution only for one particular relation of the potential difference with the ratio of diffusion constants. For the most interesting case, showing a higher diffusion in the attractive cortex, we have derived from the narrow escape expressions a condition on the potential difference to have a MFPT optimization for particles starting from a random position or from the center. The potential barrier pushing the particles towards the surface needs to be large enough to balance the slowing down of the diffusion within the central region. With this result we may better understand the presence of a minimum for the intracellular transport time since the effective diffusivity is higher in the actin cortex, and the ballistic transport on filaments favors the transport to the cell membrane. Since the cargo particles perform an intermittent search, we may imagine that the MFPT is strongly reduced compared to our passive Brownian motion and that a condition on switching rates and ballistic velocity will replace the condition on the potential difference to have a MFPT optimization.

An interesting perspective would be to derive an exact expression for the GMFPT or the CMFPT, corresponding to the one derived in Ref.~\cite{grebenkov2016}, as well as for the optimized MFPT. Moreover, it would be interesting to derive a condition for an optimized intracellular transport time by reintroducing an intermittent ballistic transport for which, to our knowledge, no analytical expression has been reported so far. Finally another interesting perspective would be the study of the distribution of first passage time, and check whether the MFPT is relevant and close to the typical first passage time.

\section{Acknowledgement}

This work was performed with financial support from the German Research Foundation (DFG) within the Collaborative Research Center SFB 1027.

%%%%%%%%%%%%%%%%%%
%%%%%%%%%%%%%%%%%%
%%% APPENDICES %%%
%%%%%%%%%%%%%%%%%%
%%%%%%%%%%%%%%%%%%

\appendix

%%%%%%%%%%%%%%%%
%%% DELTA->0 %%%
%%%%%%%%%%%%%%%%

\section{The thin cortex limit}

\subsection{Link with the surface-mediated diffusion problem}
\label{appendixSMD}

From Eqs.~(\ref{MFPT2}) and~(\ref{eqSMBulk}), the bulk MFPT $t_{\rm B}$ can be identified with the MFPT $t_2$ when $\Delta \rightarrow 0$ whereas the MFPT $t_1(r,\theta)$ can be expanded in the radial coordinate as
\begin{equation}
\label{expandt1}
t_1(r,\theta) = f_0(\theta) + (r-{\cal R}) f_1(\theta) + \frac{(r-{\cal R})^2}{2} f_2(\theta) + \cdots
\end{equation}
Using Eq.~(\ref{MFPT4}), the reflective boundary condition $\partial_r t_1({\cal R},\theta) = 0$ gives $f_1(\theta)=0$ while the absorbing boundary condition $t_1({\cal R},\theta) = 0$ yields $f_0(\theta) = 0$ for $|\theta| < \varepsilon/2$. In leading order in the $\Delta \ll {\cal R}$ expansion, we can then set $f_0(\theta) = t_\Sigma(\theta)$. Eq.~(\ref{MFPT1}) can then be written as
\begin{equation}
D_1 \left[\frac{\partial^2 t_1}{\partial r^2}(r,\theta) + \frac{d-1}{r} \frac{\partial t_1}{\partial r}(r,\theta) + \frac{1}{r^2} \Delta_\theta t_1(r,\theta) \right] = -1,
\end{equation}
which becomes in leading order
\begin{equation}
\frac{D_1}{{\cal R}^2} \Delta_\theta t_\Sigma(\theta) + D_1 f_2(\theta) = -1.
\end{equation}
With Eq.~(\ref{eqSMSurf}), the function $f_2(\theta)$ can be identified:
\begin{equation}
\label{eqSMf2}
D_1 f_2(\theta) = - \frac{\lambda}{k} \frac{\partial t_{\rm B}}{\partial r}({\cal R},\theta).
\end{equation}
Finally the boundary conditions~(\ref{MFPT3}) give
\begin{gather}
t_\Sigma(\theta) = t_{\rm B}({\cal R},\theta), \\
\frac{\partial t_{\rm B}}{\partial r} ({\cal R},\theta) = - \frac{D_1 \exp(-\beta V_1)}{D_2 \exp(-\beta V_2)} \Delta f_2(\theta),
\end{gather}
considering the relation $\partial_r t_1 ({\cal R}-\Delta,\theta) = - \Delta f_2(\theta)$ from Eq.~(\ref{expandt1}). Using the boundary condition~(\ref{eqSMBC}) and the expression of $f_2(\theta)$ given by Eq.~(\ref{eqSMf2}), we obtain the conditions $k\to \infty$ and 
\begin{equation}
\frac{kD_2}{\lambda} = \lim_{\Delta \rightarrow 0}\frac{\Delta \exp(-\beta V_1)}{\exp(-\beta V_2)},
\end{equation}
implying that the potential difference must then depend logarithmically on the width $\Delta$ to recover the SMDP.

\subsection{Derivation of the MFPT solution in the thin cortex limit}
\label{appendixDelta0}

First we derive the general solution of the MFPT in the thin cortex limit for a potential difference $\Delta V<+\infty$ (i.e. excluding the annulus geometry). In the inner shell the MFPT can be expanded as
\begin{equation}
t_2(r,\theta) = t_{\rm B}^{(0)}(r,\theta) + \frac{\Delta}{\cal R} t_{\rm B}^{(1)}(r,\theta) + \left(\frac{\Delta}{\cal R}\right)^2 t_{\rm B}^{(2)}(r,\theta) + \cdots
\end{equation}
where $t_{\rm B}^{(i)}(r,\theta)$ are independent of $\Delta$ and defined for $r\in[0,{\cal R}]$. From Eq.~(\ref{MFPT2}) these bulk functions satisfy the equations $D_2 \nabla_{\bf x}^2 t_{\rm B}^{(0)}({\bf x})=-1$ and $D_2 \nabla_{\bf x}^2 t_{\rm B}^{(j)}({\bf x})=0$ for $j\ge1$. In the cortex the relevant variable is $u = ({\cal R}-r)/\Delta$ and the MFPT can be expanded as 
\begin{equation}
t_1(u,\theta) = t_{\Sigma}^{(0)}(u,\theta) + \frac{\Delta}{\cal R} t_{\Sigma}^{(1)}(u,\theta) + \left(\frac{\Delta}{\cal R}\right)^2 t_{\Sigma}^{(2)}(u,\theta) +\cdots
\end{equation}
where $t_{\Sigma}^{(i)}(u,\theta)$ are independent of $\Delta$ and defined for $u\in[0,1]$. The Eq.~(\ref{MFPT1}) becomes
\begin{equation}
\label{eqOuter}
\frac{D_1}{{\cal R}^2} \left[ \frac{\partial^2 t_1}{\partial u^2} - \frac{(d-1)\Delta}{\cal R} \left(1 + \frac{\Delta u}{\cal R}\right) \frac{ \partial t_1}{\partial u} + \frac{\Delta^2}{{\cal R}^2} \Delta_\theta t_1 \right] =  - \frac{\Delta^2}{{\cal R}^2} + {\cal O}\left(\frac{\Delta^3}{{\cal R}^3}\right).
\end{equation}
with $\Delta_\theta = (\sin \theta)^{2-d} \partial_\theta (\sin\theta)^{d-2} \partial_\theta$ the Laplace operator on the unit hypersphere in $d$ dimension, assuming the invariance over the azimuthal angle $\varphi$ in dimension $d=3$. The boundary conditions~(\ref{MFPT3}) and~(\ref{MFPT4}) become
\begin{gather}
t_1(u=0,|\theta|<\varepsilon/2) = 0, \label{BC1}\\
\frac{\partial t_1}{\partial u}(u=0,|\theta|>\varepsilon/2) = 0, \label{BC2}\\
t_1(u=1,\theta) = t_2({\cal R},\theta), \label{BC3}\\
- D_1 \exp(-\beta V_1) \frac{\partial t_1}{\partial u}(u=1,\theta) = D_2 \exp(-\beta V_2) \Delta \frac{\partial t_2}{\partial r}({\cal R},\theta)\label{BC4}.
\end{gather}

In leading order, Eq.~(\ref{eqOuter}) writes $\partial_u^2 t_{\Sigma}^{(0)}(u,\theta) =0$ implying the general solution $t_{\Sigma}^{(0)}(u,\theta) = A_0(\theta)u + B_0(\theta)$. The boundary conditions~(\ref{BC2}) and~(\ref{BC4}) impose the unique solution $A_0(\theta) = 0$. Then from Eq.~(\ref{BC3}) we obtain $t_{\Sigma}^{(0)}(\theta) = t_{\rm B}^{(0)}({\cal R},\theta)$ with the condition $t_{\Sigma}^0(|\theta|<\varepsilon/2) = 0$.

In first order in $\Delta/{\cal R}$, the Eq.~(\ref{eqOuter}) writes $\partial_u^2 t_{\Sigma}^{(1)}(u,\theta) =0$ implying the same general solution $t_{\Sigma}^{(1)}(u,\theta) = A_1(\theta)u + B_1(\theta)$. The boundary condition~(\ref{BC2}) imposes $A_1(\theta) = 0$ for $|\theta|>\varepsilon/2$ and Eq.~(\ref{BC2}) gives
\begin{equation}
\label{BCorder1}
A_1(\theta) = - \frac{D_2 \exp(-\beta V_2)}{D_1 \exp(-\beta V_1)} {\cal R} \frac{\partial t_{\rm B}^{(0)}}{\partial r}({\cal R},\theta).
\end{equation}
The two boundary conditions imply then
\begin{equation}
\label{eqRefl}
\frac{\partial t_{\rm B}^{(0)}}{\partial r}({\cal R},|\theta|>\varepsilon/2).
\end{equation}
Finally Eqs.~(\ref{BC1}) and~(\ref{BC3}) determine the function $B_1(\theta)$ in terms of $\partial_r t_{\rm B}^{(0)}(R,\theta)$ and $t_{\rm B}^{(1)}(R,\theta)$, which is not important for the following steps.

In second order in $\Delta/{\cal R}$, Eq.~(\ref{eqOuter}) writes
\begin{equation}
\frac{D_1}{{\cal R}^2} \frac{\partial^2  t_{\Sigma}^{(2)}}{\partial u^2}(u,\theta) = \frac{D_1}{{\cal R}^2}\left( (d-1) A_1(\theta) - \Delta_\theta t_{\Sigma}^{(0)}(\theta)  \right)-1.
\end{equation}
The general solution is then
\begin{equation}
\label{eqOrder2}
\frac{D_1}{{\cal R}^2}  t_{\Sigma}^{(2)}(u,\theta) = \left[\frac{D_1}{{\cal R}^2}\left( (d-1) A_1(\theta) - \Delta_\theta t_{\Sigma}^{(0)}(\theta)  \right)-1\right] \frac{u^2}{2} + A_2(\theta) u + B_2(\theta).
\end{equation}
The boundary condition~(\ref{BC2}) imposes $A_2(\theta) = 0$ for $|\theta|>\varepsilon/2$ and Eq.~(\ref{BC2}) yields
\begin{equation}
\frac{D_1}{{\cal R}^2} \Delta_\theta t_{\Sigma}^{(0)}(\theta) -  \frac{D_2 \exp(-\beta V_2)}{D_1 \exp(-\beta V_1)} {\cal R} \frac{\partial t_{\rm B}^{(1)}}{\partial r}({\cal R},\theta) =-1,
\end{equation}
for $|\theta|>\varepsilon/2$. Finally, Eqs.~(\ref{BC1}) and~(\ref{BC3}) determine the function $B_2(\theta)$ in terms of $\partial_r t_{\rm B}^{(1)}(R,\theta)$ and $t_{\rm B}^{(2)}(R,\theta)$.

Defining $t_{\rm B} = t_{\rm B}^{(0)}$ and $t_{\Sigma} = t_{\Sigma}^{(0)}$, the leading order is then given by the bulk equations $D_2 \nabla^2 t_{\rm B}(r,\theta) = 0$ and
\begin{equation}
\frac{\partial t_{\rm B}}{\partial r}({\cal R},|\theta|>\varepsilon/2)=0, 
\end{equation}
and the surface equations $t_{\Sigma}(\theta) = t_{\rm B}({\cal R},\theta)$ and $t_{\Sigma}(|\theta|<\varepsilon/2) = 0$ while $\Delta V<+\infty$. These equations are the same as the disk geometry NEP with a diffusion constant $D_2$.

We hypothesize that the thin cortex limit is valid when the second boundary condition of Eq.~(\ref{MFPT3}) is equivalent to Eq.~(\ref{eqRefl}):
\begin{equation}
\label{condMFPT3}
\left|D_1 \exp(-\beta V_1) \frac{\partial t_1}{\partial r}({\cal R}-\Delta,\theta) \right| \ll \left|D_2 \exp(-\beta V_2) \frac{\partial t_2}{\partial r}({\cal R}-\Delta,\theta) \right|.
\end{equation}
The derivative of $t_1$ is equal to
\begin{equation}
\frac{\partial t_1}{\partial r}(r={\cal R}-\Delta,\theta) = - \frac{1}{\Delta} \frac{\partial t_1}{\partial u} (u=1,\theta) \simeq - \frac{\Delta}{{\cal R}^2} \frac{\partial t_{\Sigma}^{(2)}}{\partial u} (u=1,\theta).
\end{equation}
For $|\theta| > \varepsilon/2$, Eq.~(\ref{eqOrder2}) yields
\begin{equation}
\frac{\partial t_{\Sigma}^{(2)}}{\partial u} (u=1,\theta) = - \Delta_\theta t_{\Sigma}^{(0)}(\theta) - \frac{{\cal R}^2}{D_1}
\end{equation}
with $t_{\Sigma}^{(0)}(\theta) = t_{\rm B}^{(0)}({\cal R},\theta)$ which diverges when $\varepsilon \ll 1$. We obtain then
\begin{equation}
\frac{\partial t_1}{\partial r}(r={\cal R}-\Delta,\theta) \simeq \frac{\Delta}{{\cal R}^2} \Delta_\theta t_2({\cal R},\theta).
\end{equation}
Assuming $\Delta_\theta t_2 = {\cal O}(T_2)$ and $\partial_r t_2 = {\cal O}(T_2/{\cal R})$, Eq.~(\ref{condMFPT3}) gives the condition
\begin{equation}
\frac{\Delta}{\cal R} \ll \frac{D_2}{D_1} \exp(-\beta \Delta V),
\end{equation}
in addition to $\Delta \ll {\cal R}$ supposed initially.

For the annulus geometry ($\exp(-\beta \Delta V)=0$), Eq.~(\ref{eqRefl}) is no more satisfied since the boundary condition~(\ref{BCorder1}) is automatically verified. Then the bulk equation is $D_2 \nabla^2 t_{\rm B}(r,\theta) = 0$ while the surface equation is
\begin{equation}
\frac{D_1}{{\cal R}^2} \Delta_\theta t_{\Sigma}(\theta) =-1,
\end{equation}
with the surface equations $t_{\Sigma}(\theta) = t_{\rm B}({\cal R},\theta)$ and $t_{\Sigma}(|\theta|<\varepsilon/2) = 0$. These equations are the same as the SMDP with an infinite absorption rate.

%%%%%%%%%%%%%%%%%%%%%%%%%%
%%% GENERAL EQS OF NEP %%%
%%%%%%%%%%%%%%%%%%%%%%%%%%

\section{Derivation of the narrow escape expressions}
\label{appendixGeneralNE}

The MFPT is the solution of Eqs.~(\ref{eqMFPT}) and~(\ref{eqMFPTBC}). Denoting $P_s({\bf x}) = \exp[-\beta V({\bf x})]/Z$ the stationary probability density for the closed domain ($\varepsilon = 0$), with $Z=\int_\Omega d{\bf x}\exp[-\beta V({\bf x})]$, the bulk equation (\ref{eqMFPT}) is
\begin{equation}
\label{eqMFPT2}
\nabla_{\bf x} \cdot \left[ D({\bf x}) P_s({\bf x}) \nabla_{\bf x} t({\bf x}) \right] = - P_s({\bf x}).
\end{equation}
Integrating over the volume $\Omega$, Eq.~(\ref{eqMFPT2}) becomes
\begin{equation}
\int_\Omega d{\bf x} \ \nabla_{\bf x} \cdot \left[ D({\bf x}) P_s({\bf x}) \nabla_{\bf x} t({\bf x}) \right] = -1,
\end{equation}
since the stationary probability density is normalized. Using the divergence theorem on the left hand side and the boundary condition (\ref{eqMFPTBC}) we finally obtain a condition involving the escape region:
\begin{equation}
\label{eqCONS}
\int_{\partial \Omega_\epsilon} dS({\bf x}) \  D({\bf x}) P_s({\bf x}) {\bf n} \cdot \nabla_{\bf x} t({\bf x}) = -1,
\end{equation}
where ${\bf n}$ is the outward pointing unit vector normal to the surface $\partial \Omega$. Moreover, for the studied geometries the diffusion constant $D({\bf x})$ and the potential $V({\bf x})$ do not depend on $\varepsilon$ and may be considered as constant close to the escape region with the values $D({\bf x_0})$ and $V({\bf x_0})$, respectively.

\subsection{Two dimensions NEP}

We first derive the general narrow escape solution in 2d, following the derivation of Refs.~\cite{ward1993, pillay2010, chevalier2011}. The escape region with an angle $\varepsilon \ll 1$ is considered as a perturbation on the external boundary. Motivated by the results of the literature, we make the ansatz
\begin{equation}
\label{outer2d}
t({\bf x}) = \tau_0 \ln \frac{\varepsilon}{4} + \tau_1({\bf x}) + \cdots,
\end{equation}
where $\tau_0$ is constant while $\tau_1({\bf x})$ obeys Eqs.~(\ref{eqMFPT}) and~(\ref{eqMFPTBC}) far from the escape region.

Close to the escape region, the (inner) solution can be derived in terms of the inner variable $\widetilde {\bf x} = 2 ({\bf x} - {\bf x_0})/\varepsilon{\cal R}$ with ${\bf x_0}=({\cal R},0)$ the center of the escape region. The escape region is a segment defined by the coordinates $\widetilde x = 0$ and $|\widetilde y| \le 1$, and the reflecting boundaries are to the half-lines defined by the coordinates $\widetilde x = 0$ and $|\widetilde y| \ge 1$, in the limit $\varepsilon \to 0$. In these coordinates, the MFPT is denoted $v(\widetilde {\bf x})$. The bulk Eq.~(\ref{eqMFPT2}) writes then
\begin{equation}
\label{inner}
D({\bf x_0}) \nabla_{\widetilde {\bf x}}^2 v(\widetilde {\bf x}) = - (\varepsilon/2)^2.
\end{equation}
In leading order, the function $v(\widetilde {\bf x})$ is then solution of Laplace's equation. The boundary conditions are $v(\widetilde {\bf x}) = 0$ on the absorbing boundary and ${\bf n} \cdot \nabla_{\widetilde {\bf x}} v(\widetilde {\bf x}) = 0$ on the reflective boundary. Finally, the condition involving the escape region~(\ref{eqCONS}) becomes
\begin{equation}
\label{eqCONS2d}
D({\bf x_0}) P_s({\bf x_0}) \int_{-1}^1 d\widetilde y \ \frac{\partial v}{\partial \widetilde x} = 1.
\end{equation}
These equations can be solved by choosing elliptic coordinates ($\mu$, $\nu$) defined by $\widetilde x = \sinh \mu \sin \nu $ and $\widetilde y = \cosh \mu \cos \nu$. The Laplace's equation writes in this coordinate system: $\partial_{\mu\mu} v + \partial_{\nu\nu} v = 0$ and the boundary conditions are $v(\mu = 0,\nu) =0$ and $\partial_\nu v(\mu,\nu=0) = 0$ on the absorbing and reflecting boundaries, respectively. The $\nu$-independent solution $v(\mu,\nu)= A \mu$ satisfies this Cauchy system. On the escape region we have then $\partial v/\partial \widetilde x = A/\sin \nu$ and $\widetilde y = \cos \nu$, and the condition (\ref{eqCONS2d}) gives
\begin{equation}
A = \frac{1}{\pi D({\bf x_0}) P_s({\bf x_0})}.
\end{equation}

The outer and the inner solutions are matched in an intermediate region such that ${\bf x} \rightarrow {\bf x_0}$ and $|{\bf \tilde x}| \rightarrow \infty$ ({\it i.e.} $\mu \rightarrow \infty$ for elliptic coordinates). In this limit, the inner solution behaves like
\begin{equation}
v({\bf \tilde x}) \simeq A \ln(2|{\bf \tilde x}|) = A \ln \frac{4|{\bf x} - {\bf x_0}|}{\varepsilon{\cal R}}.
\end{equation}
and comparing it with the outer solution given by Eq.~(\ref{outer2d}), we obtain $\tau_0=-A$ and the behavior of the function $\tau_1({\bf x})$ close to the escape region is
\begin{equation}
\tau_1({\bf x} \rightarrow {\bf x_0}) = \frac{1}{\pi D({\bf x_0}) P_s({\bf x_0})} \ln \frac{|{\bf x} - {\bf x_0}|}{{\cal R}}.
\end{equation}

Following Ref.~\cite{pillay2010}, the pseudo Green's function $G({\bf x}|{\bf x_0})$ is introduced via the equations
\begin{gather}
\exp(\beta V({\bf x})) \nabla_{\bf x} \cdot \left[ D({\bf x}) \exp(-\beta V({\bf x})) \nabla_{\bf x}  G({\bf x}|{\bf x_0})  \right] = P_s({\bf x_0}) - \delta({\bf x} - {\bf x_0}), \quad {\bf x} \in \Omega \label{eqGreen1}\\
{\bf n} \cdot \nabla G({\bf x}|{\bf x_0}) = 0 , \quad {\bf x} \in \partial \Omega \label{eqGreen2}\\
\int_{\Omega} d{\bf x}\ G({\bf x}|{\bf x_0}) P_s({\bf x}) = 0 \label{eqGreen3},
\end{gather}
where $\delta({\bf x})$ represents the Dirac distribution which can be omitted here since ${\bf x_0}$ belongs to the boundary of the domain and the limit
\begin{equation}
G({\bf x} \rightarrow {\bf x_0} |{\bf x_0}) = -\frac{1}{\pi D({\bf x_0})} \ln \frac{|{\bf x} - {\bf x_0}|}{{\cal R}} + R_0, \label{eqGreen2d}
\end{equation}
where $R_0$ is the (unknown) regular part of the pseudo Green's function at the center of the escape region. The function $\tau_1({\bf x})$ is then given by
\begin{equation}
\tau_1({\bf x}) = - \frac{G({\bf x}|{\bf x_0})}{P_s({\bf x_0})}  + \frac{R_0}{P_s({\bf x_0})}.
\end{equation}

Hence the spatial average of the MFPT is
\begin{equation}
\label{expNEGMFPTapp}
\langle t \rangle =  \frac{1}{P_s({\bf x_0})} \left[ - \frac{1}{\pi D({\bf x_0})}\ln \frac{\varepsilon}{4} + R_0 \right],
\end{equation}
and the MFPT is
\begin{equation}
\label{expNEMFPT}
t({\bf x}) = \langle t \rangle - \frac{G({\bf x} |{\bf x_0})}{P_s({\bf x_0})} = \frac{1}{P_s({\bf x_0})} \left\{ \frac{1}{\pi D({\bf x_0})} \ln \frac{4|{\bf x} - {\bf x_0}|}{\varepsilon{\cal R}} +  R_0 -R({\bf x}|{\bf x_0}) \right\},
\end{equation}
where the regular part of the pseudo Green's function is
\begin{equation}
\label{RGreen2dDef}
R({\bf x}| {\bf x_0}) = G({\bf x} |{\bf x_0}) +\frac{1}{\pi D({\bf x_0})} \ln \frac{|{\bf x} - {\bf x_0}|}{{\cal R}}.
\end{equation}

\subsection{Three dimensions NEP}

We now derive the general narrow escape solution in 3d, following the derivation of Refs.~\cite{ward1993, cheviakov2010, cheviakov2012}. The calculation is slightly different from 2d, since the MFPT has a divergence as $1/\varepsilon$ whereas the subleading order behaves like $\ln \varepsilon$. One then needs to go to the second order of corrections to obtain the order $\varepsilon^0$. Motivated by the results of the literature, we make the ansatz
\begin{equation}
\label{outer3d}
t({\bf x}) = \frac{2\tau_0}{\varepsilon} + \tau_1({\bf x}) + \varepsilon \ln \frac{\varepsilon}{4} \tau_2({\bf x}) + \cdots,
\end{equation}
where $\tau_0$ is constant while $\tau_1({\bf x})$ obeys Eqs.~(\ref{eqMFPT}) and~(\ref{eqMFPTBC}) far from the escape region.

Close to the escape region, the (inner) solution can be derived in terms of the inner variable $\widetilde {\bf x} = 2 ({\bf x} - {\bf x_0})/\varepsilon{\cal R}$ with ${\bf x_0}=({\cal R},0,0)$ the center of the escape region. The escape region is a unit disk defined by the coordinates $\widetilde x = 0$ and $\widetilde y^2 + \widetilde z^2 \le 1$, and the reflecting boundary is the plane outside the unit disk defined by $\widetilde x = 0$ and $\widetilde y^2 + \widetilde z^2  \ge 1$, in the limit $\varepsilon \to 0$. In these coordinates, the MFPT is denoted $v(\widetilde {\bf x})$. The bulk Eq.~(\ref{eqMFPT2}) still verify the Eq.~(\ref{inner}) in three dimensions. The boundary conditions are $v(\widetilde {\bf x}) = 0$ on the absorbing boundary and ${\bf n} \cdot \nabla_{\widetilde {\bf x}} v(\widetilde {\bf x}) = 0$ on the reflective boundary. Finally the condition involving the escape region (\ref{eqCONS}) becomes
\begin{equation}
\label{eqCONS3d}
D({\bf x_0}) P_s({\bf x_0}) {\cal R} \int_{0}^1 d\rho (2\pi\rho) \ \frac{\partial v}{\partial \widetilde x} = 1,
\end{equation}
where $\rho = \sqrt{\widetilde y^2 + \widetilde z^2}$ on the surface $\widetilde x = 0$. The inner solution is decomposed as
\begin{equation}
\label{inner3d}
v(\widetilde {\bf x}) = \frac{2v_0(\widetilde {\bf x})}{\varepsilon} + v_1(\widetilde {\bf x}) \ln \frac{\varepsilon}{4} + v_2(\widetilde {\bf x}) + \cdots.
\end{equation}
From Eq.~(\ref{inner}), $v_0(\widetilde {\bf x})$ and $v_1(\widetilde {\bf x})$ are solution of Laplace's equation whereas $v_2(\widetilde {\bf x})$ is solution of a Poisson's equation. These equations can be solved by choosing the oblate spheroidal coordinates ($\mu$, $\nu$, $\varphi$) defined by $\widetilde x = \sinh \mu \sin \nu $, $\widetilde y = \cosh \mu \cos \nu \cos\varphi$ and $\widetilde z = \cosh \mu \cos \nu \sin\varphi$. Considering the invariance of the problem over the azimuthal angle $\varphi$, the Laplace's equation for $v_0$ writes in this coordinate system
\begin{equation}
\frac{1}{\sinh^2 \mu + \sin^2 \nu} \left[ \frac{1}{\cosh \mu} \partial_\mu (\cosh \mu \partial_\mu v_0) + \frac{1}{\cos \nu} \partial_\nu (\cos \nu \partial_\nu v_0) \right] = 0
\end{equation}
and the boundary conditions are $v_0(\mu = 0,\nu) =0$ and $\partial_\nu v_0(\mu,\nu=0) = 0$ on the absorbing and reflecting boundaries, respectively. The $\nu$-independent solution
\begin{equation}
\label{solv0}
v_0(\mu) = A_0 \int_0^\mu \frac{d\mu'}{\cosh \mu'}
\end{equation}
satisfies this Cauchy system. On the escape region we have then $\partial v_0/\partial \widetilde x = A_0/\sqrt{1-\rho^2}$ and the condition (\ref{eqCONS3d}) gives
\begin{equation}
A_0 = \frac{1}{2\pi D({\bf x_0}) P_s({\bf x_0}) {\cal R}}.
\end{equation}
Far from the escape region ({\it i.e.} $\mu \rightarrow \infty$), the inner solution behaves like
\begin{equation}
v_0({\bf \tilde x}) \simeq A_0 \left(\frac{\pi}{2} - \frac{1}{|{\bf \tilde x}|} \right) = A_0 \left( \frac{\pi}{2} - \frac{\varepsilon{\cal R}}{2|{\bf x} - {\bf x_0}|} \right).
\end{equation}
The outer and the inner solutions are matched in an intermediate region such that ${\bf x} \rightarrow {\bf x_0}$ and $|{\bf \tilde x}| \rightarrow \infty$:
\begin{equation}
\label{matching3d}
\frac{2\tau_0({\bf x})}{\varepsilon} + \tau_1({\bf x}) + \varepsilon \ln \frac{\varepsilon}{4} \tau_2({\bf x}) + \cdots \simeq  \frac{2v_0(\widetilde {\bf x})}{\varepsilon} + v_1(\widetilde {\bf x}) \ln \frac{\varepsilon}{4} + v_2(\widetilde {\bf x}) + \cdots
\end{equation}
Comparing the first orders of both sides, we obtain $\tau_0=A_0\pi/2$ and the behavior of the function $\tau_1({\bf x})$ close to the escape region is
\begin{equation}
\tau_1({\bf x} \rightarrow {\bf x_0}) = \frac{1}{2\pi D({\bf x_0}) P_s({\bf x_0})} \frac{1}{|{\bf x} - {\bf x_0}|}.
\end{equation}

Following Ref.~\cite{cheviakov2010}, the pseudo Green's function $G({\bf x}|{\bf x_0})$ is introduced via the Eqs.~(\ref{eqGreen1})-(\ref{eqGreen3}) and the limit
\begin{equation}
G({\bf x} \rightarrow {\bf x_0} |{\bf x_0}) = \frac{1}{2\pi D({\bf x_0})} \frac{1}{|{\bf x} - {\bf x_0}|} + \gamma \ln \frac{|{\bf x} - {\bf x_0}|}{2{\cal R}} + R_0, \label{eqGreen3d}\\
\end{equation}
where $\gamma$ and $R_0$ are the logarithmic diverging and regular parts, respectively, of the pseudo Green's function at the center of the escape region. Note that the logarithmic divergence of the Green's function comes from the ansatz~(\ref{outer3d}). The function $\tau_1({\bf x})$ is then
\begin{equation}
\tau_1({\bf x}) = - \frac{G({\bf x}|{\bf x_0})}{P_s({\bf x_0})}  + \chi.
\end{equation}
The expression of the constant $\chi$ is determined by the inner solutions $v_1$ and $v_2$. The form of the solution $v_1$ is identically to $v_0$ given by Eq.~(\ref{solv0}). Then $v_1$ behaves for $|{\bf \tilde x}| \rightarrow \infty$ as
\begin{equation}
v_1({\bf \tilde x}) \simeq A_1 \left(\frac{\pi}{2} - \frac{1}{|{\bf \tilde x}|} \right) = A_1 \left( \frac{\pi}{2} - \frac{\varepsilon{\cal R}}{2|{\bf x} - {\bf x_0}|} \right).
\end{equation}
Identifying the next order of Eq.~(\ref{matching3d}), the constant $\chi$ is decomposed as $\chi_0 \ln(\varepsilon/4) + \chi_1$ and the logarithmic divergence of the Green's function in the Eq.~(\ref{eqGreen3d}) as
\begin{equation}
\ln \frac{|{\bf x} - {\bf x_0}|}{2{\cal R}} = \ln \frac{2|{\bf x} - {\bf x_0}|}{\varepsilon{\cal R}} + \ln \frac{\varepsilon}{4}.
\end{equation}
We finally have
\begin{equation}
A_1 = \frac{2}{\pi} \left(- \frac{\gamma}{P_s({\bf x_0})} + \chi_0 \right).
\end{equation}
Since $\tau_2({\bf x})$ is assumed to be smooth close to the escape region, $A_1 = 0$ to remove the $1/|{\bf x} - {\bf x_0}|$ divergence, which yields
\begin{equation}
\chi_0 = \frac{\gamma}{P_s({\bf x_0})}.
\end{equation}

The expression of $\chi_1$ is deduced from the next order of corrections. In Ref.~\cite{cheviakov2010}, the solution of Poisson's equation is derived and the general solution of $v_2$ far from the escape region is
\begin{align}
v_2({\bf \tilde x}) &\simeq A_2 \left(\frac{\pi}{2} - \frac{1}{|{\bf \tilde x}|} \right) + B_2 \left[ \ln |{\bf \tilde x}| - \frac{2}{\pi |{\bf \tilde x}|} \left( 2 \ln 2 - \frac{3}{2} \right) \right] \\
&= A_2 \left( \frac{\pi}{2} - \frac{\varepsilon{\cal R}}{2|{\bf x} - {\bf x_0}|} \right) + B_2 \left[ \ln \frac{2|{\bf x} - {\bf x_0}|}{\varepsilon{\cal R}} - \frac{\varepsilon{\cal R}}{\pi|{\bf x} - {\bf x_0}|} \left( 2 \ln 2 - \frac{3}{2} \right) \right].
\end{align}
Identifying the next order of the matching condition~(\ref{matching3d}), we obtain
\begin{equation}
A_2 = \frac{2}{\pi} \left(- \frac{R_0}{P_s({\bf x_0})} + \chi_1 \right) \quad {\rm and } \quad B_2 = - \frac{\gamma}{P_s({\bf x_0})}.
\end{equation}
The next orders are assumed to be smooth close to the escape region, which gives
\begin{equation}
A_2 + B_2 \frac{2}{\pi} \left(2\ln2 -\frac{3}{2} \right)= 0
\end{equation}
to remove the $1/|{\bf x} - {\bf x_0}|$ divergence, and
\begin{equation}
\chi_1 = \frac{1}{P_s({\bf x_0})} \left[ R_0 + \gamma \left(2\ln2 - \frac{3}{2} \right) \right].
\end{equation}

Hence the spatial average of the MFPT is
\begin{equation}
\label{expNEGMFPT3dapp}
\langle t \rangle =  \frac{1}{P_s({\bf x_0})} \left[ \frac{1}{2 D({\bf x_0}) {\cal R} \varepsilon} + \gamma \ln \varepsilon + R_0 - \frac{3}{2} \gamma \right],
\end{equation}
and the expression of the MFPT is
\begin{equation}
\label{expNEMFPT3d}
t({\bf x}) = \langle t \rangle -  \frac{G({\bf x} |{\bf x_0})}{P_s({\bf x_0})} = \frac{1}{P_s({\bf x_0})} \left\{ - \frac{1}{2\pi D({\bf x_0})} \frac{1}{|{\bf x} - {\bf x_0}|} + \frac{1}{2 D({\bf x_0}) \varepsilon {\cal R}} + \gamma \ln \varepsilon + R_0 - \frac{3}{2} \gamma - R({\bf x}| {\bf x_0}) \right\},
\end{equation}
where the regular part of the pseudo Green's function is
\begin{equation}
\label{RGreen3dDef}
R({\bf x}| {\bf x_0}) = G({\bf x} |{\bf x_0}) - \frac{1}{2\pi D({\bf x_0})|{\bf x}-{\bf x_0}|}.
\end{equation}

%%%%%%%%%%%%%%%%%%%%%%%%
%%% NARROW ESCAPE 2D %%%
%%%%%%%%%%%%%%%%%%%%%%%%

\section{Derivation of the Green's function and narrow escape expression in two dimensions}
\label{appendix2d}

For the two-shell geometry, the diffusion constant and the stationary probability density close to the escape region are $D({\bf x_0})=D_1$ and $P_s({\bf x_0}) = \exp(-\beta V_1)/Z$, respectively, with the partition function
\begin{equation}
Z = \int_\Omega \exp(-\beta V({\bf x})) = \pi \left\{ {\cal R}^2 \exp(-\beta V_1) + ({\cal R}-\Delta)^2 \left[ \exp(-\beta V_2) - \exp(-\beta V_1) \right] \right\} .
\end{equation}
Eqs.~(\ref{eqGreen1})-(\ref{eqGreen3}) and Eq.~(\ref{eqGreen2d}) for the pseudo Green's function can be rewritten in the two-shell geometry to obtain its regular part $R({\bf x}|{\bf x_0})$ defined by Eq.~(\ref{RGreen2dDef}). We use the dimensionless polar coordinates $(r,\theta)$ defined by $x/{\cal R}=r\cos \theta$ and $y/{\cal R}=r\sin \theta$. The regular part of the Green's function is denoted as $R(r,\theta)$ without loss of generality, and the escape region is located at $r=1$ and $\theta=0$. Since the normal vector ${\bf n}$ is radially oriented, the normal derivative of $\ln |{\bf x} - {\bf x_0}|/{\cal R}$ at the point $(r,\theta)$ is
\begin{equation}
\label{relLogSurf2d}
{\bf n} \cdot \nabla_{\bf x} \ln \frac{|{\bf x} - {\bf x_0}|}{{\cal R}} =  \frac{r -  \cos \theta}{r^2 - 2 r \cos \theta +1}.
\end{equation}

The value of $R(r,\theta)$ in the inner shell and in the outer shell are defined by $R_1(r,\theta)$ and $R_2(r,\theta)$, respectively. From Eq.~(\ref{eqGreen1}), the bulk equations verified by $R_i(r,\theta)$ are
\begin{gather}
D_1 \nabla^2 R_1(r,\theta) = \frac{\exp(-\beta V_1)}{Z'}, \quad r>1-\delta \label{RGreen2d1} \\
D_2 \nabla^2 R_2(r,\theta) = \frac{\exp(-\beta V_1)}{Z'}, \quad r<1-\delta.  \label{RGreen2d2}
\end{gather}
with $Z'=Z/{\cal R}^2$ and $\delta = \Delta/{\cal R}$. Analogously to Eq.~(\ref{MFPT3}) the boundary conditions at $r=1-\delta$ are
\begin{gather}
R_1(1-\delta,\theta) =  R_2(1-\delta,\theta) , \label{RGreen2d3} \\
D_1 \exp(-\beta V_1) \frac{\partial R_1}{\partial r}(1-\delta,\theta)  - D_2 \exp(-\beta V_2) \frac{\partial R_2}{\partial r}(1-\delta,\theta) = \frac{D_1 \exp(-\beta V_1)-D_2 \exp(-\beta V_2)}{\pi D_1} \frac{(1-\cos\theta)-\delta}{2(1-\delta)(1-\cos\theta)+\delta^2}, \label{RGreen2d4}
\end{gather}
where the relation (\ref{relLogSurf2d}) has been used, and from Eq.~(\ref{eqGreen2}) the reflective boundary condition at $r=1$ becomes 
\begin{equation}
\frac{\partial R_1}{\partial r}(1,\theta) = \frac{1}{2\pi D_1} \label{RGreen2d5}.
\end{equation}
Finally, the condition~(\ref{eqGreen3}) gives the last relation needed to derive $R$:
\begin{equation}
\int_0^{2\pi} d\theta \left[ \exp(-\beta V_2)\int_0^{1-\delta} dr\ r R_2(r,\theta) + \exp(-\beta V_1)\int_{1-\delta}^1 dr\ r R_1(r,\theta) \right] = 0. \label{RGreen2d6}
\end{equation}

The general solution of Eqs.~(\ref{RGreen2d1}) and~(\ref{RGreen2d2}), which can be rewritten as $D_i \nabla^2 R_i(r,\theta) = \exp(-\beta V_1)/Z'$, has the form
\begin{equation}
R_i(r,\theta) = \sum_{n=0}^\infty f_{i,n}(r) \cos(n\theta)
\end{equation}
due to the periodicity on the polar coordinate $\theta$ and the symmetry of the problem under the $\theta \rightarrow - \theta$ transformation. The functions $f_{i,n}(r)$ must satisfy the differential equations
\begin{gather}
f_{i,n}''(r) +\frac{1}{r} f_{i,n}'(r) - \frac{n^2}{r^2} f_{i,n}(r) =0, \quad n\ge1\\
D_i \left[ f_{i,0}''(r) +\frac{1}{r} f_{i,0}'(r) \right] = \frac{\exp(-\beta V_1)}{Z'}.
\end{gather}
The global solution is then given by $f_{i,n}(r) = a_{i,n} r^n + b_{i,n} r^{-n}$ (for $n\ge 1$) and $f_{i,0}(r) = \exp(-\beta V_1)r^2/(4D_iZ') + a_{i,0} + b_{i,0} \ln r$. Hence the functions $R_1$ and $R_2$ have the general form
\begin{gather}
R_1(r,\theta) = \frac{\exp(-\beta V_1) r^2}{4 D_1 Z'} + a_0 + b_0 \ln r + \sum_{n=1}^\infty \left[a_n r^n + b_n r^{-n}\right]\cos(n\theta), \\
R_2(r,\theta) = \frac{\exp(-\beta V_1) r^2}{4 D_2 Z'} + c_0 + d_0 \ln r + \sum_{n=1}^\infty \left[c_n r^n + d_n r^{-n}\right]\cos(n\theta).
\end{gather}

Since the problem is not singular at $r=0$, we must impose $d_n =0$ for $n\ge0$ to have a non divergent solution $R_2(r,\theta)$. The boundary condition at $r=1$ given by Eq.~(\ref{RGreen2d5}) writes
\begin{equation}
\frac{\exp(-\beta V_1)}{2 D_1 Z'} + b_0 + \sum_{n=1}^\infty n \left(a_n - b_n\right)\cos(n\theta)  = \frac{1}{2\pi D_1}.
\end{equation}
The orthogonality of $\cos(n\theta)$ imposes $a_n=b_n$ for $n\ge1$ and
\begin{equation}
b_0 = \frac{1}{2\pi D_1} - \frac{\exp(-\beta V_1)}{2 D_1 Z'} = \frac{[\exp(-\beta V_2)-\exp(-\beta V_1)](1-\delta)^2}{2D_1 Z'}.
\end{equation}

The continuity at $r=1-\delta$, expressed by the Eq.~(\ref{RGreen2d3}), becomes
\begin{gather}
\frac{\exp(-\beta V_1)(1-\delta)^2}{4D_1Z'} + a_0 + b_0 \ln(1-\delta) + \sum_{n=1}^\infty a_n \left[(1-\delta)^n + (1-\delta)^{-n}\right]\cos(n\theta) = \nonumber \\
\frac{\exp(-\beta V_1) (1-\delta)^2}{4 D_2 Z'} + c_0 + \sum_{n=1}^\infty c_n (1-\delta)^n \cos(n\theta).
\end{gather}
The orthogonality of $\cos(n\theta)$ yields
\begin{gather}
a_0-c_0 = \frac{\exp(-\beta V_1)(D_1-D_2)(1-\delta)^2}{4 Z' D_1 D_2} - \frac{[\exp(-\beta V_2)-\exp(-\beta V_1)](1-\delta)^2}{2D_1 Z'} \ln(1-\delta), \label{EXPa0c0}\\
c_n=a_n \left[1+(1-\delta)^{-2n}\right].\label{EXPancn}
\end{gather}
Since $\int_0^{2\pi} d\theta \cos(n\theta) = 0$ for $n\ge1$, the condition (\ref{RGreen2d6}) becomes
\begin{equation}
\int_0^{1-\delta} dr\ r\left[\frac{D_1 \exp(-\beta V_1) r^2}{4 D_2 Z'} + c_0\right] + \int_{1-\delta}^1 dr\ r\left[\frac{\exp(-\beta V_1) r^2}{4 D_1 Z'} + a_0 + b_0 \ln r \right] =0,
\end{equation}
and the relation (\ref{EXPa0c0}) finally gives
\begin{gather}
a_0 = \frac{\pi \exp(-\beta V_1)}{8D_1D_2Z'^2} \{ (D_1-D_2)(1-\delta)^4 \exp(-\beta V_2) + D_2[\exp(-\beta V_1)-\exp(-\beta V_2)](1-\delta)^2[3(1-\delta)^2-2] \nonumber\\
-D_2 \exp(-\beta V_1)\} - \frac{\pi}{2D_1Z'^2}[\exp(-\beta V_1)-\exp(-\beta V_2)]^2(1-\delta)^4 \ln(1-\delta), \label{EXPa0}
\end{gather}
and the expression of $c_0$ can be directly read from Eq.~(\ref{EXPa0c0}). Finally, to determine the $a_n$ and $c_n$ coefficients for $n\ge1$, the condition (\ref{RGreen2d4}) at $r=1-\delta$ becomes
\begin{gather}
D_1\exp(-\beta V_1) \left\{  \frac{\exp(-\beta V_1) (1-\delta)}{2 D_1 Z'}  + \frac{b_0}{1-\delta} + \sum_{n=1}^\infty a_n n \left[ (1-\delta)^{n-1} - (1-\delta)^{-n-1} \right] \cos(n\theta) \right\} \nonumber\\
- D_2\exp(-\beta V_2) \left\{  \frac{\exp(-\beta V_1) (1-\delta)}{2 D_2 Z'}  + \sum_{n=1}^\infty c_n n (1-\delta)^{n-1} \cos(n\theta) \right\}  = \nonumber\\
\frac{D_1 \exp(-\beta V_1)-D_2 \exp(-\beta V_2)}{\pi D_1} \frac{(1-\cos\theta)-\delta}{2(1-\delta)(1-\cos\theta)+\delta^2}.
\end{gather}
The orthogonality of $\cos(n\theta)$ gives
\begin{equation}
D_1\exp(-\beta V_1) a_n n \left[ 1 - (1-\delta)^{-2n} \right] - D_2\exp(-\beta V_2) c_n n = -\frac{D_1 \exp(-\beta V_1)-D_2 \exp(-\beta V_2)}{2 \pi D_1},
\end{equation}
where the equality (for $n \ge 1$)
\begin{equation}
2\int_0^{2\pi} d\theta \ \frac{(1-\cos\theta)-\delta}{2(1-\delta)(1-\cos\theta)+\delta^2} \cos(n\theta) = - (1-\delta)^{n-1}
\end{equation}
has been used. The relation (\ref{EXPancn}) yields
\begin{equation}
a_n = \frac{1}{\pi D_1 n} \frac{(D_1 \exp(-\beta V_1)-D_2 \exp(-\beta V_2))(1-\delta)^{2n}}{ D_1 \exp(-\beta V_1) + D_2 \exp(-\beta V_2) - (D_1 \exp(-\beta V_1)-D_2 \exp(-\beta V_2))(1-\delta)^{2n}}, \label{EXPan}
\end{equation}
and the $c_n$ coefficients can be read using the Eq.~(\ref{EXPancn}).

We focus on two quantities: the average MFPT and the MFPT starting at the center given by Eqs.~(\ref{expNEGMFPTapp}) and~(\ref{expNEMFPT}) applied at ${\bf x} = {\bf 0}$, respectively. These quantities are equal to
\begin{gather}
\langle t \rangle = \frac{Z}{\exp(-\beta V_1)} \left[ - \frac{1}{\pi D_1} \ln \frac{\varepsilon}{4} + \frac{\exp(-\beta V_1)}{4D_1Z'} + a_0 + 2 \sum_{n=1}^\infty a_n\right],\\
t({\bf 0}) = \frac{Z}{\exp(-\beta V_1)} \left[ - \frac{1}{\pi D_1} \ln \frac{\varepsilon}{4} + \frac{\exp(-\beta V_1)}{4D_1Z'} + (a_0-c_0) + 2 \sum_{n=1}^\infty a_n\right].
\end{gather}

In the following, we express these quantities in terms of $\chi = 1-\delta$ and $\Delta V = V_2 - V_1$. The series of $a_n$ is equal to
\begin{equation}
\sum_{n=1}^\infty a_n = \frac{1}{\pi D_1}\sum_{n=1}^\infty \frac{u \chi^{2n}}{n\left[1-u \chi^{2n}\right]}, \qquad u = \frac{D_1-D_2\exp(-\beta \Delta V)}{D_1 + D_2 \exp(-\beta \Delta V)}.
\end{equation}
Since $|u|<1$ and $0 \le \chi<1$, this series can be rewritten as
\begin{equation}
\sum_{n=1}^\infty a_n = \frac{1}{\pi D_1} \sum_{n=1}^\infty \sum_{k=1}^\infty \frac{u^k \chi^{2nk}}{n} = - \frac{1}{\pi D_1} \sum_{k=1}^\infty u^k \ln(1-\chi^{2k}).
\end{equation}

Using the expression of $a_0$ given by Eq.~(\ref{EXPa0}), the GMFPT writes
\begin{gather}
T(\chi) = \frac{D_1 \langle t \rangle}{{\cal R}^2} = \left\{1-\chi^2[1-\exp(-\beta \Delta V)]\right\} \left\{- \ln \frac{\varepsilon}{4} - 2 \sum_{k=1}^\infty \left[ \frac{D_1-D_2\exp(-\beta \Delta V)}{D_1 + D_2 \exp(-\beta \Delta V)} \right]^k \ln(1-\chi^{2k}) \right\} \nonumber \\
+ \frac{1}{8} + \frac{(D_1-D_2)\chi^2 \exp(-\beta \Delta V) + 3 D_2 [\exp(-\beta \Delta V)-1](1-\chi^2)}{8D_2 \left\{1-\chi^2[1-\exp(-\beta \Delta V)]\right\}} \chi^2 - \frac{[1-\exp(-\beta\Delta V)]^2 \chi^4\ln \chi}{2\left\{1-\chi^2[1-\exp(-\beta \Delta V)]\right\}},\label{eqGMFPT2dapp}
\end{gather}
and from the expression of $a_0-c_0$ given by Eq.~(\ref{EXPa0c0}), the CMFPT is
\begin{gather}
T_0(\chi) = \frac{D_1 t({\bf 0})}{{\cal R}^2} = \left\{1-\chi^2[1-\exp(-\beta \Delta V)]\right\} \left\{- \ln \frac{\varepsilon}{4} - 2 \sum_{k=1}^\infty \left[ \frac{D_1-D_2\exp(-\beta \Delta V)}{D_1 + D_2 \exp(-\beta \Delta V)} \right]^k \ln(1-\chi^{2k}) \right\} \nonumber \\
+ \frac{1}{4} + \frac{D_1-D_2}{4D_2} \chi^2 - \frac{\exp(-\beta\Delta V) -1}{2}\chi^2\ln \chi.\label{eqCMFPT2dapp}
\end{gather}

%%%%%%%%%%%%%%%%%%%%%%%%
%%% NARROW ESCAPE 3D %%%
%%%%%%%%%%%%%%%%%%%%%%%%

\section{Derivation of the Green's function and narrow escape expression in three dimensions}
\label{appendix3d}

For the two-shell geometry, the diffusion constant and the stationary probability density close to the escape region are respectively $D({\bf x_0})=D_1$ and $P_s({\bf x_0}) = \exp(-\beta V_1)/Z$, with the partition function
\begin{equation}
Z = \int_\Omega \exp(-\beta V({\bf x})) = \frac{4\pi}{3} \left\{ {\cal R}^3 \exp(-\beta V_1) + ({\cal R}-\Delta)^3 \left[ \exp(-\beta V_2) - \exp(-\beta V_1) \right] \right\} .
\end{equation}
The Eqs.~(\ref{eqGreen1})-(\ref{eqGreen3}) and Eq.~(\ref{eqGreen3d}) for the pseudo Green's function can be rewritten in the two-shell geometry to obtain its regular part $R({\bf x}|{\bf x_0})$ defined by Eq.~(\ref{RGreen3dDef}). We use the dimensionless spherical coordinates $(r,\theta,\varphi)$ defined by $x/{\cal R}=r\cos \theta$, $y/{\cal R}=r\sin \theta\cos \varphi$ and $z/{\cal R}=r\sin \theta\sin \varphi$. The regular part of the Green's function is denoted as $R({\bf x}|{\bf x_0}) = R(r,\theta,\varphi)/{\cal R}$ without loss of generality (such that $D_1 R(r,\theta)$ is dimensionless), and the escape region is located at $r=1$ and $\theta=0$. Note that the problem is invariant under the rotation of the azimuthal angle $\varphi$, and hence $R((r,\theta,\varphi)$ is independent of $\varphi$. Since the normal vector ${\bf n}$ is radially oriented, the normal derivative of $|{\bf x} - {\bf x_0}|^{-1}$ at the point $(r,\theta,\varphi)$ is
\begin{equation}
\label{relLogSurf3d}
{\bf n} \cdot \nabla_{\bf x} \frac{{\cal R}}{|{\bf x} - {\bf x_0}|} = - \frac{r - \cos \theta}{(r^2 - 2 r \cos \theta +1)^{3/2}}.
\end{equation}

The value of $R(r,\theta)$ in the inner shell and the outer shell are defined by $R_1(r,\theta)$ and $R_2(r,\theta)$, respectively. From Eq.~(\ref{eqGreen1}) the bulk equations verified by $R_i(r,\theta)$ are
\begin{gather}
D_1 \nabla^2 R_1(r,\theta) = \frac{\exp(-\beta V_1)}{Z'}, \quad r>1-\delta \label{RGreen3d1} \\
D_2 \nabla^2 R_2(r,\theta) = \frac{\exp(-\beta V_1)}{Z'}, \quad r<1-\delta.  \label{RGreen3d2}
\end{gather}
with $Z'=Z/{\cal R}^3$. Analogously to Eq.~(\ref{MFPT3}) the boundary conditions at $r=1-\delta$
\begin{gather}
R_1(1-\delta,\theta) =  R_2(1-\delta,\theta) , \label{RGreen3d3} \\
D_1 \exp(-\beta V_1) \frac{\partial R_1}{\partial r}(1-\delta,\theta)  - D_2 \exp(-\beta V_2) \frac{\partial R_2}{\partial r}(1-\delta,\theta) = \frac{D_1 \exp(-\beta V_1)-D_2 \exp(-\beta V_2)}{2 \pi D_1} \frac{(1-\cos\theta)-\delta}{[2(1-\delta)(1-\cos\theta)+\delta^2]^{3/2}}, \label{RGreen3d4}
\end{gather}
where the relation (\ref{relLogSurf3d}) has been used, and from Eq.~(\ref{eqGreen2}) the reflective boundary condition at $r=1$ becomes 
\begin{equation}
\frac{\partial R_1}{\partial r}(1,\theta) = \frac{1}{4\pi D_1} \frac{1}{\sqrt{2(1-\cos\theta)}} \label{RGreen3d5}.
\end{equation}
Finally, the condition~(\ref{eqGreen3}) gives the last relation needed to derive $R$:
\begin{equation}
2\pi \int_0^{\pi} d\theta \sin\theta \left[ \exp(-\beta V_2)\int_0^{1-\delta} dr\ r^2 R_2(r,\theta) + \exp(-\beta V_1)\int_{1-\delta}^1 dr\ r^2 R_1(r,\theta) \right] = - \frac{1}{2\pi D_1} Z'. \label{RGreen3d6}
\end{equation}

The general solution of Eqs.~(\ref{RGreen3d1}) and~(\ref{RGreen3d2}), which can be rewritten as $D_i \nabla^2 R_i(r,\theta) = \exp(-\beta V_1)/Z'$, has the form
\begin{equation}
R_i(r,\theta) = \sum_{n=0}^\infty f_{i,n}(r) P_n(\cos\theta)
\end{equation}
where $P_n(X)$ is the Legendre polynomial of degree $n$, due to the spherical symmetry of the problem, independent of the azimuthal $\varphi$. The functions $f_{i,n}(r)$ must satisfy the differential equations
\begin{gather}
f_{i,n}''(r) +\frac{2}{r} f_{i,n}'(r) - \frac{n(n+1)}{r^2} f_{i,n}(r) =0, \quad n\ge1\\
D_i \left[ f_{i,0}''(r) +\frac{2}{r} f_{i,0}'(r) \right] = \frac{\exp(-\beta V_1)}{Z'}.
\end{gather}
The global solution is then given by $f_{i,n}(r) = a_{i,n} r^n + b_{i,n} r^{-n-1}$ (with $n\ge 1$) and $f_{i,0}(r) = \exp(-\beta V_1)r^2/(6D_iZ') + a_{i,0} + b_{i,0}/r$. Hence the functions $R_1$ and $R_2$ have the general form
\begin{gather}
R_1(r,\theta) = \frac{\exp(-\beta V_1) r^2}{6 D_1 Z'} + a_0 + \frac{b_0}{r} + \sum_{n=1}^\infty \left[a_n r^n + b_n r^{-n-1}\right]P_n(\cos\theta), \\
R_2(r,\theta) = \frac{\exp(-\beta V_1) r^2}{6 D_2 Z'} + c_0 + \frac{d_0}{r} + \sum_{n=1}^\infty \left[c_n r^n + d_n r^{-n-1}\right]P_n(\cos\theta).
\end{gather}

Since the problem is not singular at $r=0$, we must impose $d_n =0$ for $n\ge0$ to have a non divergent solution $R_2(r,\theta)$. The boundary condition at $r=1$ given by Eq.~(\ref{RGreen3d5}) writes
\begin{equation}
\frac{\exp(-\beta V_1)}{3 D_1 Z'} - b_0 + \sum_{n=1}^\infty \left[n a_n - (n+1) b_n\right] P_n(\cos\theta)  = \frac{1}{4\pi D_1} \frac{1}{\sqrt{2(1-\cos\theta)}}.
\end{equation}
The orthogonality of the Legendre polynomials and the integral value
\begin{equation}
\int_{-1}^1 dX \frac{P_n(X)}{\sqrt{2(1-X)}} = \frac{2}{2n+1},
\end{equation}
impose the relations
\begin{gather}
b_0 = \frac{\exp(-\beta V_1)}{3 D_1 Z'} - \frac{1}{4\pi D_1} = \frac{[\exp(-\beta V_1)-\exp(-\beta V_2)](1-\delta)^3}{3 D_1 Z'},\label{EXPb03d}\\
a_n = \frac{1}{4\pi D_1 n} + \frac{n+1}{n} b_n. \label{EXPanbn3d}
\end{gather}

The continuity at $r=1-\delta$, expressed by the Eq.~(\ref{RGreen3d3}), becomes
\begin{gather}
\frac{\exp(-\beta V_1)(1-\delta)^2}{6D_1Z'} + a_0 + \frac{b_0}{1-\delta} + \sum_{n=1}^\infty \left[a_n (1-\delta)^n + b_n(1-\delta)^{-n-1}\right]P_n(\cos\theta) = \nonumber\\
\frac{\exp(-\beta V_1) (1-\delta)^2}{6 D_2 Z'} + c_0 + \sum_{n=1}^\infty c_n (1-\delta)^n P_n(\cos\theta).
\end{gather}
Using the Eq.~(\ref{EXPanbn3d}), the orthogonality of the Legendre polynomials yields
\begin{gather}
a_0-c_0 =  \left[\frac{D_1-D_2}{2D_2} + \exp(-\beta V_2)-\exp(-\beta V_1) \right] \frac{\exp(-\beta V_1)(1-\delta)^2}{3 D_1 Z'}, \label{EXPa0c03d}\\
c_n= \frac{1}{4\pi D_1 n} + b_n \left[\frac{n+1}{n}+(1-\delta)^{-2n-1}\right].\label{EXPbncn3d}
\end{gather}
Since $\int_{-1}^1 dX P_n(X) = 0$ for $n\ge1$, the condition (\ref{RGreen3d6}) becomes
\begin{equation}
2\pi \int_0^{1-\delta} dr\ r^2\left[\frac{D_1 \exp(-\beta V_1) r^2}{3 D_2 Z'} + 2 c_0\right] + \int_{1-\delta}^1 dr\ r^2\left[\frac{\exp(-\beta V_1) r^2}{3 D_1 Z'} + 2a_0 + 2\frac{b_0}{r} \right] = - \frac{Z'}{2\pi},
\end{equation}
and the relation (\ref{EXPa0c03d}) finally gives
\begin{gather}
a_0 = -\frac{1}{2\pi D_1} - \frac{2\pi \exp(-\beta V_1)}{45D_1D_2Z'^2} \{ 3D_2 + [2 (D_2-D_1)\exp(-\beta V_2)-8 D_2 (\exp(-\beta V_1)-\exp(-\beta V_2)) \nonumber\\
 -10 D_2 (\exp(-\beta V_1)-\exp(-\beta V_2))^2](1-\delta)^5 +15D_2[\exp(-\beta V_1)-\exp(-\beta V_2)](1-\delta)^3 \} , \label{EXPa03d}
\end{gather}
and the expression of $c_0$ can be directly read from Eq.~(\ref{EXPa0c03d}). Finally, to determine the $a_n$ and $c_n$ coefficients for $n\ge1$, the condition (\ref{RGreen3d4}) at $r=1-\delta$ becomes
\begin{gather}
D_1\exp(-\beta V_1) \left\{  \frac{\exp(-\beta V_1) (1-\delta)}{3 D_1 Z'}  - \frac{b_0}{(1-\delta)^2} + \sum_{n=1}^\infty \left[ a_n n (1-\delta)^{n-1} - (n+1) b_n (1-\delta)^{-n-2} \right] P_n(\cos\theta) \right\} \nonumber\\
-D_2\exp(-\beta V_2) \left\{  \frac{\exp(-\beta V_1) (1-\delta)}{3 D_2 Z'}  + \sum_{n=1}^\infty c_n n (1-\delta)^{n-1} P_n(\cos\theta) \right\} = \nonumber\\
\frac{D_1 \exp(-\beta V_1)-D_2 \exp(-\beta V_2)}{2 \pi D_1} \frac{(1-\cos\theta)-\delta}{[2(1-\delta)(1-\cos\theta)+\delta^2]^{3/2}}.
\end{gather}
The orthogonality of the Legendre polynomials yields
\begin{equation}
D_1\exp(-\beta V_1) \left[ a_n  - \frac{n+1}{n} b_n (1-\delta)^{-2n-1} \right] - D_2\exp(-\beta V_2) c_n = -\frac{D_1 \exp(-\beta V_1)-D_2 \exp(-\beta V_2)}{2 \pi D_1},
\end{equation}
where the equality (for $n \ge 1$)
\begin{equation}
\int_{-1}^1 dX \  \frac{(1-X)-\delta}{2(1-\delta)(1-X)+\delta^2} P_n(X) = - \frac{2n}{2n+1}(1-\delta)^{n-1}.
\end{equation}
has been used. The Eqs.~(\ref{EXPanbn3d}) and~(\ref{EXPbncn3d}) gives
\begin{gather}
b_n = \frac{2n+1}{4\pi D_1 (n+1)}\frac{(D_1 \exp(-\beta V_1)-D_2 \exp(-\beta V_2))(1-\delta)^{2n+1}}{D_1 \exp(-\beta V_1) + \frac{n}{n+1}D_2 \exp(-\beta V_2) - (D_1 \exp(-\beta V_1)-D_2 \exp(-\beta V_2))(1-\delta)^{2n+1}}, \label{EXPbn3d}
\end{gather}
and the $a_n$ and $c_n$ coefficients can be read from Eqs.~(\ref{EXPanbn3d}) and~(\ref{EXPbncn3d}), respectively .

The expressions of $\gamma$ and $R_0$ can be derived from the limit close to the escape region $(1,0)$ of $R_1(r,\theta)$ whose the expression is
\begin{equation}
R_1(r,\theta) = \frac{\exp(-\beta V_1) r^2}{6 D_1 Z'} + a_0 + \frac{b_0}{r} + \frac{1}{4\pi D_1} \sum_{n=1}^\infty \frac{r^n}{n} P_n(\cos\theta) +  \sum_{n=1}^\infty \left[ \frac{n+1}{n} r^n + r^{-n-1}\right] b_n P_n(\cos\theta),
\end{equation}
after using Eq.~(\ref{EXPanbn3d}). The generating function of Legendre polynomials yields
\begin{equation}
\sum_{n=1}^\infty r^n P_n(\cos\theta) = \frac{{\cal R}}{|{\bf x}-{\bf x_0}|} \quad \Leftrightarrow \quad \sum_{n=1}^\infty \frac{r^n}{n} P_n(\cos\theta) = \ln \frac{2}{1-r\cos\theta + |{\bf x}-{\bf x_0}|/{\cal R}},
\end{equation}
which determines the expression of the first series. The limit when ${\bf x} \to {\bf x_0}$ is then
\begin{equation}
R({\bf x}\to{\bf x_0}|{\bf x_0}) = - \frac{1}{4\pi D_1 {\cal R}} \ln \frac{|{\bf x}-{\bf x_0}|}{2{\cal R}}+ \frac{1}{\cal R} \left[\frac{\exp(-\beta V_1)}{6 D_1 Z'} + a_0 + b_0 + \sum_{n=1}^\infty \frac{2n+1}{n} b_n \right],
\end{equation}
and the expressions of $\gamma$ and $R_0$ are
\begin{equation}
\gamma = - \frac{1}{4\pi D_1 {\cal R}} \quad {\rm and} \quad R_0= \frac{1}{\cal R} \left[\frac{\exp(-\beta V_1)}{6 D_1 Z'} + a_0 + b_0 + \sum_{n=1}^\infty \frac{2n+1}{n} b_n \right].
\end{equation}

We focus on two quantities: the average MFPT and the MFPT starting at the center given by Eqs.~(\ref{expNEGMFPT3dapp}) and~(\ref{expNEMFPT3d}) applied at ${\bf x} = {\bf 0}$, respectively. These quantities are equal to
\begin{gather}
\langle t \rangle = \frac{Z}{\exp(-\beta V_1){\cal R}} \left[ \frac{1}{2 D_1 \varepsilon} - \frac{1}{4\pi D_1} \ln \varepsilon + \frac{\exp(-\beta V_1)}{6D_1Z'} + a_0 + b_0 + \sum_{n=1}^\infty \frac{2n+1}{n}b_n + \frac{3}{8\pi D_1}\right],\\
t({\bf 0}) = \frac{Z}{\exp(-\beta V_1){\cal R}} \left[ \frac{1}{2 D_1 \varepsilon} - \frac{1}{4\pi D_1} \ln \varepsilon + \frac{\exp(-\beta V_1)}{6D_1Z'} + (a_0-c_0) + b_0 + \sum_{n=1}^\infty \frac{2n+1}{n}b_n - \frac{1}{8\pi D_1}\right].
\end{gather}

In the following, we express these quantities in terms of $\chi = 1-\delta$ and $\Delta V = V_2 - V_1$. The series of $b_n$ is equal to
\begin{equation}
\sum_{n=1}^\infty \frac{2n+1}{n} b_n = \frac{1}{4\pi D_1}\sum_{n=1}^\infty \frac{(2n+1)^2}{n(n+1)} \frac{u_n \chi^{2n+1}}{1-u_n \chi^{2n+1}}, \qquad u_n = \frac{D_1-D_2\exp(-\beta \Delta V)}{D_1 + \frac{n}{n+1}D_2 \exp(-\beta \Delta V)}.
\end{equation}
This expression cannot be further simplified. Using the expression of $a_0$ and $b_0$ given by Eqs.~(\ref{EXPa03d}) and~(\ref{EXPb03d}), respectively, the GMFPT writes
\begin{gather}
T(\chi) = \frac{D_1 \langle t \rangle}{{\cal R}^2} = \left\{1-\chi^3[1-\exp(-\beta \Delta V)]\right\} \left\{ \frac{2\pi}{3\varepsilon} - \frac{1}{3} \ln \varepsilon + \frac{1}{3} \sum_{n=1}^\infty \frac{(2n+1)^2}{n(n+1)} \frac{u_n \chi^{2n+1}}{1-u_n \chi^{2n+1}} -\frac{1}{6} \right\} \nonumber \\
+ \frac{D_2 + [(D_1-D_2)\exp(-\beta \Delta V) + 4 D_2 (1-\exp(-\beta \Delta V)) + 5 D_2(1-\exp(-\beta \Delta V))^2]\chi^5}{15 D_2 \left\{1-\chi^3[1-\exp(-\beta \Delta V)]\right\}} \nonumber\\
- \frac{\left[ 1-\exp(-\beta \Delta V) \right] \left\{1+\chi^3[1-\exp(-\beta \Delta V)]\right\}}{3 \left\{1-\chi^3[1-\exp(-\beta \Delta V)]\right\}} \chi^3,\label{eqGMFPT3dapp}
\end{gather}
and from the expression of $a_0-c_0$ and $b_0$ given by Eqs.~(\ref{EXPa0c0}) and~(\ref{EXPb03d}), respectively, the CMFPT is
\begin{gather}
T_0(\chi) = \frac{D_1 t({\bf 0})}{{\cal R}^2} = \left\{1-\chi^3[1-\exp(-\beta \Delta V)]\right\} \left\{\frac{2\pi}{3\varepsilon} - \frac{1}{3} \ln \varepsilon + \frac{1}{3} \sum_{n=1}^\infty \frac{(2n+1)^2}{n(n+1)} \frac{u_n \chi^{2n+1}}{1-u_n \chi^{2n+1}} - \frac{1}{6} \right\} \nonumber \\
+ \frac{1}{6} + \frac{D_1-D_2}{6D_2} \chi^2 - \frac{1-\exp(-\beta \Delta V)}{3} \chi^2(1-\chi).\label{eqCMFPT3dapp}
\end{gather}

%%%%%%%%%%%%%%%%%%%%%%%%%%%%%%%%%
%%% Special case by Grebenkov %%%
%%%%%%%%%%%%%%%%%%%%%%%%%%%%%%%%%

\section{Grebenkov's solution for piecewise constant diffusivity}
\label{appGreb}

In Ref.~\cite{grebenkov2016}, Grebenkov has shown an exact expression for the MFPT satisfying the equation
\begin{equation}
\nabla_{\bf y}^2 t({\bf y}) = - \frac{1}{D({\bf y})},
\end{equation}
which is compatible with Eq.~(\ref{eqMFPT}) only for the potential $\beta V({\bf y}) = \ln D({\bf y})$ for the origin of potential fixed at the reference of diffusion constants ($1 {\rm m}^2/{\rm s}$). The solution derived by Grebenkov writes in terms of complex variables:
\begin{equation}
\label{GrebEq}
t(y) = \int_\Omega \frac{dx}{D(x)} \left[ - \frac{\ln |\phi_{y}^{-1}(x)|}{2\pi} + W\left(\phi_{y}^{-1}(x) \right)\right]
\end{equation}
where $\phi_{y}(x)$ is the conformal mapping of the unit disk ${\cal D}$ on the domain $\Omega$ where the origin is fixed as $\phi_{y}(0) = y$, preserving the harmonic measure of the escape region which corresponds here to the perimeter for both domains ${\cal D}$ and $\Omega$. The explicit expression of $W(z)$ is
\begin{equation}
W(z) = \frac{1}{\pi} \ln \frac{|1-z+\sqrt{(1-z\exp(-i\varepsilon/2)(1-z\exp(i\varepsilon/2)}|}{2\sin(\varepsilon/4)} = -\frac{1}{\pi} \ln \sin(\varepsilon/4) - \frac{1}{\pi} \sum_{n=1}^\infty c_n r^n \cos(n\theta)
\end{equation}
with $z=re^{i\theta}$ and the coefficients $c_n$ written in terms of Legendre polynomials of degree $n$ of $\cos(\varepsilon/2)$:
\begin{equation}
c_n=\frac{P_{n-1}(\cos(\varepsilon/2))+P_n(\cos(\varepsilon/2))}{2n}.
\end{equation}

For our two-shell geometry, the equivalence is valid for $\beta \Delta V = \ln (D_2/D_1)$ and the conformal mapping satisfies the identities
\begin{equation}
\label{GrebCM}
\phi_{y}(x) = \frac{y-x e^{i\alpha}}{1- x \overline y e^{i\alpha}}, \qquad  \phi_{y}^{-1}(x) = \frac{y-x}{1-x \overline y} e^{-i\alpha}
\end{equation}
where $\overline y$ is the conjugate of $y$ and $\alpha$ is an angle whose value is unimportant for the following. We first calculate the expression of the CMFPT. From the Eq.~(\ref{GrebEq}), $t(0)$ depends only on $\phi_0^{-1}(x) = -x e^{-i\alpha}$. Denoting $x=re^{i\theta}$, Eq.~(\ref{GrebEq}) becomes
\begin{equation}
t(0) = \int_\Omega \frac{r dr \ d\theta}{D(r)} \left[ - \frac{\ln r}{2\pi} - \frac{\ln \sin(\varepsilon/4)}{\pi} - \frac{1}{\pi} \sum_{n=1}^\infty  (-1)^n c_n r^n \cos(n(\theta-\alpha)) \right]
\end{equation}
since $D(x)$ depends only on the radial coordinate. Performing the orthoradial integral we obtain
\begin{equation}
t(0) = -\int_0^1 \frac{r dr}{D(r)} \left[  \ln r + 2 \ln \sin(\varepsilon/4) \right].
\end{equation}
Performing the radial integral for the piecewise constant diffusivity, the dimensionless CMFPT ($T_0 =D_1 t(0)$) is then equal to
\begin{equation}
T_0 = - \left( 1 + \frac{D_1-D_2}{D_2} \chi^2 \right) \ln \sin(\varepsilon/4) + \frac{1}{4} + \frac{D_1-D_2}{4D_2} \chi^2 + \frac{D_2-D_1}{2D_2} \chi^2 \ln \chi,
\end{equation}
where $\chi = 1 - \Delta/{\cal R}$. The GMFPT is here defined by
\begin{equation}
\langle t \rangle = \overline{D^{-1}}\int_\Omega \frac{dy}{D(y)}t(y) \qquad {\rm with} \qquad \overline{D^{-1}} = \int_\Omega \frac{dy}{D(y)}.
\end{equation}
Eq.~(\ref{GrebEq}) gives
\begin{equation}
\label{GrebT}
\langle t \rangle = \overline{D^{-1}}\int_\Omega \frac{dy}{D(y)} \int_\Omega \frac{dx}{D(x)} \left[ - \frac{\ln |\phi_{y}^{-1}(x)|}{2\pi} + W\left(\phi_{y}^{-1}(x) \right)\right] \equiv I_1 + I_2,
\end{equation}
in which the two contributions $I_1$ and $I_2$ are introduced to simplify the derivation. The first contribution is independent of $\varepsilon$ and writes
\begin{equation}
I_1 = \overline{D^{-1}}\int_\Omega \frac{dy}{D(y)} t_{\rm D}(y) \qquad {\rm with} \qquad t_{\rm D}(y) = - \frac{1}{2\pi} \int_\Omega \frac{dx}{D(x)} \ln |\phi_{y}^{-1}(x)|.
\end{equation}
Denoting $x=re^{i\theta}$ and $y= r_0 e^{i\theta_0}$ the Eq.~(\ref{GrebCM}) writes
\begin{equation}
t_{\rm D}(y) = - \frac{1}{4\pi} \int_\Omega \frac{dx}{D(x)} \ln \frac{r_0^2 -2rr_0\cos(\theta-\theta_0) +r^2}{1-2rr_0 \cos(\theta - \theta_0)+r^2r_0^2}.
\end{equation}
The orthoradial integral is performed by remarking the identity
\begin{equation}
\int_0^{2\pi} d\theta \ln (1 -2 u \cos(\theta-\theta_0) +u^2)=0, \qquad \forall u \in [0,1].
\end{equation}
Hence $t_{\rm D}(y)$ depends only on the radial coordinate $r_0= |y|$ such that 
\begin{equation}
t_{\rm D}(r_0) = \frac{1}{2} \int_0^1 \frac{r dr}{D(r)} \ln \frac{2}{r^2+r_0^2+|r^2-r_0^2|}.
\end{equation}
Performing the integral for the piecewise constant diffusivity we obtain
\begin{equation}
t_{\rm D}(r_0) = 
\begin{dcases}
\frac{D_2-D_1r_0^2}{4D_1D_2} + \frac{D_1-D_2}{4D_1D_2}\chi^2 + \frac{D_2-D_1}{2D_1D_2} \chi^2 \ln \chi, &\qquad r_0<\chi\\
\frac{1-r_0^2}{4D_1} + \frac{D_2-D_1}{2D_1D_2} \chi^2 \ln r_0, &\qquad r_0>\chi
\end{dcases}
\end{equation}
for $\chi=1-\Delta / {\cal R}$. The contribution $I_1$ is thus
\begin{equation}
\label{eqI1app}
I_1 = 2 \pi \overline{D^{-1}}\int_0^{1} \frac{r_0 dr_0}{D(r_0)}t_{\rm D}(r_0) = \frac{1}{8D_1} + \frac{D_1-D_2}{8D_1D_2^2} \frac{3D_2 +(D_1-3D_2)\chi^2}{1+\frac{D_2-D_1}{D_2}\chi^2} \chi^2 - \frac{(D_1-D_2)^2}{2D_1D_2^2 } \frac{\chi^4 \ln \chi}{1+\frac{D_2-D_1}{D_2}\chi^2}.
\end{equation}
The second contribution depends on $\varepsilon$ and writes
\begin{equation}
I_2 = - \frac{\overline{D^{-1}}}{\pi} \ln \sin(\varepsilon/4) - \frac{\overline{D^{-1}}}{\pi} \int_\Omega \frac{dy}{D(y)} \int_{\cal D} \frac{dz}{D(|\phi_y(z)|)} |\phi'_y(z)|^2 \sum_{n=1}^\infty c_n r^n \cos(n\theta)
\end{equation}
where the change of variable $x = \phi_y(z)$ has been realized and $\phi'_y(z) = d\phi_y/dz$ is the complex derivation by the complex variable $z=re^{i\theta}$. The last integral can be expanded as
\begin{equation}
\int_\Omega \frac{r_0dr_0 \ d\theta_0}{D(r_0)} \int_{\cal D} \frac{dr \ d\theta}{D(|\phi_y(z)|)} \sum_{n=1}^\infty \frac{(1-r_0^2)^2c_n r^{n+1} \cos(n\theta)}{(1-2r r_0 \cos(\theta+\alpha -\theta_0)+r^2r_0^2)^2}, \quad |\phi_y(z)|^2 = \frac{r_0^2 -2 rr_0\cos(\theta+\alpha -\theta_0)+r^2}{1 -2 rr_0\cos(\theta+\alpha -\theta_0)+r^2r_0^2}.
\end{equation}
Considering the change of variable $\widetilde{\theta_0} =  \theta_0-\theta-\alpha$ and using the periodicity over this new variable, the last integral becomes
\begin{equation}
\int_\Omega \frac{r_0dr_0 \ d\widetilde{\theta_0}}{D(r_0)} \int_{\cal D} \frac{dr \ d\theta}{D(|\phi_y(z)|)} \sum_{n=1}^\infty \frac{(1-r_0^2)^2c_n r^{n+1} \cos(n\theta)}{(1-2r r_0 \cos\widetilde{\theta_0}+r^2r_0^2)^2}, \quad |\phi_y(z)|^2 = \frac{r_0^2 -2 rr_0\cos\widetilde{\theta_0}+r^2}{1 -2 rr_0\cos\widetilde{\theta_0}+r^2r_0^2}.
\end{equation}
$D(|\phi_y(z)|)$ does not depend on $\theta$ anymore and the integral over the orthoradial coordinate can be moved inside the series. Since $\int_0^{2\pi} d\theta \cos(n\theta)=0$ for $n\ge1$, this integral is equal to zero. The contribution $I_2$ is thus  
\begin{equation}
\label{eqI2app}
I_2 = -  \frac{1}{D_1} \left( 1 + \frac{D_1-D_2}{D_2} \chi^2 \right) \ln \sin(\varepsilon/4).
\end{equation}
Adding the expression of the two contributions $I_1$ and $I_2$ given by Eqs.~(\ref{eqI1app}) and~(\ref{eqI2app}), the dimensionless GMFPT ($T=D_1 \langle t \rangle$) is hence
\begin{equation}
T = -\left( 1 + \frac{D_1-D_2}{D_2} \chi^2 \right) \ln \sin(\varepsilon/4) + \frac{1}{8} + \frac{D_1-D_2}{8D_2} \frac{3D_2 +(D_1-3D_2)\chi^2}{D_2+(D_1-D_2)\chi^2} \chi^2 - \frac{(D_1-D_2)^2}{2D_2} \frac{\chi^4 \ln \chi}{D_2+(D_1-D_2)\chi^2}.
\end{equation}

\newpage

\begin{center}
{\large \bf Supplemental material: Narrow escape problem in two-shell spherical domains}
\end{center}

\section{Fully-absorbing limit}
\label{appendixF}

In this section, we study the fully-absorbing limit ($\varepsilon=2\pi$), for which the external boundary is totally absorbing. From the symmetries of this problem, the MFPT depends only on the radial coordinate $r$. From the Eqs.~(5)-(8) of the main text, the MFPT satisfy then
\begin{gather}
\frac{D_1}{r^{d-1}} \frac{d}{dr} r^{d-1} \frac{dt_1}{dr} = -1, \qquad r>{\cal R}-\Delta,\\
\frac{D_2}{r^{d-1}} \frac{d}{dr} r^{d-1} \frac{dt_2}{dr} = -1, \qquad r<{\cal R}-\Delta,\\
t_1({\cal R}-\Delta) = t_2({\cal R}-\Delta) \qquad {\rm and} \qquad D_1\exp(-\beta V_1) \frac{dt_1}{dr}({\cal R}-\Delta) = D_2\exp(-\beta V_2) \frac{dt_2}{dr}({\cal R}-\Delta), \label{BC1}\\
t_1({\cal R}) = 0.\label{BC2}
\end{gather}
The general solution of the Laplace's equations are
\begin{equation}
t_i(r) = \begin{dcases}
- \frac{r^2}{4D_i} + A_i \ln r + B_i \quad& (d=2)\\
- \frac{r^2}{6D_i} + \frac{A_i}{r} + B_i \quad& (d=3)
\end{dcases}
\end{equation}
The identification of constants $A_i$ and $B_i$ is made with Eqs.~(\ref{BC1}) and~(\ref{BC2}).

\medskip

The 2d solution for the MFPT is
\begin{gather}
\frac{D_1}{{\cal R}^2} t_1(r) = - \frac{r^2}{4{\cal R}^2} + \frac{\chi^2}{2}(1-\xi) \ln(r/{\cal R}) + \frac{1}{4} \\
\frac{D_1}{{\cal R}^2} t_2(r) = - \frac{D_1 r^2}{4 D_2 {\cal R}^2} + \frac{1}{4} + \frac{D_1-D_2}{4D_2} \chi^2 - \frac{\xi-1}{2}\chi^2 \ln\chi
\end{gather}
with $\chi=1-\Delta/{\cal R}$ and $\xi=\exp(-\beta\Delta V)$. The GMFPT expression is obtained from this solution:
\begin{gather}
\label{GMFPT2d}
T = \frac{D_1}{{\cal R}^2} \langle t \rangle = \frac{1}{8} + \frac{(D_1-D_2)\chi^2 \xi + 3 D_2 (\xi-1)(1-\chi^2)}{8D_2 \left[1-\chi^2(1-\xi)\right]} \chi^2 - \frac{(1-\xi)^2 \chi^4\ln \chi}{2\left[1-\chi^2(1-\xi)\right]},
\end{gather}
as well as the CMFPT expression:
\begin{gather}
\label{CMFPT2d}
T_0 = \frac{D_1}{{\cal R}^2} t_2(0) = \frac{1}{4} + \frac{D_1-D_2}{4D_2} \chi^2 - \frac{\xi-1}{2}\chi^2 \ln\chi.
\end{gather}

\begin{figure}[t]
\begin{center}
\includegraphics[width=16cm]{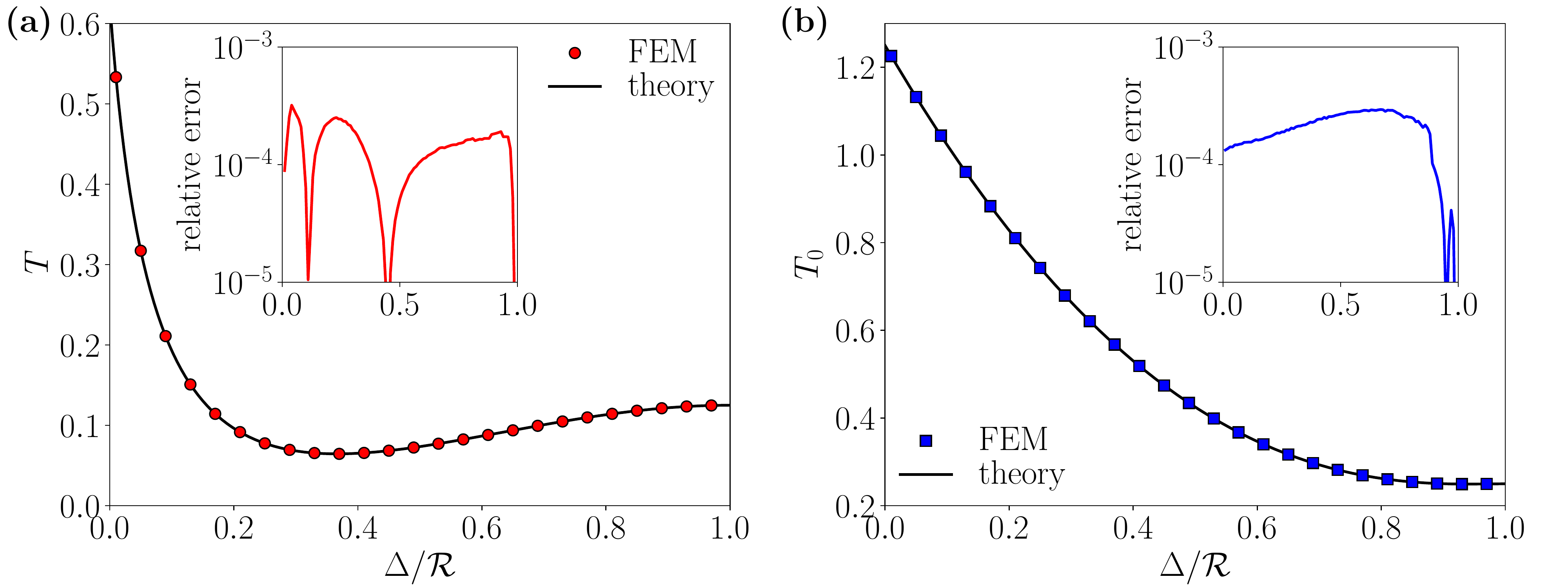}
\caption{(color online) Dependence of the dimensionless GMFPT~{\bf (a)} and CMFPT~{\bf (b)} on the outer shell width $\Delta$ in 2d, in the fully-absorbing limit, for $D_1/D_2=5$ and $\beta \Delta V=2$. The symbols show the numerical solution obtained with the finite element method using the software package FreeFem++ and the lines display the exact analytical expressions given by Eqs.~(\ref{GMFPT2d}) and~(\ref{CMFPT2d}). In insets, the relative error between numerical and analytical solutions is shown.} \label{figure2d_DELTA_SI}
\end{center}
\end{figure}

Fig.~\ref{figure2d_DELTA_SI} shows the dependence of these two MFPTs with $\Delta$, and compares these exact analytical solutions (lines) with the numerical solutions obtained with the finite element method using FreeFem++ (symbols). In insets, the relative error between these two solutions is represented, allowing us to estimate that the numerical error is of order $10^{-4}$.

\medskip

The 3d solution for the MFPT is
\begin{gather}
\frac{D_1}{{\cal R}^2} t_1(r) = - \frac{r^2}{6{\cal R}^2} - \frac{\chi^3}{3}(1-\xi) \frac{\cal R}{r} + \frac{1}{6} + \frac{\chi^3}{3}(1-\xi) \\
\frac{D_1}{{\cal R}^2} t_2(r) = - \frac{D_1 r^2}{6 D_2 {\cal R}^2} + \frac{1}{6} + \frac{D_1-D_2}{6D_2} \chi^2 - \frac{1-\xi}{3}\chi^2 (1-\chi)
\end{gather}
with $\chi=1-\Delta/{\cal R}$ and $\xi=\exp(-\beta\Delta V)$. The GMFPT expression is obtained from this solution:
\begin{gather}
\label{GMFPT3d}
T = \frac{D_1}{{\cal R}^2} \langle t \rangle = \frac{D_2 + [(D_1-D_2)\xi + 4D_2 (1-\xi) + 5 D_2(1-\xi)^2]\chi^5}{15 D_2 \left[1-\chi^3(1-\xi)\right]} - \frac{(1-\xi)\left[1+\chi^3(1-\xi)\right]}{3 \left[1-\chi^3(1-\xi)\right]}\chi^3,
\end{gather}
as well as the CMFPT expression:
\begin{gather}
\label{CMFPT3d}
T_0 = \frac{D_1}{{\cal R}^2} t_2(0) = \frac{1}{6} + \frac{D_1-D_2}{6D_2} \chi^2 - \frac{1-\xi}{3}\chi^2 (1-\chi).
\end{gather}

\begin{figure}[t]
\begin{center}
\includegraphics[width=16cm]{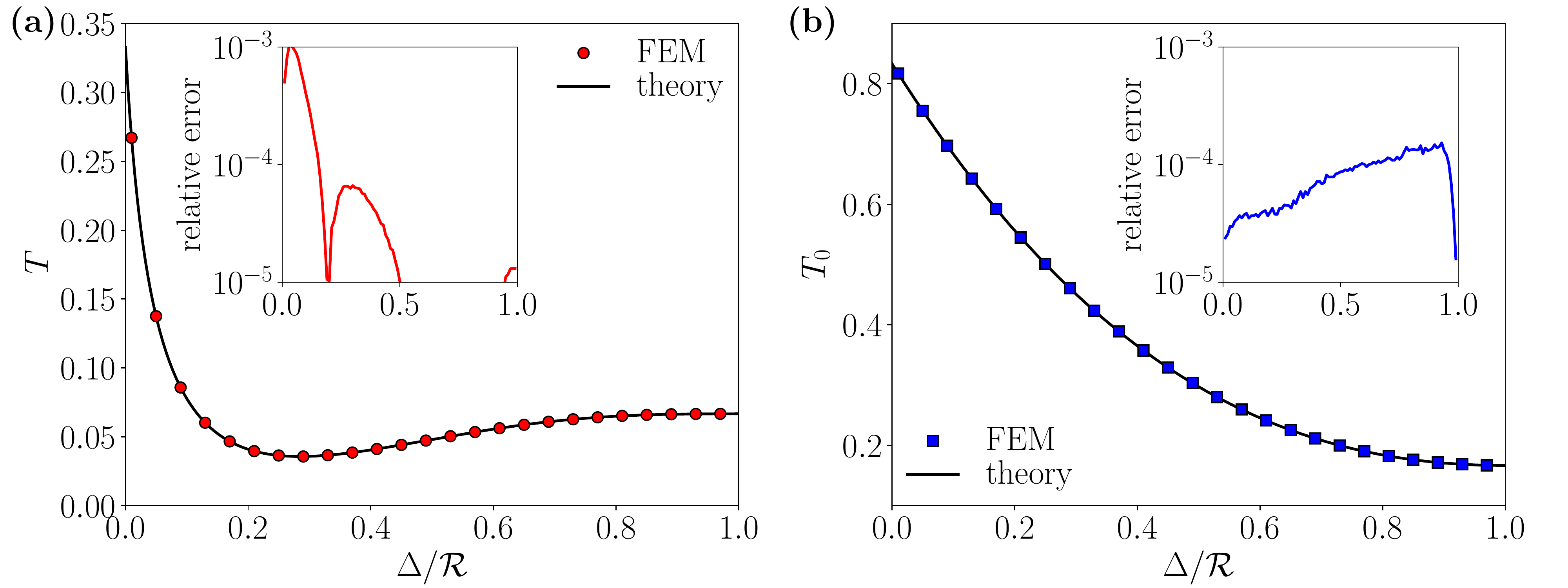}
\caption{(color online) Dependence of the dimensionless GMFPT~{\bf (a)} and CMFPT~{\bf (b)} on the outer shell width $\Delta$ in 3d, in the fully-absorbing limit, for $D_1/D_2=5$ and $\beta \Delta V=2$. The symbols show the numerical solution obtained with the finite element method using FreeFem++ and the lines display the exact analytical expressions given by Eqs.~(\ref{GMFPT3d}) and~(\ref{CMFPT3d}). In insets, the relative error between numerical and analytical solutions is shown.} \label{figure3d_DELTA_SI}
\end{center}
\end{figure}

Fig.~\ref{figure3d_DELTA_SI} shows the dependence of these two MFPTs with $\Delta$, and compares these exact analytical solutions (lines) with the numerical solutions obtained with the finite element method using the software package FreeFem++ (symbols). In insets, the relative error between these two solutions is represented, allowing us to estimate that the numerical error is of order $10^{-4}$, as for the 2d solution.	

\section{MMFPT starting point for an attractive potential difference}
\label{appendixG}

In this section, we analyze the distance of the maximal mean first passage time (MMFPT) to the center, denoted~$r_{\rm max}$. We have first analyzed the following behavior of the MMFPT in Ref.~\cite{mangeat2019} without potential barrier.

\begin{figure}[t]
\begin{center}
\includegraphics[width=16cm]{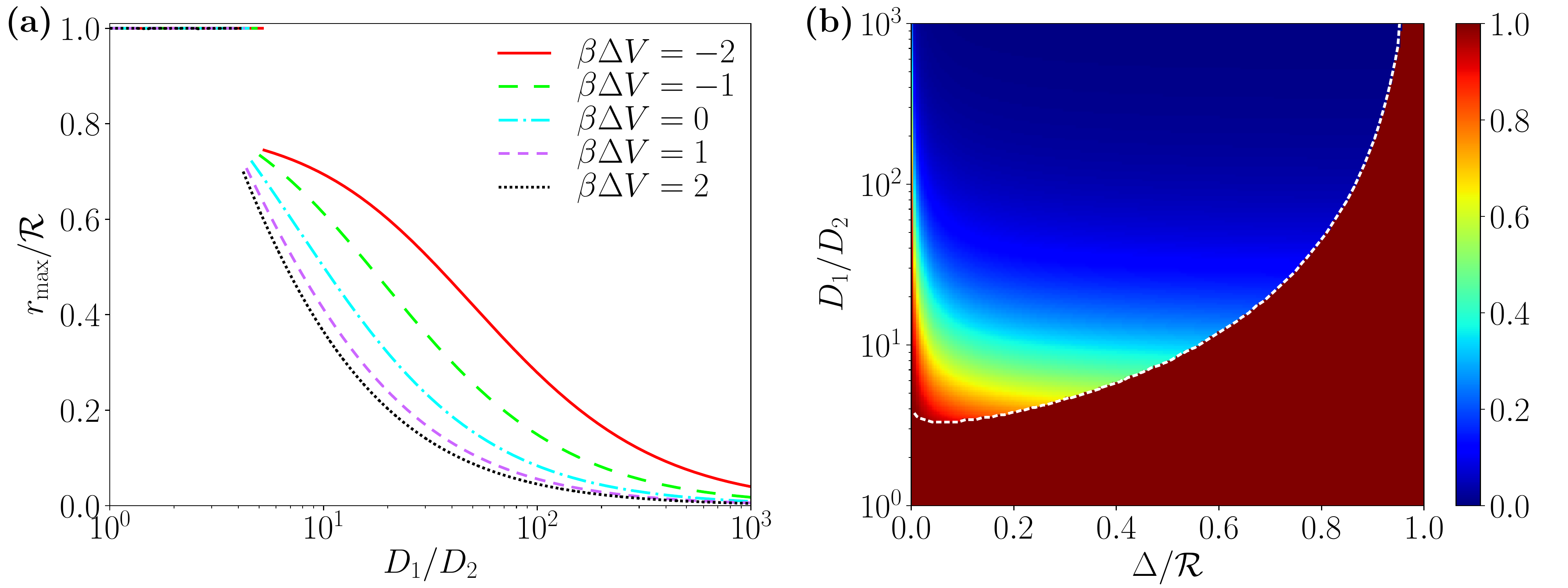}
\caption{(color online) Dimensionless distance of the MMFPT starting point to the center $r_{\rm max}/{\cal R}$, in 2d, {\bf (a)} as a function of $D_1/D_2$ for several potential difference $\beta \Delta V$, $\Delta=0.25{\cal R}$ and $\varepsilon=0.2$; {\bf (b)} as a function of $\Delta/{\cal R}$ and $D_1/D_2$ for $\beta \Delta V = 2$ and $\varepsilon=0.2$. The dashed line represents the transition line $\kappa_c(\Delta/{\cal R})$.} \label{figure2d_MMFPT2d_SI}
\end{center}
\end{figure}

Fig.~\ref{figure2d_MMFPT2d_SI} shows the numerical evaluation of this distance from the numerical solution of the MFPT obtained with the finite element method using the software package FreeFem++. Fig.~\ref{figure2d_MMFPT2d_SI}(a) displays the dependence of $r_{\rm max}$ on the ratio of diffusion constants $D_1/D_2$ for several potential difference $\beta \Delta V$, $\Delta= 0.25{\cal R}$ and $\varepsilon=0.2$. Increasing $D_1/D_2$, the distance $r_{\rm max}$ decreases discontinuously from ${\cal R}$ to $0$. For a small ratio of diffusion constants, $D_1/D_2 < \kappa_c$, the MMFPT starting position is located in the cortex, at the maximum distance to the escape region, i.e. $r_{\rm max} = {\cal R}$. For $D_1/D_2 = \kappa_c$, the MMPFT starting position jumps to the inner shell and is still located at the maximum distance to the escape region: $r_{\rm max} \lesssim {\cal R} - \Delta$. For $D_1/D_2 > \kappa_c$, $r_{\rm max}$ decreases continuously from ${\cal R} - \Delta$ to $0$. Moreover, $r_{\rm max}$ decreases with the potential difference $\beta \Delta V$, as well as the transition value $\kappa_c$.

Fig.~\ref{figure2d_MMFPT2d_SI}(b) shows the dependence of $r_{\rm max}$ on $\Delta/{\cal R}$ and $D_1/D_2$ for fixed $\beta \Delta V = 2$ and $\varepsilon=0.2$. The discontinuity is more pronounced for large cortex widths, since the distance at the transition is decreased by $\Delta/{\cal R}$. The dashed line represents the transition value of $D_1/D_2$: $\kappa_c$ as a function of $\Delta/{\cal R}$. $\kappa_c$ increases with $\Delta/{\cal R}$ and diverges for $\Delta ={\cal R}$, since $r_{\rm max}={\cal R}$ for the disk geometry.

We conclude then that the main conclusions made in Ref.~\cite{mangeat2019} stay qualitatively the same for both attractive and repulsive cortex.

%%%%%%%%%%%%%%%%%%
%%%%%%%%%%%%%%%%%%
%%% REFERENCES %%%
%%%%%%%%%%%%%%%%%%
%%%%%%%%%%%%%%%%%%

\end{document}